%% file: main.tex
\PassOptionsToPackage{dvipsnames}{xcolor}
\documentclass[twocolumn,twocolappendix]{aastex631}
\pdfoutput=1 %for arXiv submission
\usepackage[T1]{fontenc}
\usepackage[utf8]{inputenc}
\usepackage[english]{babel}
\usepackage{amsmath,amssymb,amstext}
\usepackage{apjfonts}
\usepackage{chngcntr}
\usepackage[figure,figure*]{hypcap}
\usepackage[hang]{footmisc}
\input{hyperlink-year-only-natbib-patch}
\allowdisplaybreaks
\graphicspath{{figures/}}

\newcommand*{\https}[1]{\href{https://#1}{\nolinkurl{#1}}}
\newcommand*{\http}[1]{\href{http://#1}{\nolinkurl{#1}}}

\defcitealias{Geha2017}{{Paper\,I}}
\defcitealias{Mao2021}{{Paper\,II}}
\defcitealias{saga3:Mao2024}{{Paper\,III}}
\defcitealias{saga4:Geha2024}{{Paper\,IV}}
\defcitealias{saga5:Wang2024}{{Paper\,V}}
\newcommand*{\paperone}{\citetalias{Geha2017}}
\newcommand*{\papertwo}{\citetalias{Mao2021}}
\newcommand*{\paperfour}{\citetalias{saga4:Geha2024}}
\newcommand*{\paperfive}{\citetalias{saga5:Wang2024}}

\newcommand*{\nlgpair}{five}
\newcommand*{\nlgsingle}{eight}
\newcommand*{\nhosts}{101}
\newcommand*{\nhoststotal}{203}

\newcommand*{\nsats}{378}
\newcommand*{\nsatssaga}{229}
\newcommand*{\nztotal}{75,704}
\newcommand*{\nzsaga}{45,699}
\newcommand*{\nzext}{30,005}
\newcommand*{\nzmmt}{19,795}
\newcommand*{\nzaat}{26,499}

\newcommand*{\ncandidates}{156}

\newcommand*{\ptrcompmin}{81\%}
\newcommand*{\ptrcompmed}{92\%}
\newcommand*{\skyarea}{84.7}

\newcommand*{\fnref}[1]{\ensuremath{^\text{\ref{#1}}}}
\newcommand*{\kms}{\text{km}\,\text{s}\ensuremath{^{-1}}}
\newcommand*{\msun}{\ensuremath{M_{\odot}}}
\newcommand*{\mstar}{\ensuremath{M_{\star}}}
\newcommand*{\perh}{\ensuremath{h^{-1}}}
\newcommand*{\LCDM}{\ensuremath{\Lambda}\text{CDM}}
\newcommand*{\mueff}{\ensuremath{\mu_{r_o,\text{eff}}}}
\newcommand*{\code}[1]{\ensuremath{\mathtt{#1}}}
\newcommand*{\EWHA}{\ensuremath{\text{EW}_{\text{H}_\alpha}}}

\shorttitle{The SAGA Survey.\ III.}
\shortauthors{Mao et al.}

\begin{document}
\title{The SAGA Survey.\ III.\ A Census of 101 Satellite Systems around Milky Way-mass Galaxies}

\author[0000-0002-1200-0820]{Yao-Yuan~Mao}
\affiliation{Department of Physics and Astronomy, University of Utah, Salt Lake City, UT 84112, USA}
\author[0000-0002-7007-9725]{Marla~Geha}
\affiliation{Department of Astronomy, Yale University, New Haven, CT 06520, USA}
\author[0000-0003-2229-011X]{Risa~H.~Wechsler}
\affiliation{Kavli Institute for Particle Astrophysics and Cosmology and Department of Physics, Stanford University, Stanford, CA 94305, USA}
\affiliation{SLAC National Accelerator Laboratory, Menlo Park, CA 94025, USA}
\author[0000-0002-8320-2198]{Yasmeen~Asali}
\affiliation{Department of Astronomy, Yale University, New Haven, CT 06520, USA}
\author[0000-0001-8913-626X]{Yunchong~Wang}
\affiliation{Kavli Institute for Particle Astrophysics and Cosmology and Department of Physics, Stanford University, Stanford, CA 94305, USA}
\affiliation{SLAC National Accelerator Laboratory, Menlo Park, CA 94025, USA}
\author[0000-0002-0332-177X]{Erin~Kado-Fong}
\affiliation{Department of Physics and Yale Center for Astronomy \& Astrophysics, Yale University, New Haven, CT 06520, USA}
\author[0000-0002-3204-1742]{Nitya~Kallivayalil}
\affiliation{Department of Astronomy, University of Virginia, Charlottesville, VA 22904, USA}
\author[0000-0002-1182-3825]{Ethan~O.~Nadler}
\affiliation{Carnegie Observatories, 813 Santa Barbara Street, Pasadena, CA 91101, USA}
\affiliation{Department of Physics \& Astronomy, University of Southern California, Los Angeles, CA, 90007, USA}
\affiliation{Department of Astronomy \& Astrophysics, University of California, San Diego, La Jolla, CA 92093, USA}
\author[0000-0002-9599-310X]{Erik~J.~Tollerud}
\affiliation{Space Telescope Science Institute, Baltimore, MD 21218, USA}
\author[0000-0001-6065-7483]{Benjamin~Weiner}
\affiliation{Department of Astronomy and Steward Observatory, University of Arizona, Tucson, AZ 85721, USA}
\author[0000-0002-4739-046X]{Mithi~A.~C.~de~los~Reyes}
\affiliation{Department of Physics and Astronomy, Amherst College, Amherst, MA 01002, USA}
\author[0000-0002-5077-881X]{John~F.~Wu}
\affiliation{Space Telescope Science Institute, Baltimore, MD 21218, USA}
\affiliation{Center for Astrophysical Sciences, Johns Hopkins University, Baltimore, MD 21218, USA}

\correspondingauthor{Yao-Yuan~Mao}
\email{yymao@astro.utah.edu}

\begin{abstract}
We present Data Release 3 (DR3) of the Satellites Around Galactic Analogs (SAGA) Survey, a spectroscopic survey characterizing satellite galaxies around Milky Way (MW)-mass galaxies.
The SAGA Survey DR3 includes \nsats{} satellites identified across \nhosts{} MW-mass systems in the distance range of 25--40.75\,Mpc, and an accompanying redshift catalog of background galaxies (including about 46,000 taken by SAGA) in the SAGA footprint of \skyarea{}\,sq.\,deg.
The number of confirmed satellites per system ranges from zero to 13, in the stellar mass range of $10^{6-10}\,\msun$.
Based on a detailed completeness model, this sample accounts for 94\% of the true satellite population down to $\mstar = 10^{7.5}\,\msun$.
We find that the mass of the most massive satellite in SAGA systems is the strongest predictor of satellite abundance; one-third of the SAGA systems contain LMC-mass satellites, and they tend to have more satellites than the MW.
The SAGA satellite radial distribution is less concentrated than the MW's, and the SAGA quenched fraction below $10^{8.5}\,\msun$ is lower than the MW's, but in both cases, the MW is within $1\sigma$ of SAGA system-to-system scatter.
SAGA satellites do not exhibit a clear corotating signal as has been suggested in the MW/M31 satellite systems.
Although the MW differs in many respects from the typical SAGA system, these differences can be reconciled if the MW is an older, slightly less massive host with a recently accreted LMC/SMC system.
\end{abstract}

\keywords{%
    \href{http://astrothesaurus.org/uat/290}{Companion galaxies (290)},
    \href{http://astrothesaurus.org/uat/416}{Dwarf galaxies (416)},
    \href{http://astrothesaurus.org/uat/594}{Galaxy evolution (594)},
    \href{http://astrothesaurus.org/uat/1378}{Redshift surveys (1378)},
    \href{http://astrothesaurus.org/uat/2171}{Galaxy spectroscopy (2171)}%
}

\section{Introduction}
\label{sec:intro}
The Satellites Around Galactic Analogs (SAGA) Survey is a spectroscopic galaxy survey focusing on nearby low-mass galaxies. Its primary science goal is to characterize the satellite populations, roughly down to the luminosity of the Leo I galaxy ($M_{r,o}<-12.3$), around 100 Milky Way (MW) analogs outside the Local Volume (at 25--40.75\,Mpc).
The SAGA Survey had two previous data releases. \citet[][hereafter \paperone{}]{Geha2017} introduced the survey and showed initial results based on 27 satellites around eight MW-like systems.
\citet[][hereafter \papertwo{}]{Mao2021} included 127 satellites identified around 36 MW-like systems.
In this work, we present Data Release 3 (DR3) of the SAGA Survey, which includes \nsats{} satellites identified around \nhosts{} MW-like systems. In addition, we also publish the full redshift catalog for galaxies in the SAGA fields in this data release.

\begin{figure*}[!tbp]
    \centering
    \includegraphics[width=\linewidth,clip,trim=0 1.1cm 0 0]{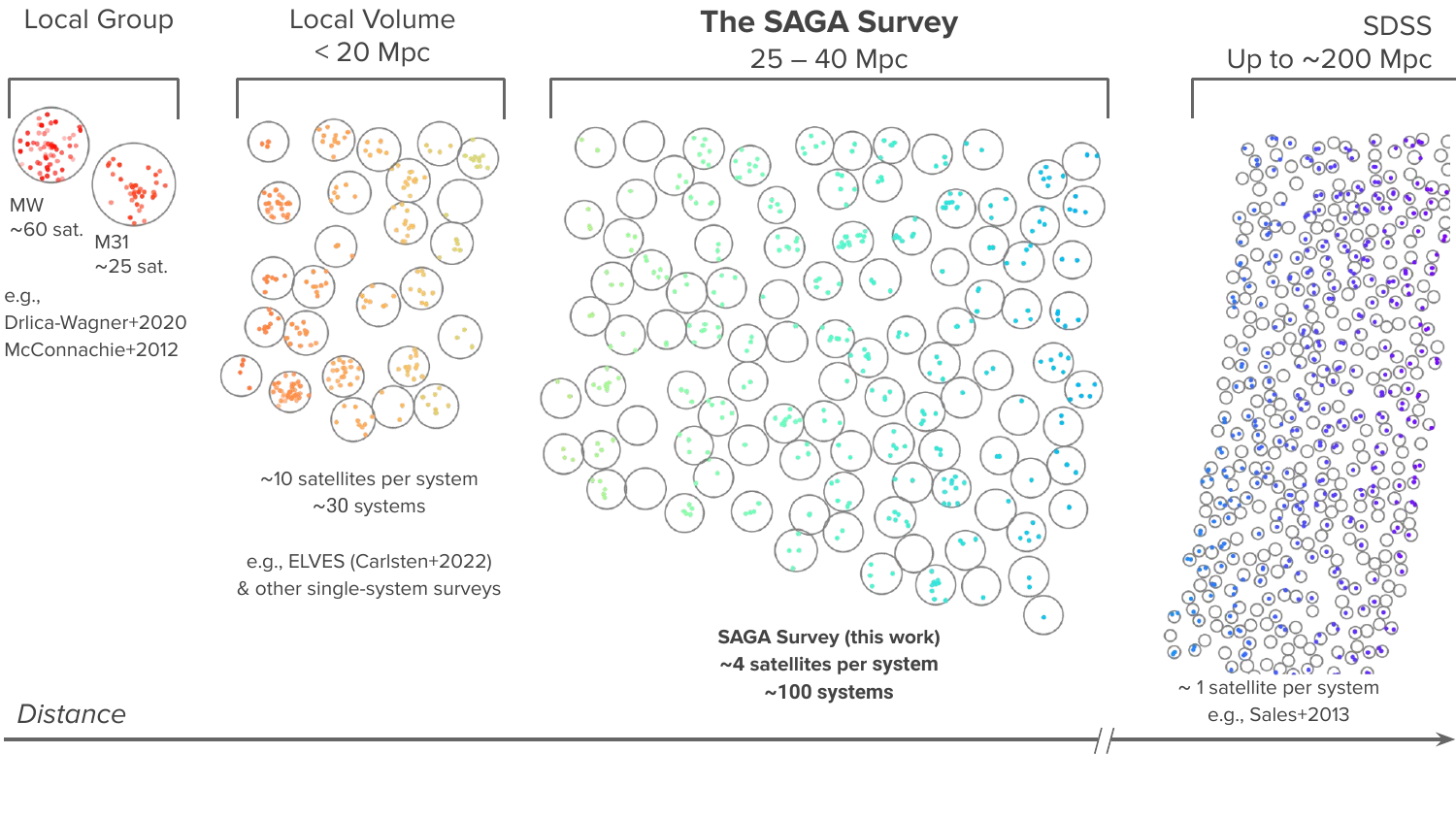}
    \caption{A schematic overview of the SAGA Survey in relation to other surveys of MW-mass satellite systems. Each gray circle represents a satellite system, and the colored points represent the satellites that are observable with current observational capacity. As the cosmological distance increases from left to right, the volume and the number of systems increase, but the depth and the number of satellites that can be surveyed in each system decrease. The SAGA Survey fills in the regime where we can survey hundreds of MW-mass satellite systems and still obtain a sizable satellite population per system.
    The distances and sizes shown are not to scale. }
    \label{fig:saga-survey}
\end{figure*}

In the past few years, significant progress has been made in characterizing MW-mass satellite systems beyond the MW and  M\,31. Generally, the distance to a system determines the technique one can use to confirm the satellite galaxies, the possible depth of the survey, and the number of systems available (see \autoref{fig:saga-survey}).  While the SAGA Survey mainly focuses on systems at 25--40.75\,Mpc, multiple satellite systems within the Local Volume ($<20$\,Mpc) have also been studied in recent years. For example, the Exploration of Local VolumE Satellites (ELVES) Survey uses surface brightness fluctuations to characterize 28 satellite systems within 12\,Mpc~\citep{2022ApJ...933...47C}.
There are multiple efforts to characterize individual nearby MW-like satellite systems, including NGC\,4258 \citep{Spencer2014}, M94 \citep{Smercina2018}, NGC\,3175 \citep{Kondapally2018}, NGC\,2950 and NGC\,3245 \citep{Tanaka2018}. Nearby satellite systems around galaxies of slightly higher masses have also been studied, such as the systems around M81 \citep{Chiboucas2013}, M101 \citep{Danieli:2017ApJ...837..136D,Bennet2019a,Carlsten:2019ApJ...878L..16C}, and Centaurus~A \citep{Crnojevic2019,Muller:1907.02012}.

Overall, these nearby satellite systems exhibit a wide range in both the number of satellites and the star-forming properties of these satellites. Comparing the MW and its bright satellite population with this statistical sample provides us with a new tool to study some unique, or even transient, aspects of the MW system.
For example, we now know that the MW has experienced specific events such as the recent accretion of the LMC and the SMC \citep[e.g.,][]{Kallivayalil2013} and a major merger that occurred around $z\sim2$, whose remnant is known as Gaia--Sausage--Enceladus \citep[GSE,][]{2018MNRAS.478..611B,2018Natur.563...85H}. These interactions can likely affect both the satellite abundance and the satellite star formation histories \citep[e.g.,][]{Simpson17:1705.03018,Hausammann:2019A&A...624A..11H}. The statistical study of nearby satellite systems provides a means to investigate the interplay between satellite populations and recent accretion and merger events.

The comparison between the MW and similar systems also provides important context for interpreting the results from simulations and models.  In particular, modern cosmological hydrodynamical simulations are often benchmarked against Local Volume observations when implementing sub-grid galaxy formation physics~\citep[e.g.,][]{Akins:2008.02805,Samuel2019a,2022MNRAS.511.1544F}. A larger sample of MW-mass satellite systems will enable a statistical comparison between observations and simulations. These recent surveys of MW-mass satellite systems such as SAGA offer rich datasets that permit detailed comparisons with hydrodynamic simulations, which are valuable for refining physical models of small-scale processes like tidal forces, ram pressure stripping, reionization, and star formation feedback mechanisms \citep[e.g.,][]{2021MNRAS.500.3776W,Samuel2022,TNG50_MWA,Greene+2023_ELVES,Karunakaran2023}.

In addition to galaxy formation physics, these satellite systems will provide critical tests of the Lambda Cold Dark Matter (\LCDM{}) paradigm. \citet{Nadler190410000,2020ApJ...893...48N,Nadler:2008.00022} have tested this approach with MW satellites to constrain the faint end of the galaxy--halo connection and alternative dark matter models, and \cite{Danieli221014233} derived complementary galaxy--halo connection constraints from Local Volume satellite populations.
While the SAGA Survey only probes the bright and classical satellite galaxies ($\mstar \sim 10^6$--$10^9\,\msun$), incorporating SAGA results in this type of analysis and comparing with dwarf galaxies in less dense environments \citep[e.g.,][]{2023A&A...676A..33H} will undoubtedly provide more powerful constraints by adding data in a regime that has been sparse to date and by breaking current degeneracies between galaxy formation and dark matter physics in this regime \citep[e.g.,][]{2401.10318}.

The main science of the SAGA Survey focuses on satellite galaxies; nevertheless, we have developed an effective targeting strategy to obtain spectroscopic redshifts for any low-redshift ($ z \lesssim 0.05$) galaxies. As SAGA Survey's survey depth goes down to $ r\sim 20.7$, it has collected a significant number of galaxy redshifts for galaxies that are low-redshift and low-mass ($10^7 < \mstar/\msun < 10^9$). This sample of galaxies is rare in past surveys, such as the Sloan Digital Sky Survey (SDSS; $r<17.77$; \citealt{Strauss2002,Sales2013}) and Galaxy And Mass Assembly (GAMA; $r < 19.8$; \citealt{Baldry2018:GAMA:DR3}), due to their limiting depths. The ongoing Dark Energy Spectroscopic Instrument (DESI; \citealt{1611.00036}) Survey will fill in this gap, but the SAGA Survey will likely have higher completeness within the SAGA Survey footprint for the foreseeable future.
These SAGA galaxy redshifts have already enabled a range of studies to train machine learning algorithms to select low-redshift or low-mass galaxy candidates beyond the SAGA Survey footprint \citep[e.g.,][]{xSAGA,DESI-LOWZ,SOM-lensing}.

This paper presents the overview of the SAGA Survey, DR3 data products, and a selection of the main science results. There are two accompanying papers: \citet[hereafter \paperfour{}]{saga4:Geha2024} will discuss in detail the star-forming properties of the SAGA satellites. 
\citet[hereafter \paperfive{}]{saga5:Wang2024} will present a detailed comparison with a theoretical prediction from an updated UniverseMachine galaxy--halo connection model \citep{2019MNRAS.488.3143B}.

%%%%%%%%%%%%%%%%%%%%%%%%%%%%%%%%%%%%%%%%%
%%%%%%%%%%%%%%%%%%%%%%%%%%%%%%%%%%%%%%%%%

This paper reviews our survey design in \autoref{sec:survey}, including host selection, photometric catalog production, and target selection.
We describe the redshift sources in \autoref{sec:spectroscopy}, including our own observations and redshifts from the literature. We present SAGA DR3 in \autoref{sec:dr3}, which includes both the satellite sample and an overall redshift catalog.
We then present the main science results in \autoref{sec:results}, including the stellar mass functions (SMFs), quenched fraction, radial distributions, satellite abundance, and co-rotating signals.
In \autoref{sec:discussion}, we discuss the MW satellite system in the context of the SAGA results, considerations for comparing SAGA results with simulations, and planned follow-up work.
Readers who are familiar with previous SAGA work or want to navigate the main results of this paper quickly can read \autoref{sec:summary}, the summary section, first.

As in \paperone{} and \papertwo{}, all distance-dependent parameters are calculated assuming $H_0 = 70$\,\kms\,Mpc$^{-1}$ and $\Omega_\text{M} = 0.27$.
Magnitudes and colors are extinction-corrected (denoted with a subscript `o,' e.g., $r_o$; using a combination of data from \citealt{schlegel98} and \citealt{Schlafly2011}; see \autoref{sec:photometry}).  Absolute magnitudes are $k$-corrected to $z=0$ using the formulas from \citet{Chilingarian2010}.

\section{The SAGA Survey}
\label{sec:survey}

\subsection{Overview of the SAGA Survey Design}

The primary goal of the SAGA Survey\footnote{\https{sagasurvey.org}\label{fn:saga}} is to characterize the satellite
galaxy populations around more than 100 Milky Way-mass galaxies down to an absolute magnitude of $M_{r,o} = -12.3$.
As of this data release, we have completed the survey for \nhosts{} systems.
To balance depth and volume, the SAGA Survey focuses on the regime slightly outside the Local Volume, selecting systems from 25 to 40.75~Mpc, as shown in \autoref{fig:saga-survey}.
The survey depth in apparent magnitude is roughly $r_o < 20.7$. The primary galaxies in these systems are selected by stellar masses and environments, which we detail in \autoref{sec:hosts}. We then identify potential satellite galaxy candidates based on photometric information, using the photometric catalogs from DESI Legacy Imaging Surveys DR9 (\autoref{sec:photometry}, \autoref{sec:targets}). We obtain redshifts for the candidates, including redshifts from existing literature, but mostly with new observations (\autoref{sec:spectroscopy}), to confirm whether the candidates have comparable redshifts as the primary galaxies.

\subsection{SAGA Host Selection and SAGA Footprint}
\label{sec:hosts}

%%%%%%%%%%%%%%%%%%%%%%%%%%%%%%%%%%%%%%%%%
\begin{figure}[!tbp]
    \centering
    \includegraphics[width=\linewidth]{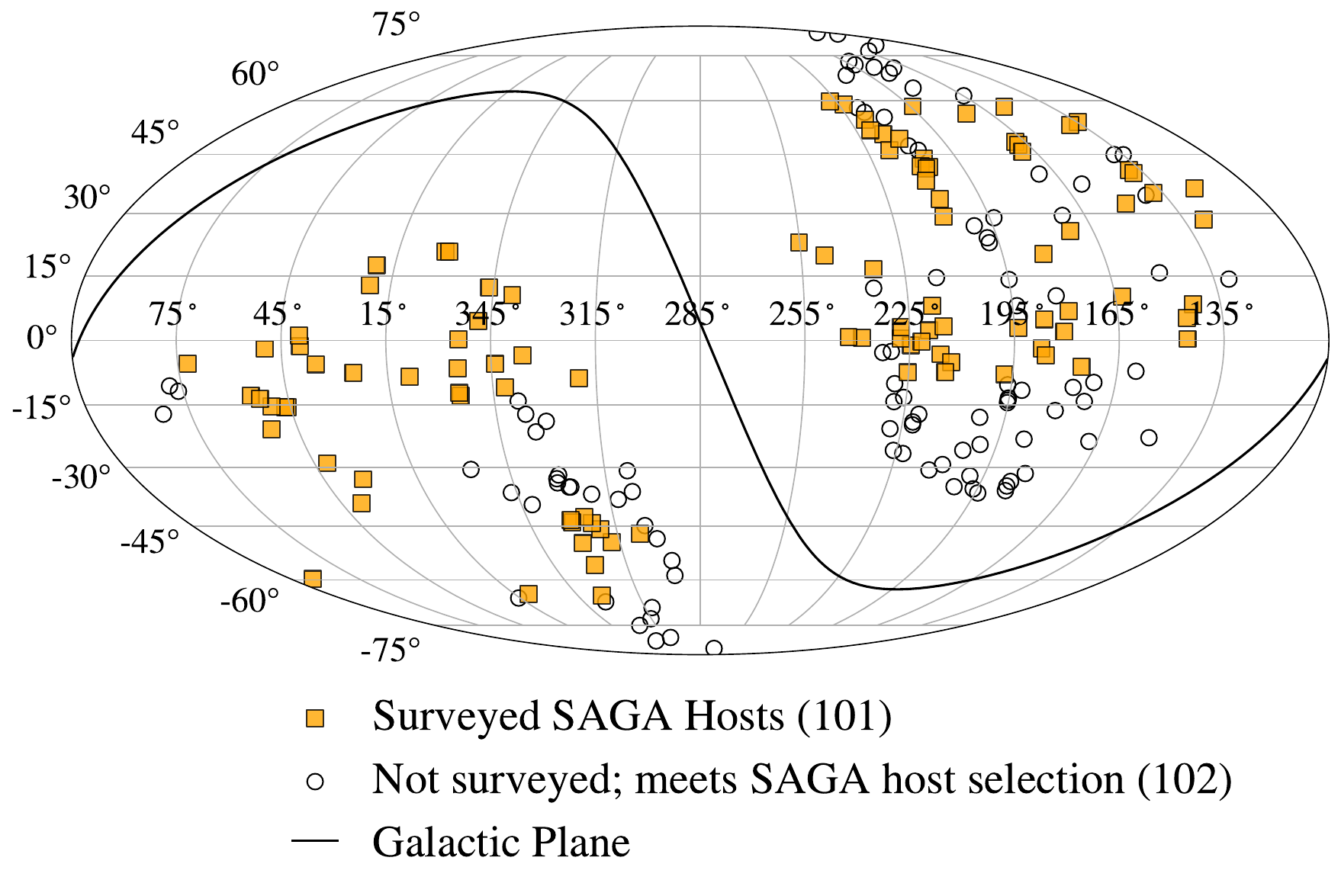}
    \caption{Sky distribution of \nhoststotal{} SAGA Milky Way analog systems as defined in \autoref{sec:hosts}. Orange-filled squares show the \nhosts{} systems that have been surveyed and are presented in this work. Empty circles show systems that match our host selection (i.e., from the ``Grand List'' with criteria \ref{eq:host-MK-cut} through \ref{eq:host-vel-cut} applied) but have not been surveyed.}
    \label{fig:hosts-coverage}
\end{figure}
%%%%%%%%%%%%%%%%%%%%%%%%%%%%%%%%%%%%%%%%%

The selection of the SAGA MW analogs (``hosts'') remains the same as described in Section 2.1 of \papertwo{}.  We briefly summarize the selection criteria here and refer the readers to \papertwo{} for details. Our selection is primarily based on distance, $K$-band luminosity, and local environment. The selection criteria are
\begin{subequations}%
\begin{align}
     & \text{Stellar Mass: }  -23 > M_K > -24.6; \label{eq:host-MK-cut}\\
     & \text{Stellar Foreground: }  |b| \geq 25^\circ;\\
     & \text{Stellar Foreground: } \text{\it Hp}^{(< 300\,\text{kpc})}_\text{brightest star} > 5; \label{eq:host-stellar-cut}\\
     & \text{Environment: }  K < K^{(< 300\,\text{kpc})}_\text{brightest gal.} - 1.6; \label{eq:env-cut-K}\\
     & \text{Environment: }  M_\text{halo} < 10^{13} \msun; \label{eq:env-cut-group}\\
     & \text{Distance: } 25-40.75\,\text{Mpc; and}  \label{eq:host-dist-cut}\\
     & \text{Distance: } v_\text{helio} > 1400\,\kms. \label{eq:host-vel-cut}
\end{align}\label{eq:hostlist_cuts}%
\end{subequations}%

The selection is applied on a nearby galaxy catalog that we constructed from HyperLEDA \citep{Makarov2014}\footnote{\http{leda.univ-lyon1.fr}}, and referred to as the ``Grand List'' in \papertwo{}. The Grand List is complete out to 60 Mpc in distance and down to $K = 11.9$ (corresponding to $M_K = -22$ at 60 Mpc).
The $K$-band magnitude and redshift measurements were adopted from the 2MASS Redshift Survey (2MRS; \citealt{2012ApJS..199...26H}). The distance measurements come from a combination of the NASA--Sloan Atlas (NSA v1.0.1; \citealt{Blanton2011}) and redshift-independent distance measurements compiled by HyperLEDA.
One of the stellar foreground selection criteria (Eq.~\ref{eq:host-stellar-cut}) uses the Hipparcos-2 catalog \citep{Hipparcos}.

We also include the halo mass ($M_\text{halo}$) estimates from the \citet{2017MNRAS.470.2982L} group catalog.
\citet{2017MNRAS.470.2982L} estimated these halo masses using a group finder based on the work of \citet{2005MNRAS.356.1293Y}, but with improvements to allow application on four galaxy redshift surveys.
The halo mass estimation is based on both the sum of the luminosities of member galaxies and the difference in luminosity between the central and member galaxies \citep{2016ApJ...832...39L}, and is unbiased with a typical uncertainty of $\sim$\,0.2\,dex when compared with hydrodynamical simulations.
In the SAGA redshift range ($z<0.015$), they estimated their group catalog is complete down to a halo mass of $10^{10.8}\,\msun$, which is sufficient for our use since we focus on MW-mass systems.
We use this halo mass estimate in Eq.~\ref{eq:env-cut-group} and our analysis.

\autoref{fig:hosts-coverage} shows the sky location of the SAGA hosts. Throughout this work, we use the term \textit{SAGA footprint} to refer to the sky area that is between 10 and 300 kpc in projection to one of the \nhosts{} SAGA hosts. The innermost 10\,kpc is excluded because the light from the host galaxy dominates that region, and we did not target any photometric sources in that region.
The outer boundary of 300\,kpc was designed to approximate the virial radius of MW-mass halos ($M_\text{halo} \sim 10^{12}\,\msun$), and corresponds to an angular radius of 25$'$--41$'$ at the distances of SAGA systems. Because we do not have precise measurements of the halo masses of the SAGA systems, we choose to use a fixed radius of 300\,kpc for all systems.
The SAGA footprint has a total sky area of \skyarea{}\,sq.\,deg.

\begin{figure}[!tbp]
    \centering
    \includegraphics[width=\linewidth]{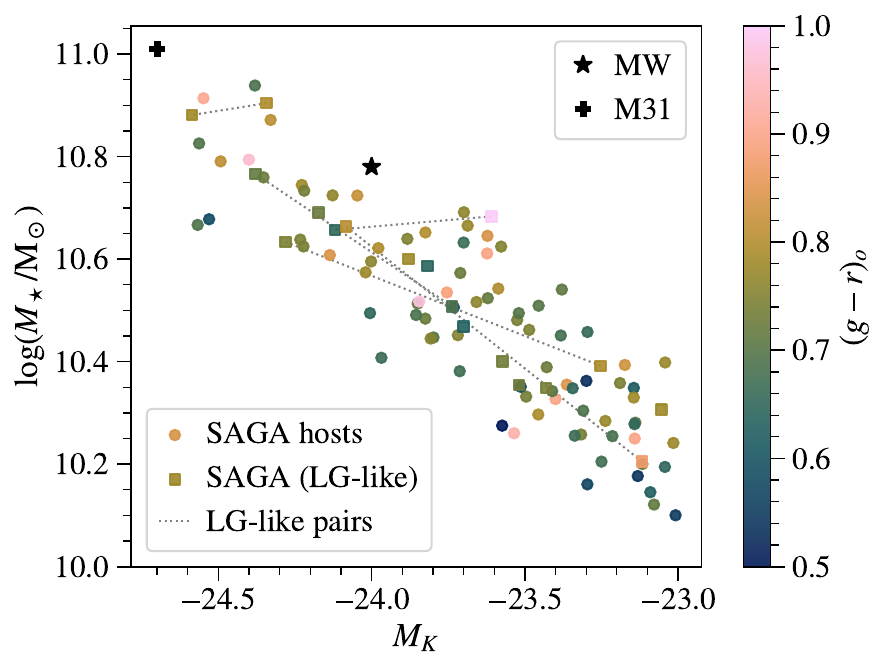}
    \caption{Distribution of the $K$-band absolute magnitudes ($x$-axis) \citep{2mass_xsc}, stellar masses ($y$-axis), and $g-r$ colors (color of the markers) of \nhosts{} SAGA host galaxies. The filled circle and square markers indicate regular and Local Group-like hosts, respectively (here Local Group-like means having exactly one MW-mass companion within 1\,Mpc). Filled square markers that are connected by a thin dotted line are Local Group-like pairs. The black star and plus sign show the luminosities and stellar masses of the Milky Way and M\,31, respectively.}
    \label{fig:hosts-properties}
\end{figure}

\autoref{fig:hosts-properties} shows the $K$-band luminosity and stellar mass of the SAGA hosts from the 2MASS Extended Source Catalog \citep{2mass_xsc}.
The Milky Way's luminosity \citep{2015ApJ...806...96L} and stellar mass \citep{2015MNRAS.451..149J}, and M\,31's luminosity \citep{2003AJ....125..525J} and stellar mass \citep{2015IAUS..311...82S} are also shown for comparison.

All SAGA hosts are less massive than M\,31, and most are less massive than the MW.
Among the \nhosts{} hosts presented here, we identify \nlgpair{} ``Local Group--like'' pairs (exactly two MW-mass hosts within 1\,Mpc of each other), and an additional \nlgsingle{} hosts that have precisely one MW-mass companion within 1\,Mpc with the companion not being one of the \nhosts{} SAGA hosts. We indicate these hosts in \autoref{fig:hosts-properties}.

The \nhosts{} SAGA hosts presented in this work are listed in \autoref{tab:hosts} (only schema is shown; table contents are available online in machine-readable format).

%%%%%%%%%%%%%%%%%%%%%%%%%%%%%%%%%%%%%%%%%
%%%%%%%%%%%%%%%%%%%%%%%%%%%%%%%%%%%%%%%%%

\subsection{DESI Imaging DR9 Photometry}
\label{sec:photometry}

After SAGA \papertwo{}, we fully updated our photometric catalog to the DESI Legacy Imaging Surveys \citep{Dey2019} Data Release 9 (DR9), released in January 2021. All SAGA systems are within the footprint of DESI Imaging DR9. This is a major upgrade compared to our first data release, which used only SDSS DR12, and to our second data release, which used a combination of SDSS DR14, Dark Energy Survey \citep[DES][]{1801.03181} DR1, and DESI Imaging DR6/7. In particular, DESI Imaging DR9 includes DECam data from other surveys, such as the DES.
DESI Imaging DR9 has a single-frame median $5\sigma$ point-source detection limit in the $r$ band of $r\sim 23.3$ \citep[slightly different for the two sub-surveys, DECaLS and BASS,][]{Dey2019}; this depth is comparable to DES and is about half magnitude deeper than SDSS.

Similar to the procedure followed in \papertwo{}, for each SAGA host, we obtain the DESI Imaging catalog within a 1$^\circ$ radius from the ``sweep'' files to ensure the virial radius (300~kpc) is always covered.
We use \code{FLUX} and \code{MW\_TRANSMISSION} to calculate de-reddened magnitudes, where \code{MW\_TRANSMISSION} is calculated using the coefficients in \citet{Schlafly2011} and the map from \citet{schlegel98}.\footnote{\https{www.legacysurvey.org/dr9/catalogs/\#galactic-extinction-coefficients}}
We only ingest objects with $r_o < 23$, $g_o < 22.5$, or $z_o < 22.5$ (any one condition).

DESI Legacy Imaging Surveys include three different surveys: the DECam Legacy Survey (DECaLS), the Beijing-Arizona Sky Survey (BASS), and the Mayall z-band Legacy Survey (MzLS). The flux calibration for each survey is done independently (with AB natural systems). Hence, there is a  slight but noticeable magnitude offset between DECaLS and BASS/MzLS, which we corrected with the following formula:
\begin{align}
r_\text{DECaLS} =  -0.0382 (g-r)_\text{BASS/MzLS} + 0.0108 + r_\text{BASS/MzLS}.
\end{align}
With this correction, the magnitudes used in SAGA DR3 all correspond to DECam filters, which differ slightly from the SDSS filters as used in \paperone{} and \papertwo{}.

We mostly use the \code{TYPE} flag to separate stars and galaxies: all objects whose $\code{TYPE} \neq \code{PSF}$ are considered galaxies. However, we do correct this flag for some bright stars that are mislabeled as galaxies and some faint galaxies that are mislabeled as stars. This correction procedure is implemented in the function \code{objects.build3.update\_star\_galaxy\_flag} in our code base.\footnote{\https{github.com/sagasurvey/saga/blob/master/SAGA/objects/build3.py}\label{fn:code-build3}}.   It involves checking the flux signal-to-noise value and specific photometric flags (\code{FITBITS}, \code{FRACFLUX}).

While the DESI Imaging DR9 catalog is fairly clean, there still exist some shredded objects. For shredded objects that are part of a large, typically spiral, galaxy, we use the better-measured galaxy radii from the Siena Galaxy Atlas 2020 \citep[SGA;][]{2023ApJS..269....3M} catalog to remove objects that are well within the light of a large galaxy.
For shredded objects from a faint source, we use a friends-of-friends algorithm\footnote{\https{github.com/yymao/FoFCatalogMatching}} to find close faint sources that potentially come from a single galaxy. We visually inspect those cases, and refit the photometry with GALFIT \citep{2002AJ....124..266P}.

Besides shredded objects, there are also spurious objects in the catalog. The most common situations include Galactic cirri and faint stars or quasars mistaken as large galaxies when the sky background is not correctly subtracted. We remove these spurious objects with both catalog properties (mainly the flux signal-to-noise values and the \code{FRACFLUX} flags; see the function \code{objects.build3.set\_remove\_flag} in our code base\fnref{fn:code-build3}). We also remove spurious objects via visual inspection. On average, we remove about 25 spurious objects per sq.\,deg., which is about 0.86\% of all galaxy sources down to $r_o = 20.7$.

In summary, our photometric catalog is based on the DESI Imaging DR9 catalog but with corrections for shredded, spurious, and misclassified objects. We ingest the photometric information, including grz and WISE (W1, ..., W4) bands, and morphology, including half-light semi-major axis (\code{SHAPE\_R}) and ellipticity parameters (\code{SHAPE\_E1}, \code{SHAPE\_E2}).

Because low-redshift, low-mass galaxies tend to also have lower surface brightness, we compare our photometric catalog with the SMUDGes catalog \citep{2023ApJS..267...27Z} to ensure that low-surface-brightness galaxies (LSBGs) and ultra-diffuse galaxies (UDGs) are not systematically missing from our photometric catalog. We found 82 objects in the SMUDGes catalog that are also in the SAGA footprint. Note that the SAGA footprint is a subset of the SMUDGes footprint, but only low-surface-brightness objects are included in the SMUDGes catalog. Among the 82 SMUDGes objects, only one (0912528+351112) is not present in the SAGA photometric catalog (which is based on the DESI Imaging DR9 catalog); however, this missing object has a magnitude ($r_o = 20.82$) that is fainter than SAGA's survey depth ($r_o < 20.7$). All other SMUDGes objects are present in the SAGA photometric catalog, and 14 of these are confirmed SAGA satellite galaxies (see definition in \autoref{sec:dr3-satellites}).

While we believe the SAGA photometric catalog is not missing LSBGs or UDGs given our intended survey depth ($r_o < 20.7$), the fitted flux for very low surface brightness objects could still be underestimated.  When comparing with the SMUDGes catalog, we found only 10 of the low-surface-brightness objects that are brighter than our survey depth in the SMUDGes catalog but fainter in the SAGA catalog (and the original DESI Imaging DR9 catalog).

\subsection{Galaxy Evolution Explorer photometry}
\label{sec:galex}
We also ingest photometry from  {\it Galaxy Evolution Explorer} \citep[GALEX;][]{Martin2005} and match these sources to the DESI Imaging DR9 catalog. The match is done by sky coordinate (the closest source within 3'' or within the optical radii).  To assess upper limits and improve photometric fits, we refit the GALEX photometry for all satellites and galaxies with $z < 0.2$ and $\mstar < 10^{10}\,\msun$.    See Section~2.3 of \paperfour{} for details.

\subsection{Target Selection}
\label{sec:targets}

In \papertwo{}, we improved our target selection strategy by adapting a two-part approach. We devote about half of our spectroscopic resources to obtain redshifts for almost all galaxy targets in the \textit{primary targeting region} in the photometric space. Based on the data we obtained in \papertwo{}, the primary targeting region contains nearly all the satellite galaxies in the SAGA systems.
We use the other half of our spectroscopic resources to target galaxies outside the primary targeting region.
The primary targeting region was defined in \papertwo{} as the following:
\begin{subequations}%
\begin{align}%
\mueff + \sigma_{\mu} - 0.7 \, (r_o - 14) &> 18.5, \label{eq:targeting-cuts-sb-r} \\
(g-r)_o - \sigma_{gr} + 0.06\,(r_o - 14) &< 0.9, \label{eq:targeting-cuts-gr-r}
\end{align}\label{eq:targeting-cuts}%
\end{subequations}%
where \mueff{} is the effective surface brightness (defined in \autoref{eq:sb}), $\sigma_{\mu}$ is the error on \mueff{}, and $\sigma_{gr} \equiv \sqrt{\sigma_g^2 + \sigma_r^2}$ is the error on the $(g-r)_o$ color.

Since \papertwo{}, we have continued this approach. For the \nhosts{} systems presented in this work, the median redshift rate (fraction of galaxy targets with confirmed redshifts) in the primary targeting region is \ptrcompmed{}. The host system with the lowest redshift rate in the primary targeting region still has a redshift rate of \ptrcompmin{}.

As we demonstrated in \papertwo{} (see the left panel of Figure 3 in \papertwo{}), our primary targeting region selects an almost complete sample for galaxies within $z < 0.015$. The typical redshift for galaxies in the primary targeting region is around $z=0.1$. The redshifts we have obtained since \papertwo{} continue to confirm the effectiveness of this strategy. Nevertheless, we also continue to obtain redshifts outside the primary targeting region, especially in the region close to the boundaries of the primary targeting region.   As shown in \cite{DESI-LOWZ}, low-redshift galaxies ($z < 0.05$) tend to be found near these boundaries.

In summary, this target selection strategy allows us to reach a high redshift rate in the primary targeting region, where almost all $z < 0.015$ galaxies reside. At the same time, we sample the rest of the photometric space, and these redshifts allow us to build a model to estimate how likely a galaxy is to be a satellite, given its photometric properties. We describe this model in \autoref{app:model}, and we use it to correct for our survey incompleteness (here, by incompleteness we refer to the redshift rate less than 100\%).

\section{Spectroscopic Data}
\label{sec:spectroscopy}

Including our previous data releases, the SAGA team has obtained in total \nzsaga{} galaxy redshifts (\autoref{sec:spec-multifiber} and \ref{sec:spec-single}) in the SAGA footprint.
Combined with \nzext{} galaxy redshifts from the literature (\autoref{sec:spec-literature}), we have compiled a catalog of \nztotal{} galaxy redshifts in the SAGA footprint.
We describe the association of spectroscopic and photometric objects in \autoref{sec:spec-combine}.

\subsection{Multifiber Spectra: Anglo-Australian Telescope and MMT}
\label{sec:spec-multifiber}

Building on our work in \paperone{} and \papertwo{}, we obtained data from both the MMT/Hectospec and Anglo-Australian Telescope (AAT)/2dF multifiber spectrographs.   In both cases, data reduction and redshift inspection are identical to that described in \papertwo{}.

MMT/Hectospec provides 300 fibers over a 1$^\circ$ diameter field of view  \citep{2005PASP..117.1411F}.   Each fiber has a diameter of $1.''5$.   We used Hectospec with the 270 line\,mm$^{-1}$ grating, resulting in wavelength coverage of 3650--9200\,\AA{} and spectral resolution of 1.2\,\AA{}\,pixel$^{-1}$ ($R\sim1000$).  Typical redshift errors are 60\,\kms.   SAGA DR3 contains \nzmmt{} unique galaxy redshifts obtained with MMT/Hectospec between 2013 and May 2022.

AAT/2dF deploys 400 fibers over a 2$^\circ$ diameter field \citep{2dfinstrument}.  Each fiber has a diameter of $2.''1$.   We used the 580V and 385R gratings in the blue and red arms, respectively, both providing a resolution of $R = 1300$ (between 1 and 1.6\,\AA{}\,pixel$^{-1}$) over a maximum wavelength range of 3700--8700\,\AA{}. Typical redshift errors are 60\,\kms.  SAGA DR3 contains \nzaat{} unique galaxy redshifts obtained with AAT/2dF between 2014 and Jan 2023.

\subsection{Single-slit Spectra: Palomar, Keck, and SALT}
\label{sec:spec-single}

Single-slit spectroscopy is used to measure redshifts for galaxies that were missed in the multifiber work due to fiber collisions, low signal-to-noise, or other issues.   Our primary single-slit facility is the 200-inch Hale Telescope at the Palomar Observatory and the Double Spectrograph \citep[DBSP; ][]{oke1982}.    We used the 316/7500 red channel grating  and 600/4000 blue channel grating, resulting in wavelength coverage of 3800--9500\,\AA{} and spectral resolution of 1.1 and 1.5\,\AA{}\,pixel$^{-1}$, respectively, for the blue and red sides.   The data were reduced using {\tt PypeIt} \citep{pypeit2020}; redshifts were measured using a modified version of {\tt Marz} \citep{10.1016/j.ascom.2016.03.001}.    SAGA DR3 contains 458 unique redshifts obtained with Palomar/DBSP between 2019 and Sept 2022.

We obtained 28 redshifts using the Deep Extragalactic Imaging Multi-Object Spectrograph \citep[DEIMOS; ][]{faber03a} on the Keck-II 10-meter telescope. Data were obtained between 2016 and 2022 with the 1200G grating with wavelength coverage 6500--8900\,\AA{} and 0.33\,\AA{}\,pixel$^{-1}$.   An additional 12 redshifts were obtained between 2020--2021 with the Southern African Large Telescope (SALT)'s RSS spectrograph using the 900 l/mm grating \citep{salt_telescope, salt_instrument}.

\subsection{External Spectroscopic and HI Data Sets}
\label{sec:spec-literature}

We include spectroscopic data in the regions of our hosts from a variety of publicly available surveys. While these surveys do not specifically target the very low-redshift ($z<0.015$) galaxies that the SAGA Survey focuses on, they help reduce the number of targets that require redshifts for our survey.

We list these surveys in decreasing order of the number of redshifts in the SAGA footprint:
SDSS \citep[DR16;][]{1912.02905},
DESI \citep[EDR1;][]{2306.06308},
GAMA \citep[DR3;][]{Baldry2018:GAMA:DR3},
the Prism Multi-object Survey \citep[PRIMUS DR1;][]{2011ApJ...741....8C,2013ApJ...767..118C},
NSA \citep[v1.0.1;][]{Blanton2011},
the VIMOS Public Extragalactic Redshift Survey \citep[VIPERS PDR2;][]{1611.07048},
the WiggleZ Dark Energy Survey \citep{WiggleZ},
the 2dF Galaxy Redshift Survey \citep[2dFGRS;][]{2dFGRS},
the HectoMAP Redshift Survey \citep[DR2;][]{2210.16499},
the HETDEX Source Catalog \citep{2301.01826},
the 6dF Galaxy Survey \citep[6dFGS;][]{Jones2004,6dF},
the Hectospec Cluster Survey, \citep[HeCS;][]{Rines2013},
the Australian Dark Energy Survey \citep[OzDES, Data Release 2;][]{OzDES},
the 2-degree Field Lensing Survey \citep[2dFLenS;][]{2dFlens}, and
the Las Campanas Redshift Survey \citep[LCRS;][]{1996ApJ...470..172S}.

Across the full SAGA footprint around \nhosts{} hosts, SDSS contributed about 13,000 galaxy redshifts. DESI, GAMA, PRIMUS, and NSA each contribute around 3,500--5,000 galaxy redshifts. The remaining surveys listed above each contribute hundreds to 1,500 galaxy redshifts, and around 6,000 in total.
We only ingested literature redshifts with high confidence; however, in some rare cases, a literature redshift can be inaccurate.
When a literature redshift indicates a satellite galaxy (about 40\% of the SAGA satellites have literature redshifts), we retarget the galaxy to verify the literature velocity measurement and to ensure homogeneity, but we do not count this case as a SAGA discovery.

We also ingested \textsc{Hi} data for galaxies that are covered by
the Arecibo Legacy Fast ALFA Survey \citep[ALFALFA;][]{ALFALFA100} and
the FAST all sky HI Survey \citep[FASHI;][]{2312.06097}.
Due to the large beam size (3$'$) of \textsc{Hi} measurement, we use the photometric counterpart information available in these catalogs to assist the match.
When the photometric counterpart information is not available or has a low confidence, we adopt the \textsc{Hi} measurements if a nearby source in the SAGA catalog has a consistent redshift.
We also matched our catalog to the HI Parkes All-Sky Survey \citep[HIPASS;][]{2004MNRAS.350.1195M}; however, due to an even larger beam size (15$'$), we only search around confirmed satellites and included HIPASS data for two of the SAGA satellites.

\subsection{Combining Spectroscopic and Photometric Data Sets}
\label{sec:spec-combine}

As the photometric catalog has been updated a few times during the course of the SAGA Survey, we combine the spectroscopic  and photometric measurements by matching the sky coordinates.
For each spectroscopic measurement, we search the photometric catalog within a 20$''$ radius.
Given the SAGA Survey depth, the photometric sources are not too crowded, and in most cases, only one unique match exists.
When multiple photometric sources are possible matches, we identify the best match based on the luminosities and sizes of those photometric objects. The specific order by which a photometric source is selected as the best match to a spectroscopic measurement is described in the list
\code{objects.build3.SPEC\_MATCHING\_ORDER} in the SAGA code base\fnref{fn:code-build3}.

In addition, if one object has multiple spectroscopic measurements with consistent redshifts, we opt to use the redshift value from the most confident measurement (usually based on the quality flag) and, when the quality flag is the same, from the measurement made with the telescope with a larger aperture. This procedure is described in the function
\code{objects.build2.}\linebreak\code{match\_spectra\_to\_base\_and\_merge\_duplicates} in the SAGA code base\footnote{\https{github.com/sagasurvey/saga/blob/master/SAGA/objects/build2.py}}.

\section{SAGA DR3}
\label{sec:dr3}

\subsection{SAGA Redshift Catalog}
\label{sec:dr3-redshifts}

%%%%%%%%%%%%%%%%%%%%%%%%%%%%%%%%%%%%%%%%%
\begin{figure}[!tbp]
    \centering
    \includegraphics[width=\linewidth,clip,trim=0 0 0 0]{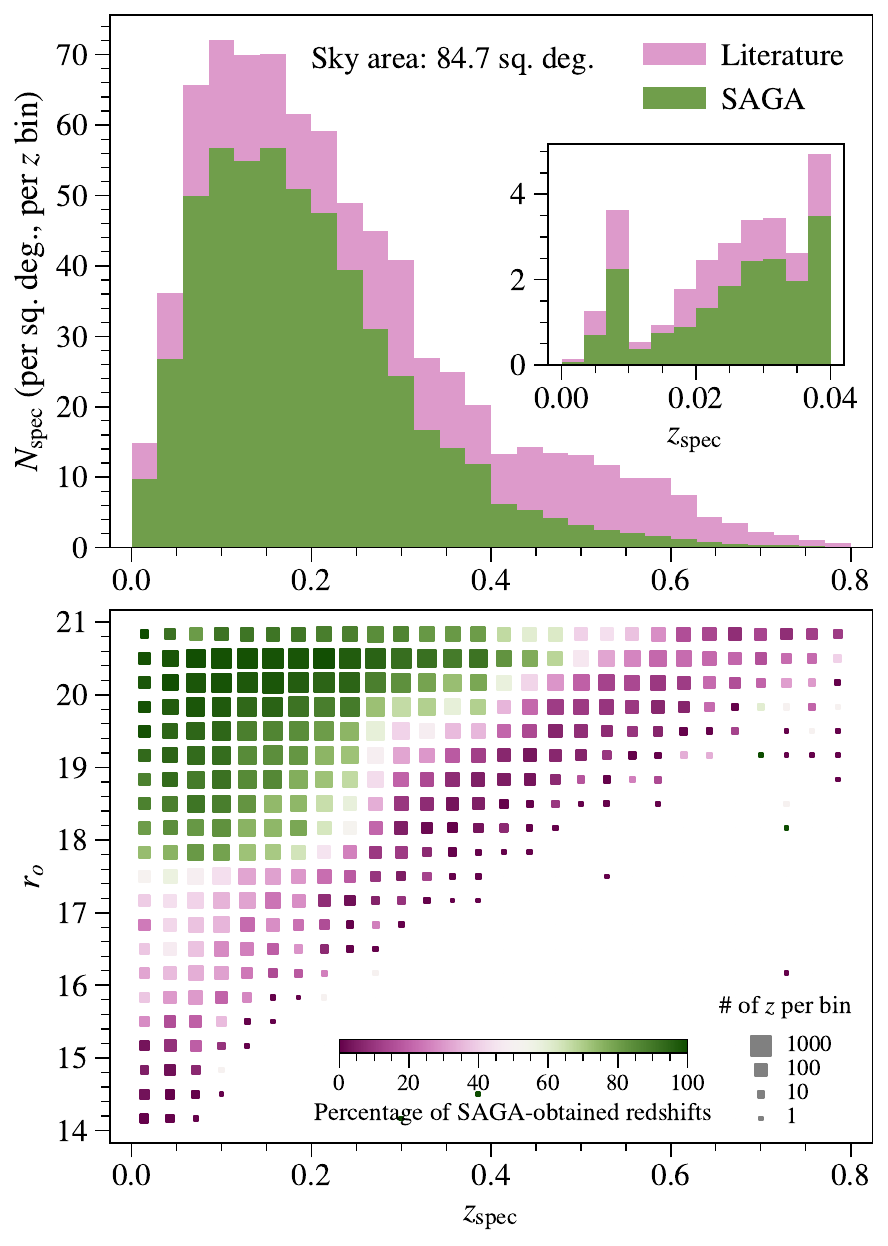}
    \caption{\textbf{Top}: A stacked histogram of confirmed redshifts of galaxies in the SAGA footprint (defined in  \autoref{sec:hosts}). The olive and pink colors in the stacked histogram show the number of spectra collected by the SAGA team and from the literature, respectively. The inset shows the same data, but zoomed in on the $z<0.04$ region. \textbf{Bottom}: The percentage of galaxy redshifts first obtained by the SAGA Survey, in bins of redshift ($x$-axis) and $r$-band magnitude ($y$-axis), represented by the color of the cell (darker olive indicates more SAGA-obtained redshifts, darker pink indicates more literature redshifts). The size of each cell represents the total number of redshifts in the corresponding bin.}
    \label{fig:z_distribution}
\end{figure}
%%%%%%%%%%%%%%%%%%%%%%%%%%%%%%%%%%%%%%%%%

We present \nztotal{} galaxy redshifts around \nhosts{} SAGA host galaxies.
About 60,000 of these redshifts are for galaxies brighter than the SAGA Survey's nominal magnitude limit, $r_o = 20.7$, but we include the fainter galaxies' redshifts in the catalog when available.
\autoref{fig:z_distribution} shows the redshift distribution of this catalog, with redshift sources and magnitude highlighted.
Overall, the \nzsaga{} SAGA redshifts (purple histogram) peak around $z=0.2$. Most SAGA-obtained redshifts are for faint galaxies $19 < r_o < 20.7$. Conversely, almost all redshifts for dim, low-redshift galaxies ($19 < r_o < 20.7$ and $z<0.2$) were obtained by SAGA.
The redshift catalog is provided in a machine-readable format, with its schema listed in \autoref{tab:redshifts}.

\subsection{Derived Properties}
\label{sec:dr3-properties}

In addition to the photometric properties available from Legacy Surveys, we also include the following derived properties in our galaxy redshift catalog.
\begin{enumerate}
    \item Radius and surface brightness: For most galaxies, the circularized half-light radius is derived directly from the axis ratio and semi-major axis from the Legacy Surveys catalog measured in the $r$ band. However, for galaxies available in the Siena Galaxy Atlas, we use $\code{D26} / 3$ as the half-light radius (one-third of the isophotal radius at 26\,mag\,arcsec$^{-2}$). We also use GALFIT to re-fit the photometry for a handful of galaxies that clearly have a bad photometric fit in the original catalog by visual inspection.  We then calculated the effective surface brightness with the following formula:
    \begin{equation}
        \label{eq:sb}
        \mueff = r_o + 2.5 \log \left(2\pi R_{r_o,\text{eff}}^2\right).
    \end{equation}
    \item Stellar mass: As in \papertwo{}, we estimate stellar mass using the \citet{bell2003} relation, but calibrated the relation to recent estimates from \citet{Zibetti2009} and \citet{Taylor2011}. We also assume a \citet{kroupa2001} initial mass function and an absolute solar $r$-band magnitude of 4.65 \citep{Willmer2018}. The resulting stellar mass conversion is
    \begin{equation}
        \label{eq:sm}
        \log[\mstar/\msun] = 1.254 + 1.098 \,(g-r)_o - 0.4 \, M_{r,o},
    \end{equation}
    where $M_{r,o}$ is the absolute magnitude $k$-corrected to $z=0$ using \citet{Chilingarian2010}.
    We expect a systematic error of 0.2\,dex on the estimated stellar mass, which is larger than the random errors due to propagating errors in photometry and distance.
\end{enumerate}

For SAGA satellite galaxies (\autoref{sec:dr3-satellites}), we also measure the following quantities:
\begin{enumerate}
    \item Star Formation Rate (SFR): we calculate the star formation rate based on the both H$\alpha$ equivalent width (\EWHA) and near-UV (NUV) flux. See Section~2.4 of \paperfour{} for details.
    \item ``Quenched'' flag: we label a galaxy as quenched if it has no significant H$\alpha$ emission ($(\EWHA - \sigma_{\EWHA})  < 2\,\mbox{\AA}$) and if it has a specific SFR (sSFR) in NUV below $-11\,\msun\,\text{yr}^{-1}$. See Section~3.1 of \paperfour{} for more details.
    \item Line flux measurements: we provide the fluxes and uncertainties of selected emission lines for SAGA satellites. See Section~2.2 of \paperfour{} and Section~3.1 of \citet{2401.16469} for more details.
\end{enumerate}

\subsection{SAGA Satellites}
\label{sec:dr3-satellites}

%%%%%%%%%%%%%%%%%%%%%%%%%%%%%%%%%%%%%%%%%
\begin{figure}[!tbp]
    \centering
    \includegraphics[width=\columnwidth,clip,trim=0 0 0 0]{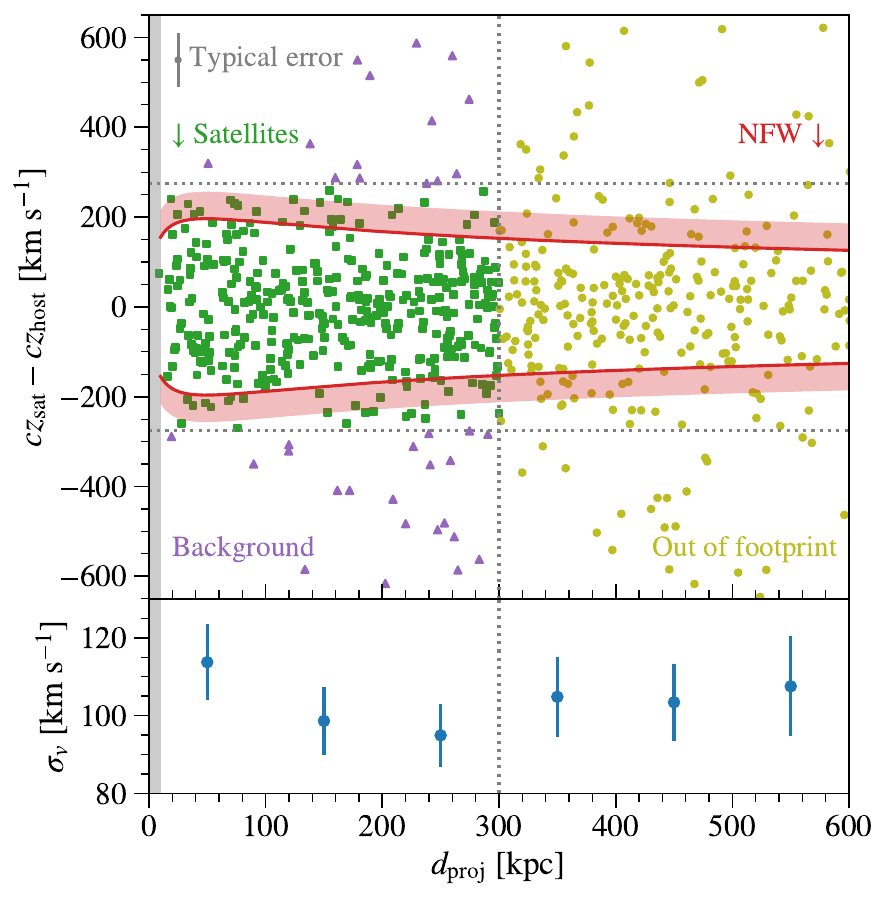}
    \caption{\textbf{Top}: Heliocentric-corrected velocities ($cz$) and projected distances of galaxies in the proximity of SAGA host galaxies. The velocity and distance plotted for each galaxy are calculated with respect to their individual hosts.
    The gray errorbars in the upper left corner show the typical velocity errors.
    We define ``satellites'' (green squares) as galaxies that lie in the SAGA footprint (within 300\,kpc in projection) and within $\pm 275\,\kms$ of their host galaxies; these criteria are indicated by dotted lines. Galaxies in the SAGA footprint that do not meet the velocity criterion are considered to be ``background'' galaxies (purple triangles). In this plot, we also show galaxies that are beyond 300\,kpc in projection (i.e., outside the SAGA footprint, yellow circles).
    The red solid lines show the circular velocity of an NFW profile with $M_\text{vir} = 2\times 10^{12} \msun{}$ and $c=10$. Note that our satellite definition does not guarantee the ``satellites'' are gravitationally bound to the host.
    The survey did not attempt to obtain any redshifts within 10\,kpc of the host galaxy (indicated as the gray vertical band on the left).
    \textbf{Bottom}: The velocity dispersion of objects within the satellite velocity selection window, in bins of projected distance.}
    \label{fig:velocity_distance}
\end{figure}
%%%%%%%%%%%%%%%%%%%%%%%%%%%%%%%%%%%%%%%%%

%%%%%%%%%%%%%%%%%%%%%%%%%%%%%%%%%%%%%%%%%
\begin{figure}[!tbp]
    \centering
    \includegraphics[width=\columnwidth,clip,trim=0 0 0 0]{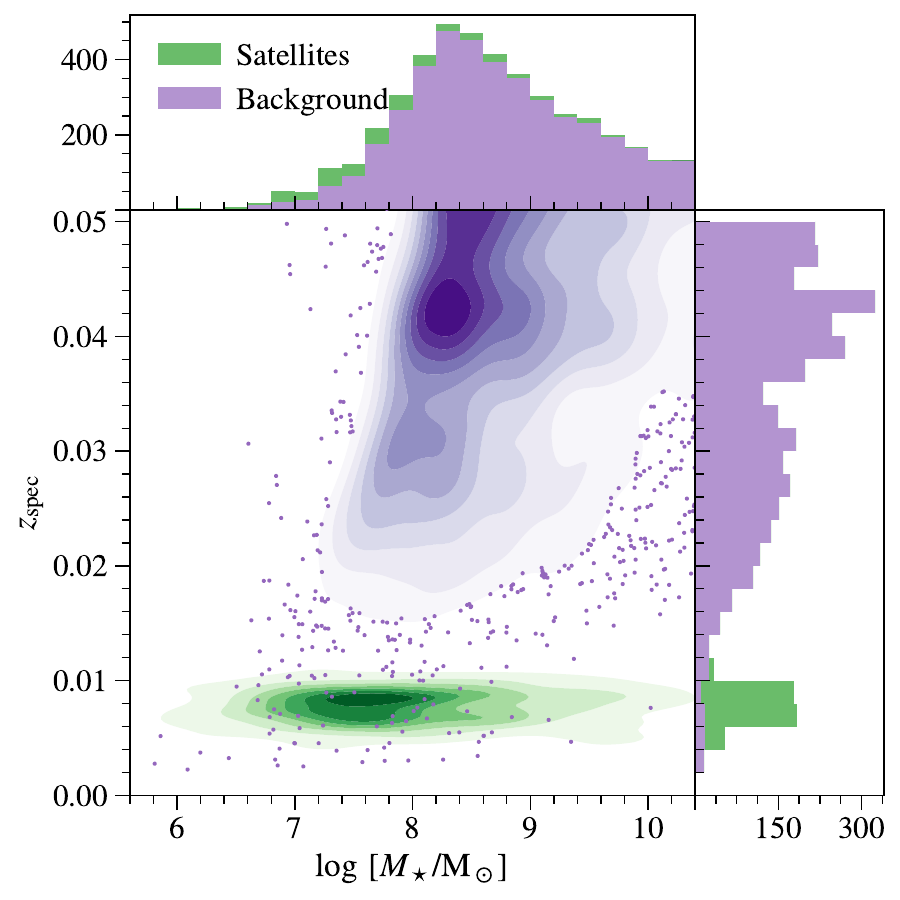}
    \caption{Distribution of stellar masses ($x$-axis; shown as $\log [\mstar/\msun{}]$) and spectroscopic redshifts ($y$-axis) of galaxies in the SAGA footprint and within $z < 0.05$. The ``satellites'' (as defined in \autoref{sec:dr3-satellites}) are shown in green and all other galaxies (``background'') are shown in purple. The main panel shows the joint distribution with contours indicating density and points indicating individual galaxies in low-density regions. The two side panels show the stacked histograms of stellar masses (top panel) and redshifts (right panel).}
    \label{fig:redshift-mass}
\end{figure}
%%%%%%%%%%%%%%%%%%%%%%%%%%%%%%%%%%%%%%%%%

As in \papertwo{}, we define a ``satellite'' as a galaxy that satisfies the following criteria: (i) fainter than the host galaxy, (ii) within 300 kpc in projection to the host galaxy, and (iii) has a heliocentric velocity that is within 275~\kms{} of the host galaxy. From the SAGA redshift catalog around \nhosts{} hosts, we identified \nsats{} satellites in total, with \nsatssaga{} whose redshifts were first obtained by the SAGA Survey.
Galaxies that do not meet these criteria are considered to be ``background'' galaxies (a handful of these galaxies are actually foreground galaxies; however, we will use the term background galaxies to refer to any non-satellites).

\autoref{fig:velocity_distance} demonstrates this definition, with the satellite galaxies being in the middle left part of the plot.  It is important to note that this observationally motivated satellite definition is not an attempt to strictly select objects that lie within the three-dimensional physical virial radius of their dark matter host halo, nor to strictly select objects that are gravitationally bound to the host. With only line-of-sight velocities and sky positions available, we chose a definition that is straightforward to implement in both observational and simulation data.

With simulations, we can estimate the fraction of interlopers (galaxies that sit outside the three-dimensional radius in the line-of-sight direction) and potential fly-by galaxies \citep[galaxies that sit within the three-dimensional radius but have a speed greater than the escape velocity;][]{Sinha11031675,An191111782}. Based on gravity-only simulations, the SAGA satellite definition gives an average interloper fraction of 30\% for MW-mass halos. This fraction varies with projected distance, and drops to 15\% for galaxies within 150~kpc in projection, as demonstrated in \papertwo{}.
Note that the radius considered here is always fixed to be 300\,kpc, even though the individual SAGA systems could have different virial radii, which are not measurable observationally.

As for potential fly-by galaxies, \autoref{fig:velocity_distance} highlights the fact that a small fraction ($\sim$4\%) of SAGA satellites may be unbound, indicated by purple squares (satellites) sitting outside the red solid bands, which represent the circular velocity of an Navarro--Frenk--White (NFW; \citealt{nfw1996}) profile with $M_\text{vir} = 2\times 10^{12} \msun{}$ and $c=10$ (red solid lines) plus the typical spectroscopic velocity error of 60\,\kms{} in our measurement. The fraction of unbound objects observed is consistent with the gravity-only simulations when we use the SAGA satellite definition to select the halos in the simulations.

\autoref{fig:velocity_distance} also shows the velocity dispersion measured from all SAGA satellites. Satellites are binned in projected distances.   For each bin, we fit a single Gaussian distribution, with its mean and variance being the free parameters, convolved with a constant velocity measurement error of 60\,\kms{}. Because we use all satellites to measure the velocity dispersion profile, it represents the stacked profile where systems with more satellites are weighted more.    The velocity dispersion declines with the projected radius to the virial radius, suggesting we are measuring a true dynamical mass.  The increase beyond 300\,kpc is likely due to increased contribution from interlopers.  We are not able to measure the velocity dispersion for individual systems due to the small number of satellites in each system.

The \nsats{} SAGA satellites with confirmed redshifts are listed in \autoref{tab:sats} (only schema is shown; table contents are available online in machine-readable format).

%%%%%%%%%%%%%%%%%%%%%%%%%%%%%%%%%%%%%%%%%
\begin{figure*}[!tbp]
    \centering
    \includegraphics[width=\linewidth,clip,trim=0 0 0 0]{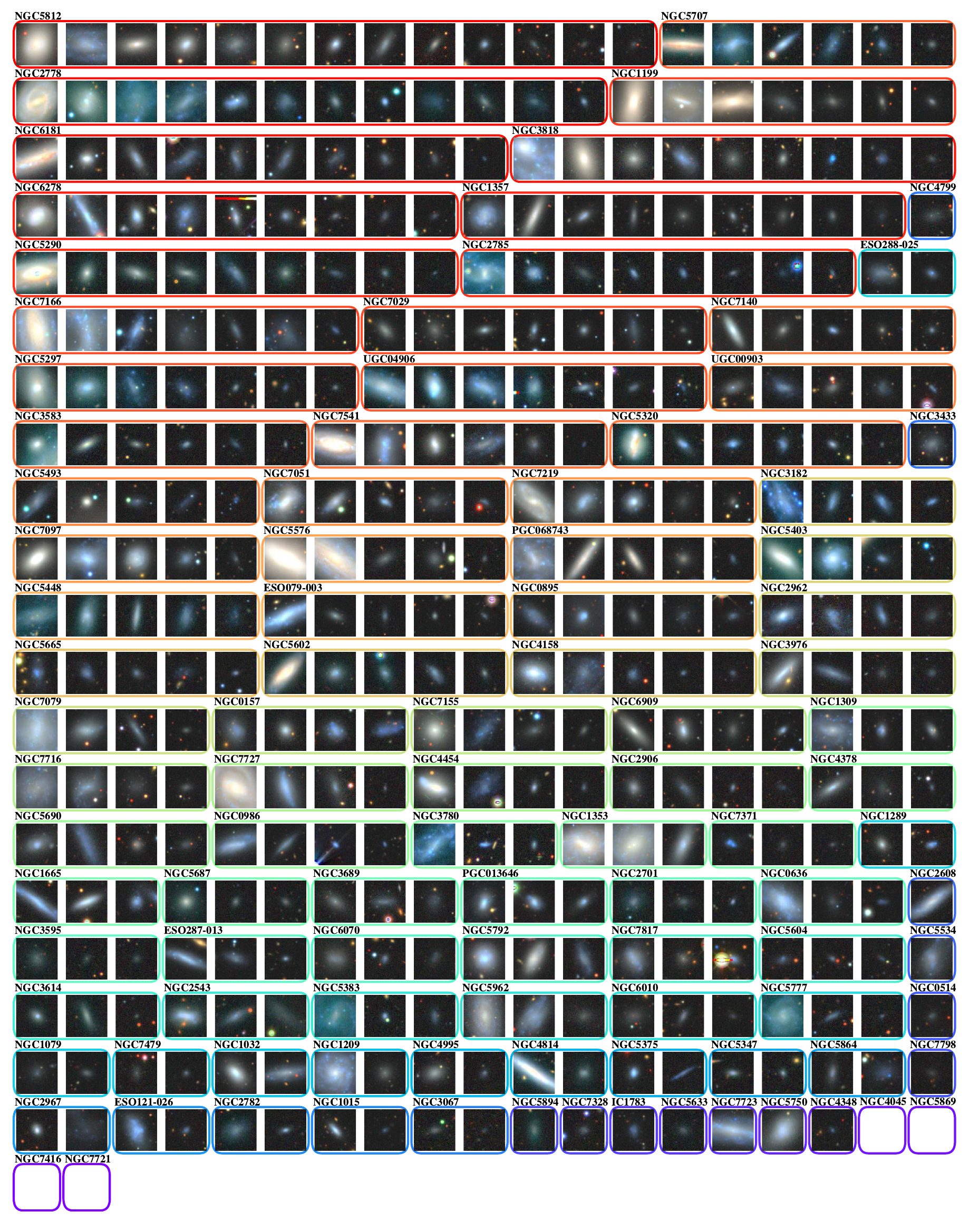}
    \caption{Images of \nsats{} satellites, sorted by $M_{r,o}$ within each of the \nhosts{} complete systems that have satellites. Four of the \nhosts{} systems have the same level of survey completeness but do not have any confirmed satellites. The images of host galaxies are not included. The colors of the frames are assigned by the number of satellites.
    Images are taken from DESI Imaging DR9 \citep{Dey2019}. The width and height of each image correspond to 40$''$.}
    \label{fig:saga-all-sats}
\end{figure*}
%%%%%%%%%%%%%%%%%%%%%%%%%%%%%%%%%%%%%%%%%

\subsection{SAGA Background Sample}
\label{sec:dr3-background}

The SAGA DR3 redshift catalog (\autoref{tab:redshifts}) includes all background galaxies (defined in \autoref{sec:dr3-satellites}) that are within the SAGA footprint and have confirmed redshift measurements. The redshift measurements include both SAGA and literature redshifts, and about 60\% of the galaxies have redshifts first obtained by SAGA. In the low-redshift ($z<0.05$), low-mass ($\mstar < 10^{9}\, \msun$) regime, 85\% of the redshifts were first obtained by SAGA.
\autoref{fig:redshift-mass} shows the distribution of the background galaxies in this low-redshift low-mass regime compared to the SAGA satellites.

Note that while these background galaxies are not satellites of one of the \nhosts{} SAGA systems, we do not know whether they are satellite galaxies of another system that is not part of the SAGA host sample. As such, we chose the term ``background galaxies'' rather than ``non-satellites'' or ``field galaxies.''

This sample of low-redshift and low-mass background galaxies, although not the main focus of the SAGA Survey, has enabled a range of studies, including optimizing machine learning algorithms to select low-mass galaxies \citep{xSAGA,DESI-LOWZ}, calibrating photometric sources for low-mass galaxy weak lensing measurements \citep{SOM-lensing}, and measuring the mass loading factor in low-mass galaxies \citep{2401.16469}.

\subsection{Likely Satellite Candidates and Incompleteness Correction}
\label{sec:dr3-candidates}

As part of SAGA DR3, we also provide a list of likely satellite candidates in \autoref{tab:candidates}, purely based on their photometric properties. We have targeted most of these candidates but were not able to obtain high enough signal-to-noise measurements to derive reliable redshifts, typically due to their low surface brightness.
An example of a likely satellite candidate is a gas-rich satellite with detected HI line emission in the UGC 903 system \citep{2311.02152}. This galaxy was not presented in SAGA DR2 and is considered a likely satellite candidate in SAGA DR3.

These likely satellite candidates contribute to our survey incompleteness; however, we do not expect all the candidates to be satellites. To fully quantify the total number of SAGA satellites for which we have not obtained redshifts, we apply an incompleteness correction by assigning a weight to each photometric object, with the weight representing the probability of being a SAGA satellite. This probability is produced by a model that fits the rate of satellite occurrence as a function of photometric properties, similar to the approach used in \papertwo{}.\footnote{Compared to \papertwo{}, the model is updated so that it produces more realistic probabilities below $\mstar = 10^{7.5} \msun$.} As an example, the satellite in UGC 903 mentioned above has a probability of being a SAGA satellite of 75\% based on this model. For details of this model, please refer to \autoref{app:model}.

Typically, galaxies with fainter surface brightness are more likely to be very low-redshift, and hence tend to have higher probabilities of being SAGA satellites.
However, our survey incompleteness does not correspond to a specific surface brightness limit (e.g., as suggested by \citealt{2022MNRAS.511.1544F}).
Our survey is highly complete down to $\mstar = 10^{7.5} \msun$. Below $\mstar = 10^{7.5} \msun$, the sheer amount of galaxy targets contributes to our survey incompleteness, in addition to the very low surface brightness objects.
As such, besides listing candidates with a probability of being a SAGA satellite above 25\% in \autoref{tab:candidates}, we also provide the aggregated survey incompleteness correction as a function of stellar masses in \autoref{tab:smf}.

Because we do not have redshifts for any unconfirmed satellite candidates, their stellar masses and quenched labels are estimated from photometric information only. The stellar mass is estimated the same way as for confirmed satellites (Eq.~\ref{eq:sm}) with the assumption that the candidate would be at the same distance as its host galaxy if it were indeed a satellite.

The quenched labels for unconfirmed candidates are determined by the following color--magnitude relation:
\begin{equation}
    \label{eq:color-quench}
    (g-r)_o = -0.041247 \, M_{r,o} - 0.089068.
\end{equation}
Unconfirmed candidates with a $(g-r)_o$ color redder than this relation are labeled as quenched. This relation was converted from the relation used by the ELVES Survey to classify their early- and late-type galaxies in \cite{2022ApJ...933...47C}. The original relation in \cite{2022ApJ...933...47C} uses $(g-i)$ color, and we convert it to DECam photometric filters as all SAGA photometry is based on DECaLS (see \autoref{sec:photometry}).

\section{Results}
\label{sec:results}

\subsection{Science Samples of SAGA Satellites}
\label{sec:samples}

We define three ``science samples'' of SAGA satellites for all analyses presented in the series of SAGA papers using SAGA DR3.
While the SAGA Survey is a magnitude-limited survey, quenched and star-forming galaxies typically have different stellar masses at a fixed magnitude. Hence, the three science samples are defined in stellar mass bins. The three bins roughly correspond to samples with (1) high completeness in both quenched and star-forming galaxies, (2) high completeness in star-forming galaxies only, and (3) lack of high completeness. The specific definitions are as follows.

%%%%%%%%%%%%%%%%%%%%%%%%%%%%%%%%%%%%%%%%%
\setlength{\belowdeluxetableskip}{-20pt}
\begin{deluxetable}{rccc}[t]
    \tablecaption{SAGA Science Samples \label{tab:samples}}
    \tablehead{\colhead{} & \colhead{Gold} & \colhead{Silver} & \colhead{Participation}}
    \startdata
    Stellar mass & $ \ge 10^{7.5}\,\msun$ & $10^{6.75}$--$10^{7.5}\,\msun$ & $ < 10^{6.75}\,\msun$ \\
    Completeness & High & High for s.f. sats & Low \\
    Confirmed satellites & 243\phantom{.7} & 117 & 18 \\
    Potential satellites & \phn16.7 & 226 & $\cdots$
    \enddata
\end{deluxetable}
%%%%%%%%%%%%%%%%%%%%%%%%%%%%%%%%%%%%%%%%%

\begin{enumerate}
\item The \textit{Gold} sample comprises satellites with a stellar mass greater than $10^{7.5}\,\msun$ and has the highest survey completeness. In this sample, our survey reaches a high completeness for both quenched and star-forming satellite galaxies. Based on our incompleteness correction model (\autoref{sec:dr3-candidates} and \autoref{app:model}), the number of confirmed satellites in the Gold sample accounts for 94\% of the true population. As such the Gold sample can be analyzed even without incompleteness correction.
There is little difference in the incompleteness correction for the Gold sample between \papertwo{} and this work, and we miss few, if any, low surface brightness galaxies in this mass range.
Even though the Gold sample only probes the more massive end of the satellite mass function, the large number of SAGA systems still makes this sample a statistically powerful tool.

\item The \textit{Silver} sample comprises satellites with a stellar mass between $10^{6.75}$--$10^{7.5}\,\msun$. This sample has high completeness for star-forming galaxies, but we expect a significant fraction of quenched satellites in this stellar mass range do not have confirmed redshifts from our survey. However, we have obtained enough redshifts in this sample to calibrate our incompleteness correction model to produce a reasonable estimate of the true number of satellites in this mass range.
According to the model prediction, the number of confirmed satellites in the Silver sample accounts for about only one-third of the estimated true population, but the majority of the missing satellites are expected to be quenched satellites in this mass range.
When compared with \papertwo{}, the incompleteness correction to the Silver sample reported in this work is significantly higher, because the larger data sets allow us to calibrate the incompleteness correction model better. Hence, some quantitative numbers about the Silver sample may differ from those in \papertwo{}, but our main scientific findings remain consistent, as we will discuss in later sections.

\item The \textit{Participation} sample comprises satellites with a stellar mass below $10^{6.75}\,\msun$. In this sample, our survey is highly incomplete due to the survey magnitude limit ($ r_o < 20.7$). However, the existence of these low-mass satellite galaxies is still of great scientific interest. The Participation sample will be included only in analyses where the lack of completeness would not impact the interpretation.
While our incompleteness correction model can still produce a prediction for the number of potential satellites in the Participation sample, this number may not be accurate because quenched satellites with a mass below $10^{6.75}\,\msun$ may have luminosities fainter than the designed survey depth where we have not attempted to obtain any data. As such, we do not have confidence in the model prediction for the Participation sample.
\end{enumerate}

These three science samples and the numbers of confirmed and potential satellites in each sample are summarized in Table 1. The cutout images of the 378 confirmed satellites in all three samples are shown in Figure 7.

\subsection{Satellite Stellar Mass Functions}
\label{sec:smf}

\subsubsection{Individual SAGA stellar mass functions}

%%%%%%%%%%%%%%%%%%%%%%%%%%%%%%%%%%%%%%%%%
\begin{figure}[!tbp]
    \centering
    \includegraphics[width=\linewidth,clip]{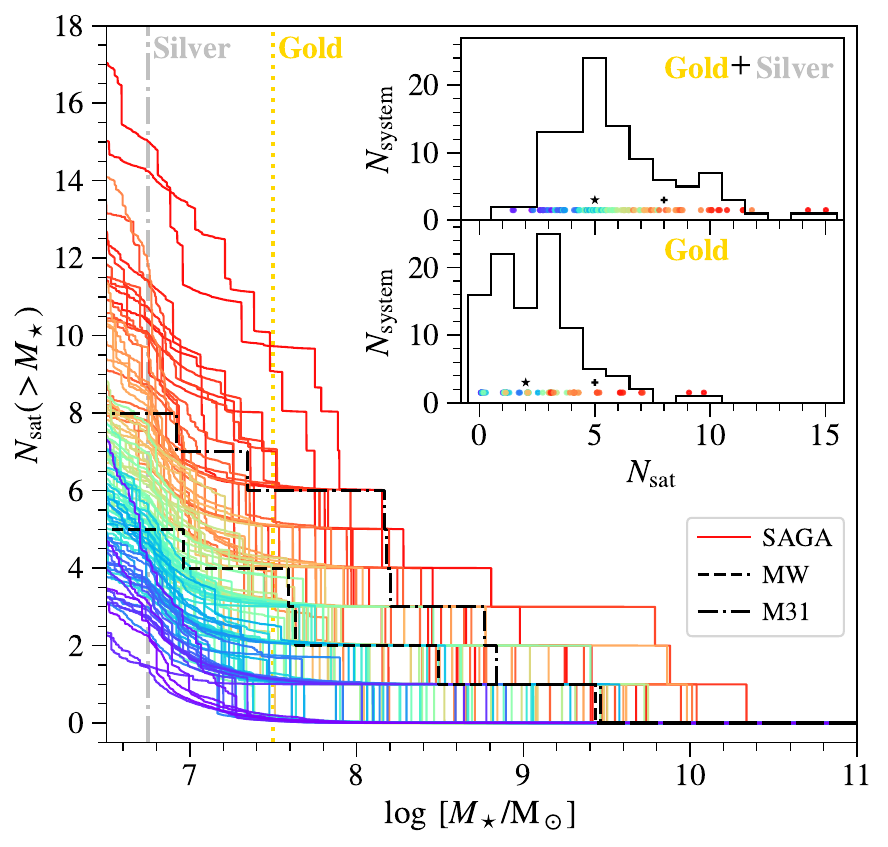}
    \caption{Cumulative satellite stellar mass function (SMF) of individual SAGA systems. Each colored curve represents one SAGA system and the color is assigned based on the number of satellites per system (same color as in \autoref{fig:saga-all-sats}). With \nhosts{} systems presented in one figure, the main purpose of this figure is to demonstrate the spread of the satellite SMF among these systems. The incompleteness correction is included in these SMFs (corresponding to the fractional vertical jumps). The gold dotted and silver dot-dashed vertical lines show the corresponding stellar mass limit of the Gold and Silver samples.
    The inset shows the distribution of the number of satellites per system for the Gold sample (lower inset) and the Gold$+$Silver sample (upper inset). The colored points in the inset indicate the satellite numbers for individual systems; hence, they correspond to the locations where the colored curves cross the vertical lines in the main panel. The black stars and plus signs in the inset show the numbers of satellites for MW and M\,31 respectively. }
  \label{fig:sat_smf_rainbow}
\end{figure}
%%%%%%%%%%%%%%%%%%%%%%%%%%%%%%%%%%%%%%%%

We first show the cumulative satellite stellar mass function (SMF) for each of the \nhosts{} SAGA systems in \autoref{fig:sat_smf_rainbow}.
While it can be difficult to discern individual SMFs in this figure, the main purpose of this figure is to demonstrate the system-to-system spread of the SMFs. The inset of this figure also allows us to quickly see the distribution of the number of satellites for the Gold and Gold$+$Silver samples.
Note that the SMFs and the number of satellites shown here already include the incompleteness correction described in \autoref{sec:dr3-candidates} and \autoref{app:model}.
Four of the \nhosts{} systems have the same level of survey completeness but do not have any confirmed satellites; their incompleteness-corrected SMFs are shown in this figure as well.

The colors of the curves in \autoref{fig:sat_smf_rainbow} are assigned based on the number of confirmed satellites in the Gold$+$Silver sample for each host. The color gradient across the curves can be observed at different mass limits, demonstrating that the high- and low-mass ends of cumulative SMFs are highly correlated. This correlation can also be seen from the inset \autoref{fig:sat_smf_rainbow}, where the color gradient of the points exists in a similar fashion in both panels. The correlation between the high- and low-mass ends of the SMF is consistent with the expectation from gravity-only simulations, as subhalo mass functions are self-similar \citep[e.g.,][]{Sales2013,Mao150302637,2019MNRAS.486.2440Z,Nadler:2209.02675}.

We include a figure showing the satellite luminosity function as \autoref{fig:sat_lf} in \autoref{app:additional-analyses}.

\subsubsection{Comparison with MW and M31 SMFs}

The satellite SMF of MW and M\,31 are also shown in \autoref{fig:sat_smf_rainbow}, and they are both well within the system-to-system spread of the SAGA SMFs.
In particular, the existence of an LMC-mass satellite is not uncommon; about one-third (34/101) of the SAGA systems have at least one confirmed satellite with a stellar mass above $10^{9}\,\msun$.
In the inset, one can see that the number of satellites in M\,31 is on the high end of the SAGA systems, which is expected given M\,31's higher mass compared to SAGA hosts (\autoref{fig:hosts-properties}).
On the other hand, the number of satellites in the MW is just slightly below the median number of the SAGA systems for the Gold$+$Silver sample. If one only considers the expected correlation between the host galaxy's stellar mass and the number of satellites, the MW would appear to have a lower satellite number, given that the majority of SAGA hosts are less massive than the MW. 

Note that in the main panel, the low-mass end of the MW SMF is closer to the blue-colored SAGA SMFs, while the high-mass end of the MW SMF is closer to the yellow-colored SAGA SMFs.
In other words, the number of satellites in the MW (to the SAGA mass limit; i.e., Gold$+$Silver sample) is lower than that of the SAGA systems that also have massive satellites like the LMC.
This observation suggests that the MW satellites do not follow the strong correlation between the number of satellites and the most massive satellite mass seen in SAGA; we will demonstrate this correlation quantitatively in \autoref{sec:correlation}.
This discrepancy suggests that the MW may have an atypical formation history compared to the SAGA hosts, which we will discuss in detail in \autoref{sec:mw-context}.

\subsubsection{Average SAGA stellar mass functions}

%%%%%%%%%%%%%%%%%%%%%%%%%%%%%%%%%%%%%%%%%
\begin{figure*}[!tbp]
    \centering
    \includegraphics[width=\linewidth,clip]{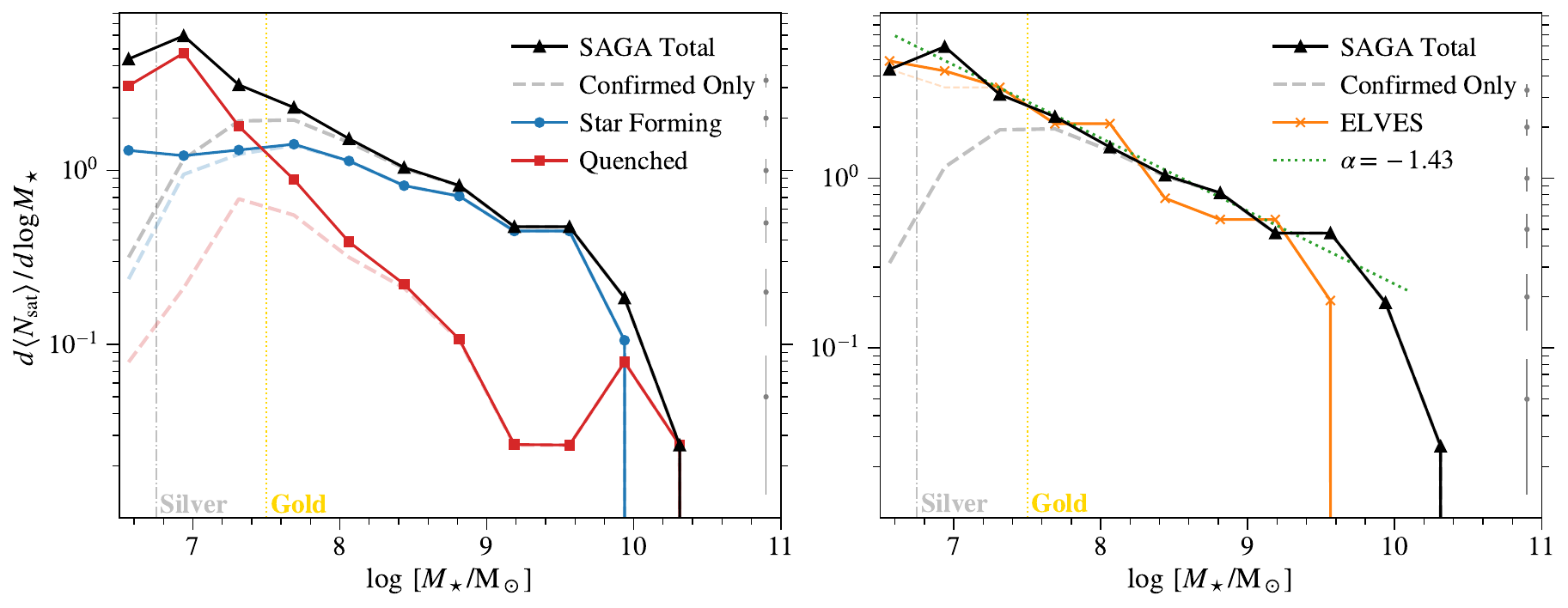}
    \caption{\textbf{Left}: The average satellite stellar mass function (SMF) of \nhosts{} SAGA systems, shown in the differential form ($d\,\langle N_\text{sat} \rangle/d\,\log\mstar$). The darker solid curves show the SMF when survey incompleteness is taken into account, while the fainter dashed curves show only confirmed satellites. The curves with black triangles, blue circles, and red squares show all satellites, only star-forming satellites, and only quenched satellites, respectively. The gold dotted and silver dot-dashed vertical lines show the corresponding stellar mass limits of the Gold and Silver samples. These SMF data points are provided in \autoref{tab:smf}.
    \textbf{Right}: Same as the left panel, but comparing the SAGA satellite SMF (black triangles) with the ELVES Survey (orange crosses; \citealt{2022ApJ...933...47C}) and a constant slope of $-0.43$ (green dotted line; typically quoted as $\alpha=-1.43$, cf. \citealt{2020ApJ...893...48N}). Both SAGA and ELVES SMFs are corrected for survey incompleteness; the uncorrected SMFs are shown as gray and faint orange dashed lines respectively.
    The error bars near the right border of both panels show the Poisson error on the SAGA SMF.}
    \label{fig:sat_smf}
\end{figure*}
%%%%%%%%%%%%%%%%%%%%%%%%%%%%%%%%%%%%%%%%

%%%%%%%%%%%%%%%%%%%%%%%%%%%%%%%%%%%%%%%%%
\begin{figure}[!tbp]
    \centering
    \includegraphics[width=\columnwidth,clip,trim=0.1cm 0.3cm 0cm 0cm]{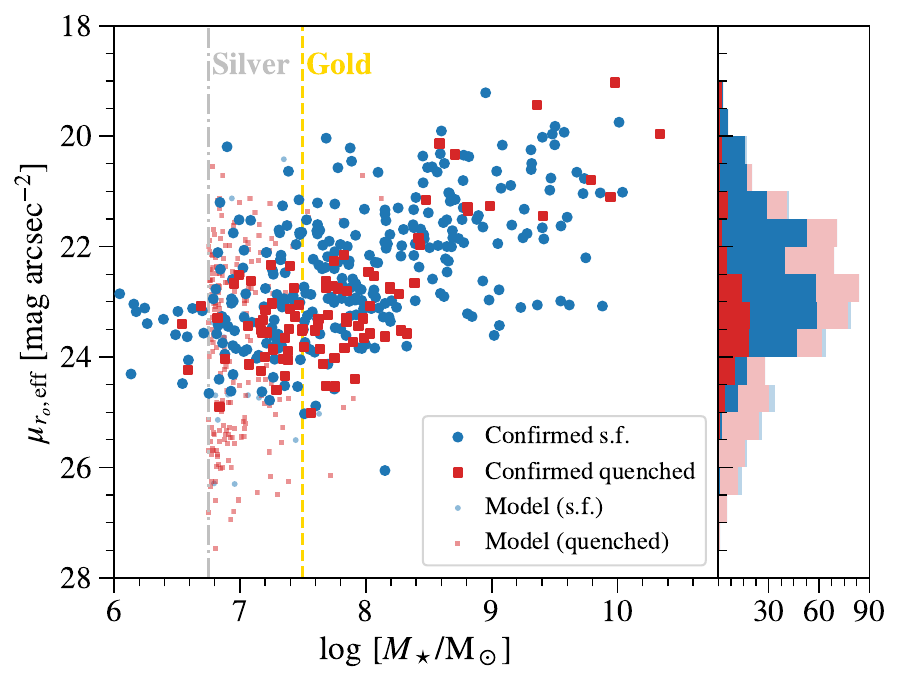}
    \caption{\textbf{Left:} The effective surface brightness--mass distribution of confirmed SAGA satellites; the large blue circles and large red squares show star-forming and quenched satellites, respectively. The small and fainter markers show a random realization of satellites that we may have missed in the Gold and Silver samples, according to our incompleteness model; the color of the small markers indicates whether the missed satellites are star forming (blue) or quenched (red). The gold and silver vertical lines show the corresponding  stellar mass limit of the Gold and Silver samples. \textbf{Right:} Stacked histograms, in bins of effective surface brightness, of the confirmed (dark) and potentially missed (faint) satellites in the Gold and Silver samples, separated by whether the satellites are star forming (blue) or quenched (red).}
    \label{fig:sb-mag}
\end{figure}
%%%%%%%%%%%%%%%%%%%%%%%%%%%%%%%%%%%%%%%%%

The left panel of \autoref{fig:sat_smf} shows the average satellite SMF in the differential form for all satellites and also for the quenched and star-forming satellites, respectively. The definition of quenched satellites is described in \autoref{sec:dr3-properties}.
The total satellite SMF can be well described by a power law, except for the most massive end.
We also observe that most satellites in the Gold Sample (down to $\mstar \sim 10^{7.5} \,\msun$) are star-forming. The quenched satellite SMF exceeds the star-forming SMF below $\mstar \sim 10^{7.5} \,\msun$.
These SMF data points are also provided in \autoref{tab:smf}.

Note the survey incompleteness has been corrected in all the satellite SMFs shown in solid curves, while the faint dashed curves show the SMFs calculated with confirmed satellites only.
The difference between the corrected and confirmed SMFs represents the estimated population of satellites that we do not have redshifts based on the incompleteness correction model (\autoref{app:model}).
We can see that our survey is very complete in the Gold Sample, and the star-forming sample is very complete in the Silver Sample as well ($ 10^{6.75} \,\msun < \mstar < 10^{7.5} \,\msun$).

In \autoref{fig:sb-mag}, we show how the  satellites that may have been missed distribute in the effective surface brightness--stellar mass space. The small dots in \autoref{fig:sb-mag} represent the population that contributes to the difference between the corrected and confirmed SMFs in the left panel of \autoref{fig:sat_smf}.
Most missed satellites are in the Silver Sample and are quenched (based on their color).
Note that our incompleteness model predicts a population of missed satellites with very low surface brightness in the Silver Sample.

\subsubsection{Comparison with ELVES SMFs}

The right panel of \autoref{fig:sat_smf} shows a comparison with the ELVES sample \citep[Table 9]{2022ApJ...933...47C} and an SMF with a constant slope (here we follow the convention used in the Schechter function, with $\alpha=-1$ being flat).
The completeness-corrected SMFs of the ELVES and SAGA Surveys match each other fairly well, and the faint-end slope of both SMFs are consistent with $\alpha=-1.43$, the best-fit value inferred from the MW satellite population \citep{2020ApJ...893...48N}.
This faint-end slope is slightly shallower than, though consistent within $2\sigma$ of, the faint-end slope of the global SMF obtained by the GAMA Survey \citep[$\alpha=-1.48 \pm 0.01$,][]{2022MNRAS.509.5467K}.
We note that the incompleteness correction model applied to the SMF (described in \autoref{app:model}) is not tuned to a specific faint-end slope.
On the high-mass end, the SAGA SMF is higher than the ELVES SMF; however, the small number of satellites in this range likely dominates the effect.

Note that the ELVES data shown here only include 14 ELVES systems.\footnote{The 14 ELVES systems used in our comparison are: NGC~253, 628, 891, 1023, 1291, 2683, 2903, 3115, 4258, 4736, 5055, 5236, 5457, and 6744.} In order to make the ELVES and SAGA samples comparable, we address the difference in host selection criteria between the two surveys. We apply the SAGA environment (Eqs.~\ref{eq:env-cut-K} and \ref{eq:env-cut-group}) and stellar mass (Eq.~\ref{eq:host-MK-cut}) cuts to produce the SMF in the right panel (orange line). We opt to apply only these basic SAGA host cuts to the ELVES sample for the sake of not reducing the ELVES sample size too drastically. If all SAGA host cuts are applied to the ELVES sample, the resulting SMF is only marginally different.

\subsection{Satellite Quenched Fraction}
\label{sec:qf}

%%%%%%%%%%%%%%%%%%%%%%%%%%%%%%%%%%%%%%%%%
\begin{figure}[!tbp]
    \centering
    \includegraphics[width=\columnwidth,clip,trim=0.1cm 0.3cm 0cm 0cm]{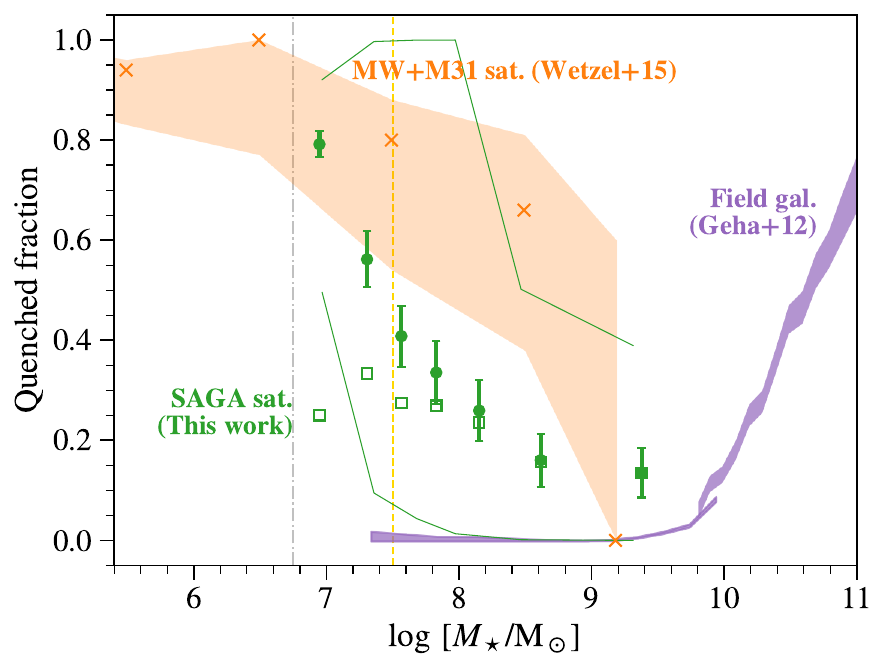}
    \caption{Satellite quenched fraction as a function of stellar mass for SAGA satellites (green points with error bars), MW and M\,31 satellites (orange crosses and shaded region; \citealt{Wetzel:1503.06799}), and isolated SDSS field galaxies (purple shaded region;  \citealt{geha2012}). See \autoref{sec:qf} for quenching definition. We also show the quenched fraction with only confirmed SAGA satellites (i.e., no incompleteness correction) as empty squares.
    The green error bars show binomial (Poisson) error on the fraction values. 
    The solid green curves indicate the $[16\%, 84\%]$ system-to-system scatter of the SAGA quenched fractions.
    The gold and silver vertical lines show the corresponding  stellar mass limit of the Gold and Silver samples.
    The SAGA quenched fraction data points are also provided in \autoref{tab:fquench}.
    }
    \label{fig:quenched_frac}
\end{figure}
%%%%%%%%%%%%%%%%%%%%%%%%%%%%%%%%%%%%%%%%%

Shown in \autoref{fig:quenched_frac} is the satellite quenched fraction, defined as the ratio of the quenched and total satellite populations. We show the averaged quenched fraction as a function of satellite stellar mass. The definition of quenched satellites corresponds to a specific star formation rate below $10^{-11}\,\text{yr}^{-1}$, estimated from a combination of NUV detection and H$\alpha$ measurement. The specific definition of quenched satellites can be found in \autoref{sec:dr3-properties}, with more detailed explanation in Section~3.1 of \paperfour{}.

As \autoref{fig:sat_smf} demonstrates, we expect that a subset of low-mass quenched satellites exist in these systems for which we do not have a confirmed redshift. Since we have a good estimate of the expected number of these unconfirmed satellites, we include them in the calculation of the quenched fraction. For unconfirmed satellites, the quenched definition is based on their color and absolute magnitude, as described in Eq.~\ref{eq:color-quench}.  Note that the incompleteness-corrected and uncorrected results are very similar in the Gold sample (right of the vertical gold dashed line), and deviate from each other at lower masses.

The satellite quenched fraction in the \nhosts{} SAGA systems is marginally higher than, but statistically consistent with, what we reported in our Stage II results (\papertwo{}).  When compared with the satellite quenched fraction in the Local Group, SAGA systems' average satellite quenched fraction is lower for stellar mass below $\mstar = 10^9\,\msun$. This difference between SAGA systems and the Local Group is unlikely to be  due to observational incompleteness for three reasons:
\begin{enumerate}
    \item First, if the SAGA Gold sample had an overall satellite quenched fraction of 70\%, like the Local Group, it would imply there were more than 200 quenched satellites with $\mstar{} > 10^{7.5}\,\msun{}$ that were not identified by our survey. This is extremely unlikely given our completeness estimates, and would significantly alter the stellar mass function if it were true.
    \item Second, we have confidence in the survey incompleteness correction (\autoref{app:model}) that has been included in the quenched fraction shown. In particular, \autoref{fig:sb-mag} shows that the population of potentially missed satellites, including many low surface brightness ones, are accounted for by the incompleteness correction model, hence included in the quenched fraction calculation.
    \item Third, regardless of the quenched fraction, the existence of star-forming satellites below $\mstar = 10^7\,\msun$ in SAGA systems differs from the Local Group satellite population.
\end{enumerate}

The most likely explanation for the difference in satellite quenched fraction is system-to-system scatter. The satellite quenched fraction shown in \autoref{fig:quenched_frac} is averaged over all SAGA satellites and the error bars show only the Poisson errors. However, the quenched fraction of individual SAGA systems has significantly larger variation from system to system. Considering the system-to-system scatter, the Local Group's quenched fraction is only about $1\sigma$ higher than that of the SAGA systems.
The system-to-system scatter in SAGA does not have a strong dependence on whether the system is in a Local Group-like pair; we will show this explicitly in \paperfour{}.

Note that the system-to-system scatter in the quenched fraction at different stellar masses is not highly correlated between mass bins (unlike the case in SMF).
On the higher-mass end ($\mstar = 10^9\,\msun$), the SAGA systems actually have a slightly higher quenched fraction when compared to that of the Local Group. This difference is not statistically significant with respect to the system-to-system scatter; however, it is interesting to note that the steeper change in the satellite quenched fraction around $\mstar = 10^{8.5}\,\msun$ in the Local Group. Statistically, this steeper change in the Local Group is only marginally significant, as we will demonstrate in \autoref{sec:mw-context}.

We also compare the SAGA quenched fraction with the ELVES Survey, and find them consistent when the difference in host selection and quenched definition are taken into account; we leave the analysis to \paperfour{}. We also further investigate the quenched fraction's dependence on projected radius and which host system properties correlate with their satellite quenched fraction in \paperfour{}.

\subsection{Satellite Properties}
\label{sec:sats-properties}

%%%%%%%%%%%%%%%%%%%%%%%%%%%%%%%%%%%%%%%%%
\begin{figure*}[!tbp]
    \centering
    \includegraphics[width=\linewidth,clip]{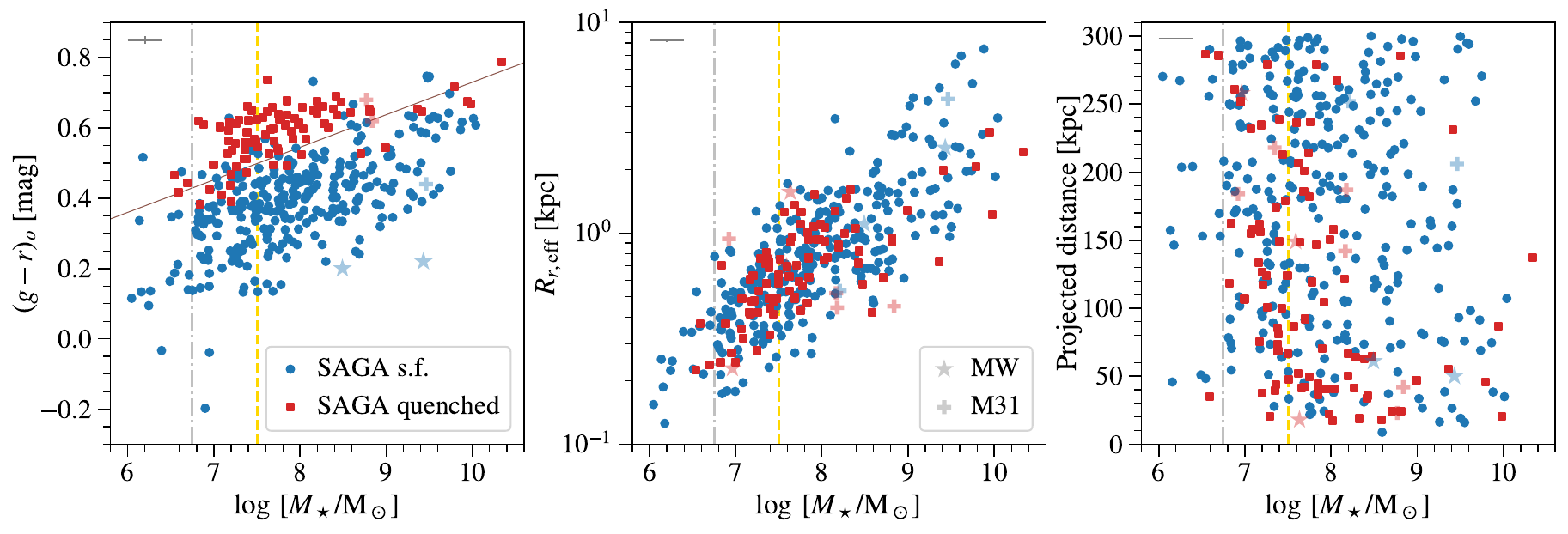}
    \caption{The color--mass (left), size--mass (middle), and projected distance--mass (right) relations of confirmed SAGA satellites. The blue circles and red squares show star-forming and quenched satellites, respectively. Large stars and plus signs indicate the satellites of the Milky Way and M\,31, respectively, with faint blue and red colors indicating star-forming and quenched. Here the color shown is $(g-r)_o$, size is the physical effective radius in kpc, distance is the projected physical separation between the satellite and its host galaxy in kpc, and mass is the estimated stellar mass. The gold and silver vertical lines show the corresponding stellar mass limit of the Gold and Silver samples. Typical errors are shown at the upper left corner of each panel; the typical errors in projected distances do not include the uncertainties in system distance measurement. In the leftmost panel, the thin diagonal brown solid line shows the color--mass relation that roughly separates star-forming and quenched satellites (i.e., combining Eqs.\ref{eq:sm} and \ref{eq:color-quench}, the latter of which is derived from \citealt{2022ApJ...933...47C}).}
    \label{fig:size-color-mass}
\end{figure*}
%%%%%%%%%%%%%%%%%%%%%%%%%%%%%%%%%%%%%%%%%

\autoref{fig:size-color-mass} presents relationships between various observational properties for confirmed SAGA satellites, split into star-forming and quenched populations. Some Local Group satellites that meet the SAGA criteria (i.e., have $M_{r,o}<-12.3$ within 300~kpc of the Milky Way or M\,31) are also shown for comparison.
For Local Group satellites, stellar masses are from the Local Volume database compiled by A.~Pace\footnote{\https{github.com/apace7/local_volume_database}}, except for the LMC/SMC \citep{2009IAUS..256...81V}. Projected distances to host are obtained from the catalog of \cite{McConnachie2012}.
Physical effective radii $R_{r,\textrm{eff}}$ are from the Local Volume database, except for the SMC \citep{Munoz2018} and M33 \citep{Smercina2023}.
Due to the proximity of these Local Group galaxies, integrated color measurements are only available for the most massive satellites (LMC and SMC among MW satellites, and M33, M110, and M32 among M\,31 satellites); we obtain these from \cite{Tollerud2011} and \cite{RC3}.

The color--mass relation (left panel) shows a visible separation between red quenched satellites and blue star-forming satellites, although some star-forming satellites have redder $(g-r)_o$ colors than expected for their mass. As a result, there is no simple color--mass cut that can cleanly separate quenched and star-forming satellites. The color--mass cut used by \citet{2022ApJ...933...47C} to separate quenched and star-forming satellites is shown as a thin brown line on the plot.
Notably, the LMC and the SMC (the two pink stars with the highest stellar masses) are bluer than any of the SAGA satellites of similar mass, possibly as a result of star formation triggered by their ongoing infall into the MW \citep{2018MNRAS.479..284S}. We refer readers to Section~7.2 of \paperfour{} for further discussion.

Both star-forming and quenched SAGA satellites appear to follow the same size--mass relation (middle panel), suggesting that the processes that drive quenching do not affect a satellite's size \citep[see also][]{10.3847/1538-4357}.
While not directly shown in this figure, we found no dependence of the size--mass relation on the satellites' projected distance to their hosts.
SAGA satellites' size--mass relation is also consistent with that of the Local Group satellites, and we observe no indication that our survey misses low-surface brightness galaxies in the Gold sample, as demonstrated in \autoref{fig:sb-mag}.

There is no correlation between SAGA satellites' stellar masses and their projected distances to their respective host galaxy (right panel). However, the quenched SAGA satellites appear to be distributed closer to their hosts than star-forming satellites (see \paperfour{}). The difference in distribution is most apparent for higher-mass satellites.
This is in contrast with the LMC and the SMC, both of which are massive, star forming, yet in proximity to the MW Galaxy. This contrast, nonetheless, is consistent with our knowledge that LMC and SMC are on their first infall and their proximity to the MW may have a short timescale \citep{Kallivayalil2013,2017MNRAS.464.3825P,cautun2019}.

We also explore the star-forming properties of SAGA satellites as a function of their projected distances further in \paperfour{}.

\subsection{Satellite Radial Distribution}
\label{sec:radial}

%%%%%%%%%%%%%%%%%%%%%%%%%%%%%%%%%%%%%%%%%
\begin{figure}[!tbp]
    \centering
    \includegraphics[width=\columnwidth,clip,trim=0.2cm 0.3cm 0 0]{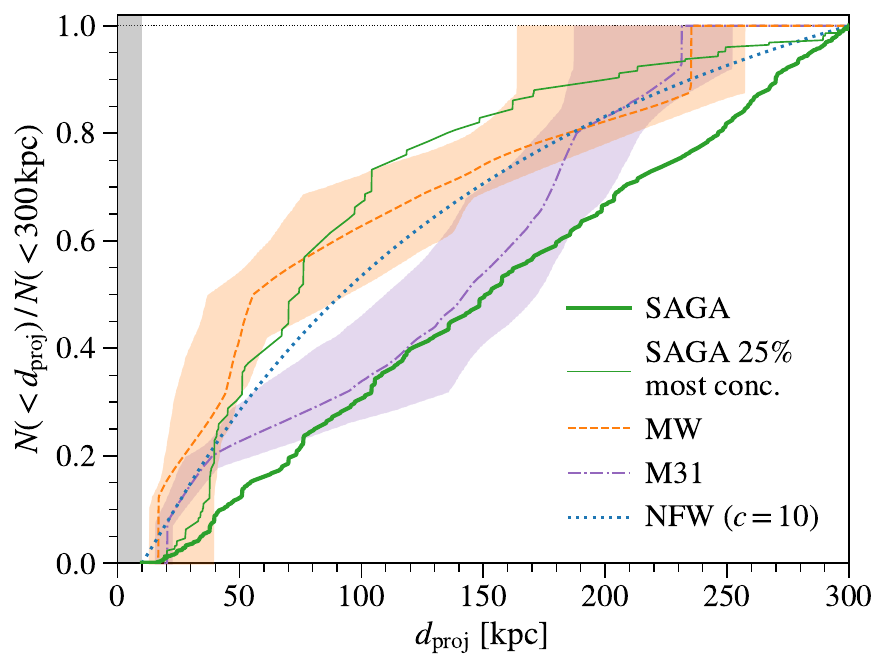}
    \caption{Average cumulative radial distribution of confirmed satellites around \nhosts{} SAGA hosts (thick green line), and satellites of MW and M\,31 (dashed orange and dot-dashed purple lines).
    For the latter, we calculated the projected distance using a random set of sight lines, and the 1$\sigma$ range is shown as shaded areas around the median lines. The thin green line shows the average radial distribution of the 25\% of SAGA hosts (25 hosts) that have the most concentrated distribution. The blue dotted line shows a projected NFW profile with a concentration parameter $c=10$.
    We did not attempt to obtain any redshifts within 10\,kpc of the host galaxy and that inner region (gray vertical band on the left) is excluded from the radial profile calculation.}
    \label{fig:sat_radial}
\end{figure}
%%%%%%%%%%%%%%%%%%%%%%%%%%%%%%%%%%%%%%%%

The average, normalized, cumulative radial distribution of satellites around SAGA hosts is presented in \autoref{fig:sat_radial}.
As in \papertwo{}, this distribution is approximately linear in projected distance, consistent with a $1/r_{\mathrm{3D}}$ spherical distribution of satellites. This finding is also consistent with the satellite radial distribution from both the ELVES systems \citep{2020ApJ...902..124C} and the measurements from xSAGA \citep[which extends the SAGA Survey with machine learning on NASA-Sloan Atlas central galaxies;][]{xSAGA}.

To compare the radial distribution with that of the MW and M\,31, we follow the same procedure as \papertwo{} to project the three-dimensional satellite distributions around the MW and M\,31 to a distribution in projected distances. The colored bands around the  MW and M\,31 lines shown in  \autoref{fig:sat_radial} denote the variability from the projection effect. The satellite distribution of the MW is significantly more concentrated (i.e., closer to the host galaxy), while M\,31's satellite distribution is more concentrated only at the innermost radii.

However, as in \papertwo{}, the difference is consistent with the strength of system-to-system variation. The SAGA hosts with the 25\% most concentrated satellite distributions (as measured by the median projected distance of satellites) have, on average, a satellite radial distribution that is just as concentrated as that of the MW.
It is well known that the MW has a more concentrated radial profile when compared with simulations \citep[e.g.,][]{2004MNRAS.353..639W,10.1093/mnras/stt2058}, likely due to a combination of the proximity of LMC and SMC to the MW center \citep{2020ApJ...893...48N}, a more concentrated dark matter distribution due to earlier halo formation, and potential incompleteness in satellite census in the outskirts of the MW. The last reason is unlikely to be an issue when we compare the MW and the SAGA systems given the SAGA mass limit, but both the LMC/SMC locations and the MW's dark matter halo can contribute to the difference we observe.

Observational constraints on satellite radial distributions probe the underlying subhalo distribution (e.g., \citealt{Nagai0408273}) and the efficiency of subhalo and satellite disruption due to the central galaxy (e.g., \citealt{Samuel2019a}). Furthermore, the radial distribution can be used to test CDM physics itself \citep[e.g.,][]{2021ApJ...920L..11N,2104.03322}. 
While the dark matter's radial distribution in a halo typically follows the NFW profile \citep{nfw1996}, the satellite radial distribution can differ due to dynamical friction, tidal effects, and, in the presence of baryons, enhanced stripping by the central galaxy \citep[e.g.,][]{Samuel2019a,Webb200606695,Green210301227,2022ApJ...933..161M}. 
These effects tend to make satellite radial distribution more extended than the dark matter's distribution \citep[e.g,][]{Nagai0408273}.
When comparing SAGA's projected satellite radial distribution with a projected NFW profile (with a concentration parameter $c=10$, which is typical for a MW-mass halo in the absence of baryons) in \autoref{fig:sat_radial}, we find that SAGA's satellite distributions are indeed more extended than the NFW profile, consistent with our expectation. A detailed comparison with simulations can help distinguish the contribution of the various effects that impact the satellite radial profile.

In addition, we find little variation in the shape or concentration of the observed satellite radial distribution with host color or satellite magnitude gap, even though these properties are potential proxies of host halo concentration \citep[e.g.,][]{More2012}. This finding is consistent with the findings in xSAGA \citep{xSAGA}.

\subsection{Comparing Satellite Abundance with Simulations}
\label{sec:nsat}

%%%%%%%%%%%%%%%%%%%%%%%%%%%%%%%%%%%%%%%%%
\begin{figure}[!tbp]
    \centering
    \includegraphics[width=\linewidth,clip]{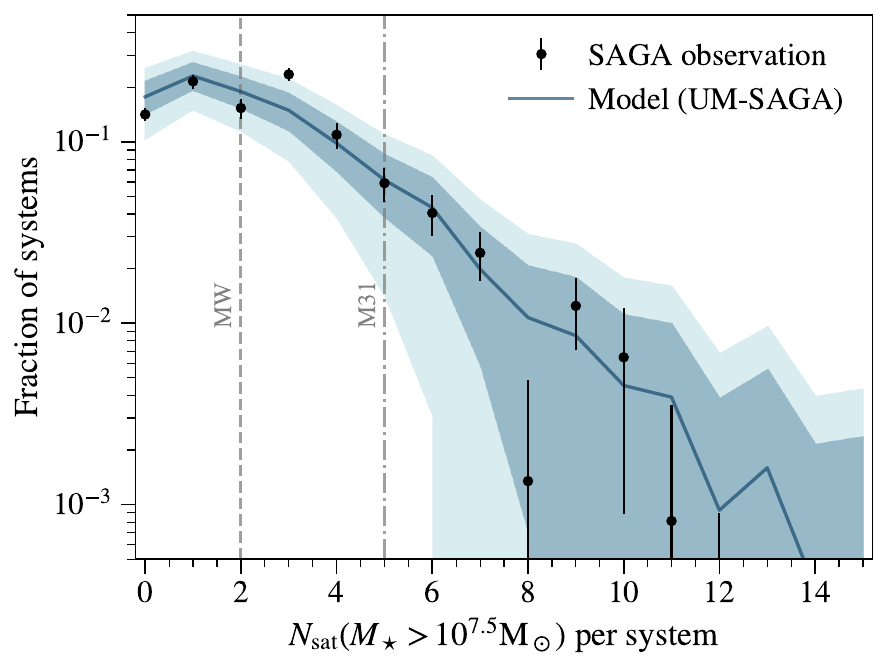}
    \caption{Distribution of satellite abundance, shown as the fraction of systems ($y$-axis) that have a specific number of satellites down to $M_{\star}>10^{7.5}\,\msun$ (gold sample, $x$-axis). The black points show the incompleteness-corrected SAGA observations, where the error bars indicate the scatter among the realizations of the incompleteness correction.
    The gray solid line shows the predictions from the best-fit galaxy--halo connection model ``UM-SAGA'', as presented in \paperfive{}. The associated dark (light) gray bands indicate the $1\sigma$ ($2\sigma$) intervals of the prediction. The satellite abundance of MW (M\,31) is shown as a dashed (dot-dashed) gray line.}
    \label{fig:nsat-dist}
\end{figure}
%%%%%%%%%%%%%%%%%%%%%%%%%%%%%%%%%%%%%%%%%

We present the distribution of satellite abundance (number of satellites per system) across our \nhosts{} SAGA hosts in \autoref{fig:nsat-dist}. This figure shows a normalized histogram of the incompleteness-corrected SAGA Gold sample's satellite number ($\mstar \geq 10^{7.5}\, \msun$) in all SAGA hosts, similar to the lower inset of \autoref{fig:sat_smf_rainbow}. Here, for any fractional satellite count (due to incompleteness correction), we draw a uniform random number in $[0,1)$ to set it to 0 or 1. The error bars indicate the scatter among these random realizations of the incompleteness correction. The vertical dashed line indicates the satellite numbers of the MW and M\,31.

The gray solid curve and shaded bands are the mean, $1\sigma$, and $2\sigma$ intervals for the predictions of the satellite abundance distribution from the empirical galaxy--halo connection model \textsc{UniverseMachine}~(UM, \citealt{2019MNRAS.488.3143B}). The predictions we show here are from the latest version, UM-SAGA, which has a newly added dwarf galaxy quenching module (motivated by findings in \citealt{2021ApJ...915..116W}) constrained by SAGA's average satellite quenched fraction and average stellar mass function. Note that the satellite abundance distribution is a derived quantity that depends on the cosmological distribution of host and subhalo populations, and was not directly fitted by the model.
The $1\sigma$ and $2\sigma$ intervals were obtained by randomly selecting \nhosts{} hosts from a parent sample of 2,500 SAGA-like hosts in a 125\,Mpc\,\perh{} cosmological simulation. More details on the model update of UM and insights we gain on satellite quenching are introduced in \paperfive{}.

Overall, the new UM-SAGA model predicts a satellite abundance distribution that agrees with SAGA observations for satellites with $\mstar \geq 10^{7.5}\, \msun$. We have conducted two-sample Kolmogorov--Smirnov tests using the 100 realizations of \nhosts{} SAGA-like hosts versus the actual SAGA data for their $N_\text{sat}$ distributions, and we found that 94 out of 100 realizations have $p$-values $>0.05$ and all realizations have $p>0.01$, meaning that the UM-SAGA model predictions are statistically indistinguishable from SAGA in terms of $N_{\rm sat}$ distributions. This is consistent with our findings in \papertwo{}, where we compared SAGA Stage II observations to the predictions of the \cite{Nadler180905542,2020ApJ...893...48N} subhalo abundance matching model. Despite the different empirical assumptions underlying the two predictions, we find that these models yield very similar predictions for SAGA satellite counts. A more detailed comparison between these predictions may inform modeling degeneracies, e.g., related to satellite disruption due to baryonic effects and modeling treatment of orphan galaxies in the simulation.

\subsection{Correlation between Satellite Abundance and System Properties}
\label{sec:correlation}

%%%%%%%%%%%%%%%%%%%%%%%%%%%%%%%%%%%%%%%%%
\begin{figure}[!tbp]
    \centering
    \includegraphics[width=\linewidth]{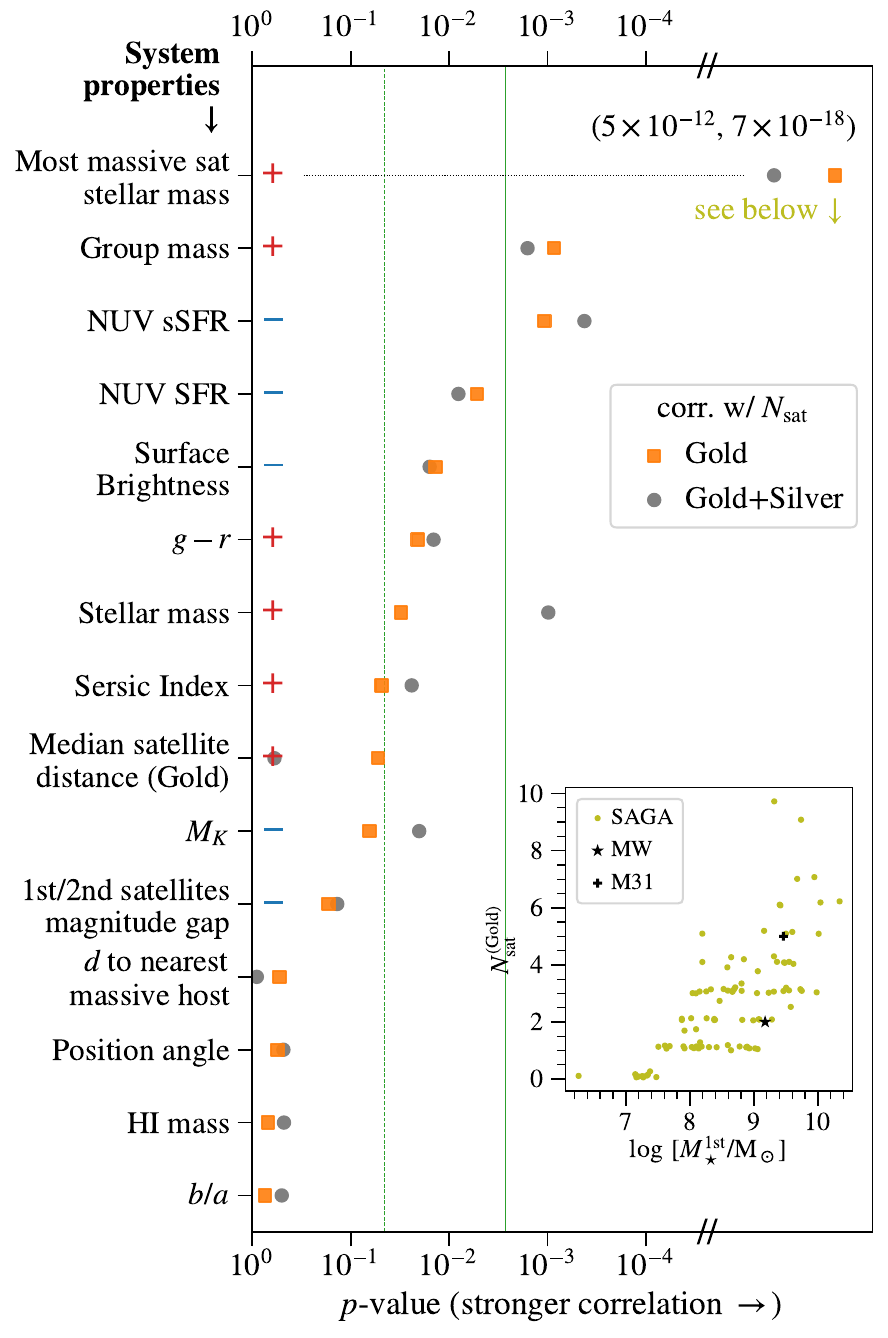}
    \caption{The correlation strength (shown as $p$-values) between each host property shown on the $y$-axis and the number of satellites per host in the Gold (shown as orange-filled squares) and Gold$+$Silver (shown as gray-filled circles) samples, presented as $p$-values from the Spearman rank correlation test. The plus and minus signs on the left show the sign of the correlation. The green vertical dashed and solid lines indicate $p$-values of 0.046 and 0.0027 (corresponding to 2$\sigma$ and 3$\sigma$ criteria, and correlation coefficients of $~$0.2 and $~$0.3 with 101 systems). The $p$-values for the correlation between the most massive satellite mass and the number of satellites in both samples are extremely low, and are not shown to scale. The inset shows a scatter plot between the most massive satellite mass and the number of satellites in the Gold sample to demonstrate the observed correlation; the black star in the inset shows where the MW system would be located in this plot. See \autoref{sec:correlation} for a more thorough discussion of the host properties' meaning.}
    \label{fig:nsat_correlation}
\end{figure}
%%%%%%%%%%%%%%%%%%%%%%%%%%%%%%%%%%%%%%%%%

%%%%%%%%%%%%%%%%%%%%%%%%%%%%%%%%%%%%%%%%%
\begin{figure}[!tbp]
    \centering
    \includegraphics[width=\linewidth]{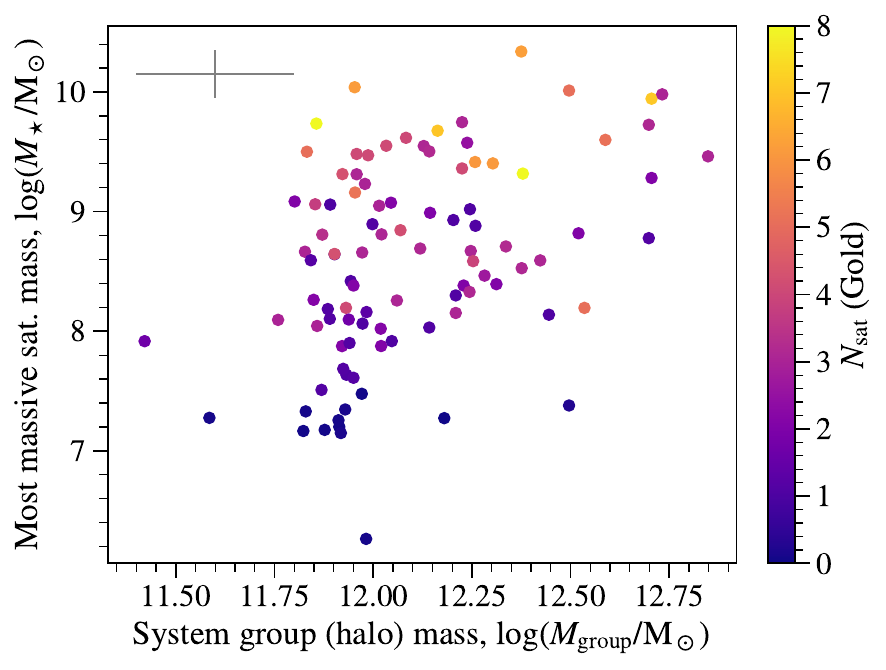}
    \caption{Dependence of satellite abundance (color) on the estimated group halo mass ($x$-axis) and the stellar mass of the most massive confirmed satellite ($y$-axis) of each system. The group mass was estimated by \citet{2017MNRAS.470.2982L}. Typical errors are shown in the upper left corner. The direction of the color gradient of the points shows that the satellite abundance primarily varies with the stellar mass of the most massive satellite. At a fixed most massive satellite mass, there is no strong dependence of the satellite abundance on group mass.}
    \label{fig:nsat_two_properties}
\end{figure}
%%%%%%%%%%%%%%%%%%%%%%%%%%%%%%%%%%%%%%%%%

A notable strength of the SAGA Survey is that the large number of systems surveyed allows us to investigate if any properties of the satellite systems (or their host galaxies) have a strong correlation with the number of satellites in the system. We tested a wide range of properties, including:
\begin{enumerate}
    \item Photometric properties of the host galaxies: $K$-band absolute magnitude $M_K$, $g-r$ color, surface brightness, Sersic index, the ratio of semi-minor to semi-major axes $b/a$, and position angle.
    \item Other measured or inferred properties of the host galaxies: Stellar mass, group mass \citep[][see \autoref{sec:hosts} for a brief description of how the halo mass estimate was obtained]{2017MNRAS.470.2982L}, star formation rate and specific star formation rate from NUV, HI mass, distance to the nearest galaxies with $M_K < -23$.
    \item Summary statistics of the satellite population: Stellar mass of the most massive satellite, magnitude gap (difference) between the first and the second brightest satellites, median projected distance of confirmed satellites in the Gold sample.
\end{enumerate}

Given this long list of properties, the look-elsewhere effect can impact the interpretation of the statistical significance of any observed correlation. While we report the measured correlation strengths below, we limit our discussion to a few most significant cases, where the $p$-value is below or of the order of $10^{-3}$.
Note that for 101 systems, a $p$-value of 0.046 (2$\sigma$) corresponds to a Spearman correlation coefficient of 0.2 and a $p$-value of $2.7 \times 10^{-3}$ (3$\sigma$) corresponds to a coefficient of 0.3.

We calculate the Spearman's rank correlation coefficient between the number of satellites and each of the properties listed above.
Some properties listed above are not available or not defined for certain systems. For example, a small number of the host galaxies do not have HI or NUV measurements, and systems with no confirmed satellites will not have any of the summary statistics of the satellite population defined. When a system does not have a property defined, the system is excluded from the correlation calculation for that property.
The number of satellites used here is the number of confirmed satellites plus any survey incompleteness correction as described in \autoref{app:model}. We repeat the calculation for both the Gold and the Gold+Silver samples.

\autoref{fig:nsat_correlation} summarizes the results of our correlation calculation. Each row shows the correlation strength (right is stronger) between the property listed on the left and the number of satellites in the Gold (orange-filled squares) or Gold+Silver (gray-filled circles) samples. The sign on the leftmost edge shows the sense of the correlation. We express the correlation strength in terms of $p$-value because the number of systems that enter the calculation in each row can be different, and the $p$-values already take the number of systems into account.

\subsubsection{Notable strong correlations}

Among the properties that we tested, the stellar mass of the most massive satellite has the strongest correlation with satellite abundance, and the correlation strength is significantly stronger than that of any other properties, including the halo mass (obtained from the \citealt{2017MNRAS.470.2982L} group catalog; see \autoref{sec:hosts}). One might expect that the halo mass would correlate most strongly with satellite abundance; however, gravity-only simulations have shown that even at the fixed halo mass, satellite abundance will vary with halo properties such as halo formation time \citep[e.g.,][]{Mao150302637}.
On the other hand, if the satellites are simply random draws of some underlying stellar mass function \citep[see, e.g., the Poisson point process model proposed by][]{Mao150302637}, then both the most massive satellite mass and the satellite abundance at a fixed mass limit are directly dictated by the overall amplitude of the satellite mass function, resulting in a strong correlation between those two quantities.

This effect can be seen most clearly in \autoref{fig:nsat_two_properties}, where colors of the points indicate satellite abundance, and hence, the direction of the color gradient tells us how the satellite abundance depends on the two quantities plotted, group mass and most massive satellite mass.
We can see that the color gradient goes vertically, supporting the hypothesis that the overall amplitude of the satellite mass function drives the correlation between satellite abundance and the most massive satellite mass.
A similar correlation is also observed in nearby MW-mass systems \citep{2022ApJ...930...69S}.
\autoref{fig:nsat_two_properties} also shows that at a fixed group (halo) mass, the amplitude of the satellite mass function still varies, likely with other halo properties that we are not able to measure here.

Another correlation that stands out from \autoref{fig:nsat_correlation} is the negative correlation between the host galaxy sSFR and the satellite abundance.  At a given host galaxy luminosity, galaxies with higher sSFR are bluer and have lower host halo mass \citep[e.g.,][]{bell2003}.
As such, it is not surprising that we observe this negative correlation between host sSFR and satellite abundance. In \autoref{app:additional-analyses}, we include a figure similar to \autoref{fig:nsat_two_properties} but with $y$-axis replaced with the host sSFR as \autoref{fig:nsat_two_properties_ssfr} for readers who wish to inspect this correlation further.

We plan to explore these correlations in more detail in follow-up work, where we more carefully take into account the inter-correlations among the different properties, which will allow us to distinguish the primary correlation.
In addition, in \autoref{app:additional-analyses}, we also show how the satellite SMF as a whole changes with a selection of the system properties presented here.

\subsection{Co-rotating Pairs of Satellites}
\label{sec:sat-pairs}

%%%%%%%%%%%%%%%%%%%%%%%%%%%%%%%%%%%%%%%%%
\begin{figure}[!tbp]
    \centering
    \includegraphics[width=\linewidth]{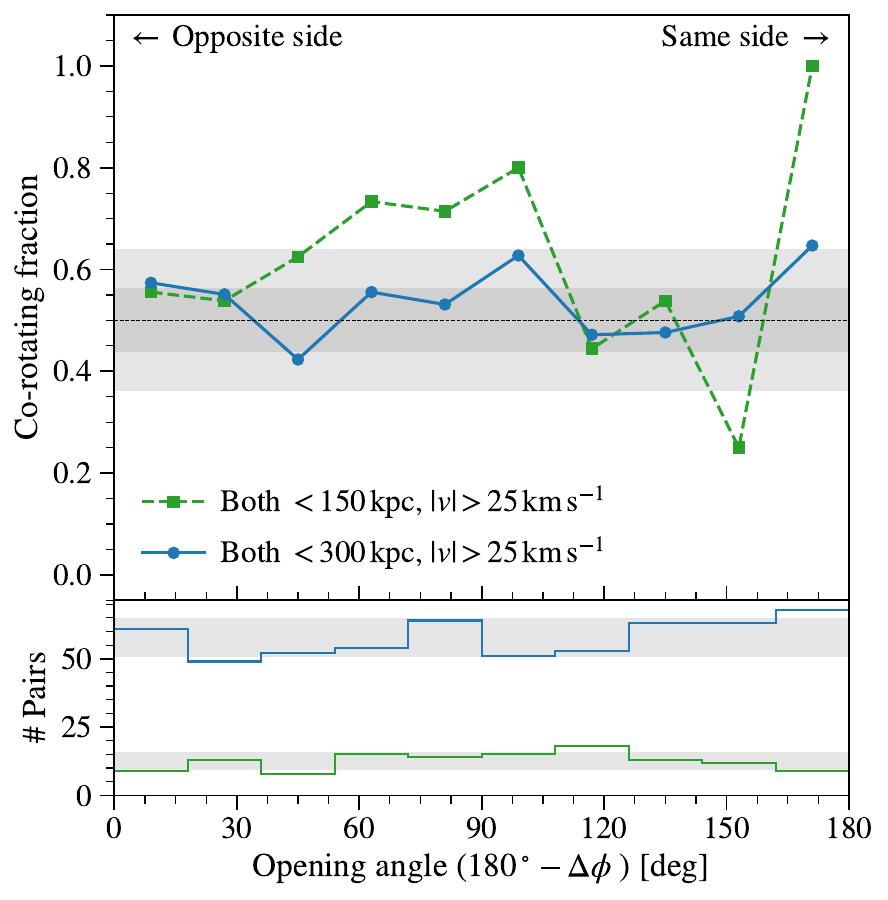}
    \caption{\textbf{Upper}: Fraction of corotating satellite pairs in bins of opening angles between the pairs. We follow the convention of \citep{Phillips2015},  such that a pair of satellites on opposite sides of a host have an opening angle of 0$^\circ$. The green squares show the corotating fraction from pairs of satellites that are within 150\,kpc of their hosts.
    The blue points show the same but including all confirmed satellites up to 300\,kpc from their hosts.
    The dark and light gray bands show the $1\sigma$ and $2\sigma$ of co-rotation fraction when all satellites are assigned a position angle (relative to their hosts) uniformly at random.
    \textbf{Lower}: The number of pairs in bins of opening angles. Same as above, green and blue lines indicate satellites that are within 150 and 300\,kpc, respectively.
    The gray bands show the $1\sigma$ distribution of the number of pairs when all satellites are assigned a position angle (relative to their hosts) uniformly at random.
    See \autoref{sec:sat-pairs} for details about how the corotating fraction is calculated.}
    \label{fig:sat_coroating_pairs}
\end{figure}
%%%%%%%%%%%%%%%%%%%%%%%%%%%%%%%%%%%%%%%%%

We update the analysis presented in \papertwo{} to identify corotating pairs of satellites, which follows the procedure presented in \citet{Phillips2015}. We identify all pairs of SAGA satellites, where a pair is any two confirmed SAGA satellites in the same host. For each pair, we calculate the ``opening angle'' with respect to their host galaxy and check whether their spectroscopic velocities have a corotating signature. Consistent with the notation in \citet{Phillips2015}, the opening angle is defined as $180^\circ$ minus the difference between position angles. Two satellites are said to be exhibiting a corotation signature if they are moving toward the opposite (same) direction with respect to the host when the opening angle is less (greater) than 90$^\circ$. Only satellites with a velocity difference with respect to their hosts greater than 25\,\kms are included in this analysis; this exclusion removes satellites whose velocity difference may be attributed with a wrong sign due to the velocity measurement uncertainty.

The upper left panel in \autoref{fig:sat_coroating_pairs} shows the fraction of corotating pairs in bins of opening angle. We repeat the analysis with all confirmed satellites passing the velocity cut (blue-filled circles) and with only satellites within 150\,kpc in projection to the host galaxies (same velocity cut applies).
We calculate the expected errors by repeating our analysis 5,000 times, but each time, we assign each satellite a random position angle sampled uniformly between 0$^\circ$ and 360$^\circ$. This procedure preserves the satellite radial and velocity distributions but would erase any corotating signal. The 1$\sigma$ interval from this random analysis is shown as the gray band. The two shades of gray band corresponds to the random analyses with two different projected distance cuts. The numbers of pairs in each opening angle are shown in the lower panel.

Overall, the signals are consistent with no corotation (the horizontal line at 0.5). In a few bins, the signal shows a 1$\sigma$--2$\sigma$ deviation, which is expected given the number of bins used in the analysis. We do not observe an excess of opposed pairs (small opening angles) that have the same sign of velocity offsets, as recently reported in SDSS data \citep{2401.10342}.
The only notable deviation is the rightmost bin: when the satellites in a pair have very similar position angles, there is an excess in the corotation fraction. The excess is more significant (3$\sigma$) in the analysis when we only include satellites within 150\,kpc, which hints that the excess comes from pairs with small separation. In \autoref{app:additional-analyses}, a follow-up analysis demonstrates there is a small excess of close satellite pairs, and the satellites in close pairs tend to have a small velocity difference (see \autoref{fig:sat_close_pairs}).

\section{Discussion}
\label{sec:discussion}

\subsection{Milky Way in the Cosmological Context}
\label{sec:mw-context}

A major science goal of the SAGA Survey is to put the Milky Way system, the most studied satellite system, in a cosmological context. First, we note that the stellar masses of the SAGA host galaxies span about 0.8 dex, with a median mass about 0.3 dex lower than that of the MW (see \autoref{fig:hosts-properties}).  Despite the host mass difference, the average satellite abundance of the SAGA systems and the shape of the average satellite SMF both agree with those of the MW.

The agreement in the average satellite abundance highlights an intriguing aspect of the MW, as one might naively expect the MW to have a higher satellite abundance compared to the SAGA systems given the host mass difference. In addition, the MW's satellite abundance is at the very low end among SAGA systems that have a satellite as massive as the LMC, as shown in the inset of \autoref{fig:nsat_correlation}.
While the existence of an LMC-mass satellite is by itself not uncommon (\autoref{fig:sat_smf_rainbow}), the MW system may differ from these systems in that the LMC joined the MW system only very recently \citep[e.g.,][]{besla2007, Kallivayalil2013}.

If we consider the MW satellite system to be composed of older, less-recently accreted satellites plus the recently accreted LMC/SMC and their satellites \citep[e.g.,][]{Kallivayalil2018, cautun2019, Patel2020, Santos-Santos2021}, many of the apparent discrepancies between the MW and SAGA systems are readily explained. In this picture, the MW system would be expected to have a low satellite abundance\footnote{Here we exclude the satellites that came in with the LMC/SMC, which will be less massive than those we probe in the SAGA Survey.} due to the MW halo's lower mass and earlier formation time before LMC/SMC accretion.
We would also expect MW satellites to have a more concentrated radial distribution due to the earlier formation time, and the fact that LMC and SMC happen to be at their pericenters makes the MW satellite radial distribution even more concentrated.
This picture is consistent with the findings of recent zoom-in simulations of Milky Way-like systems that accrete LMC analogs at late times \citep{10.1093/mnras/stt2058,2021MNRAS.504.5270D,10.1093/mnras/stad1395,Buch240408043}.

The MW also appears to have a sharp transition in the satellite quenched fraction around the stellar mass of $10^{8.5}\,\msun$: the two satellites (LMC and SMC) above this stellar mass are both star forming, and the satellites below this stellar mass are all quenched. In contrast, the average quenched fraction of the SAGA systems (\autoref{fig:quenched_frac}) has a much smoother transition, and seems to plateau around 17\% at the high-mass end ($\sim 10^9\,\msun$) and 67\% at the low-mass end ($\sim 10^7\,\msun$). It is worth noting that we have identified several star-forming satellites even below $10^7\,\msun$ in SAGA systems. This result is consistent with the ELVES Survey \citep{2022ApJ...933...47C}; however, these low-mass star-forming satellites are not observed in the Milky Way.

At the high-mass end, the apparent difference in satellite quenched fraction may be explained by the recent accretion of the LMC and the SMC. At the low-mass end, the lack of low-mass star-forming satellites in the Milky Way may be due to earlier satellite infall or the influence of the Local Group (see further discussion in \paperfour{}). However, the discrepancy in quenched fraction does not have a strong statistical significance, considering that the MW is a single host. If we draw a random sample from SAGA's average quenched fraction using the five most massive MW satellites' stellar masses, about 16\% of the draws will result in the two most massive satellites being star forming and the rest being quenched.

Simulations have shown that the recent infall of the LMC may contribute to the corotating plane of satellites in the MW \citep{10.3847/1538-4357/ac2c05,10.1093/mnras/stab955}. If the recent accretion of an LMC-mass satellite is a necessary factor for the formation of a corotating satellite plane, then the lack of corotating pairs among the SAGA systems can be understood given that the massive satellites in the SAGA systems seem to fall in earlier, based on their colors (\autoref{fig:size-color-mass}, left panel).

Overall, these comparisons highlight the impact of the merger history on the present-day satellite population. In the specific case of the LMC, studies have shown that it has significant impact on the MW's satellite population \citep{2015ApJ...813..109D,2019ApJ...885L...8W,2019MNRAS.486.2440Z,2020ApJ...893...48N}, the MW's potential as a whole \citep{2104.09515, 2021ApJ...919..109G}, and potentially the orbital properties of the satellites \citep[e.g.,][]{Patel2020}. The results from the SAGA Survey tell a similar story about the impact of massive satellites.

\subsection{Recommendations for Simulation Comparisons}
\label{sec:sim-comparison}

The \nhosts{} satellite systems presented here will greatly improve the statistical significance in studies that compare simulation predictions with observed satellite systems. Here we list specific considerations that should be implemented or discussed when comparing the SAGA data with simulations.

\begin{enumerate}
    \item Host mass: While the SAGA systems are generally MW-mass systems, the host galaxies of the SAGA systems span a range of stellar masses (see \autoref{fig:hosts-properties}), with a median slightly lower than the MW. We do not have precise observational constraints on the halo masses. Based on the estimated halo masses from the \citet{2017MNRAS.470.2982L} group catalog (see \autoref{sec:hosts}), the distribution of the halo masses of SAGA systems peaks around $10^{12}\,\msun$ but has the 16th and 84th percentiles at $10^{11.9}$ and $10^{12.3}\,\msun$, respectively.\footnote{This halo mass distribution is consistent with subhalo abundance matching estimate (see Figure 2 of \paperone{}).}
    Hence, care should be taken when comparing the SAGA systems with simulated halos, especially if the latter were selected with a fixed halo mass. If possible, it is recommended to draw a stellar mass sample from the simulation that mimics the SAGA hosts' stellar mass distribution.
    \item Host isolation/environment: We apply isolation criteria when selecting SAGA systems (see \autoref{sec:hosts}), and the relatively isolated environment can have an impact on the satellite populations. Similar isolation criteria should be applied when comparing to simulated systems. If the observational isolation criteria are not easy to implement in the simulation, an approximation would be to select host halos that do not have other MW-mass (or larger) companions within about 2 virial radii (or 600\,kpc). We did not impose direct constraints on the host galaxy's color, morphology, or star formation rate when selecting the SAGA systems.
    \item Satellite definition: Since we only have redshift measurements for the SAGA satellites, the satellites are defined to have consistent redshifts (velocities) as their respective host galaxies and to be within 300\,kpc in projection (see \autoref{sec:dr3-satellites}). This definition differs from the typical satellite (or subhalo) definition used in simulations (i.e., within the virial radius). The observational definition is straightforward to implement in simulations and should be used in comparisons between SAGA data and simulations.
    \item Incompleteness correction: Based on the redshift data we have obtained, we have an accurate estimate of the number of satellites for which we have not obtained redshifts. This incompleteness correction (see \autoref{app:model}) has been incorporated in our satellite SMF and quenched fraction analyses, and the correction is non-negligible below $\mstar = 10^{7.5}\,\msun$. When comparing SAGA data with simulations, we recommend incorporating the incompleteness correction using the table we provided (\autoref{app:tables}). Forward modeling the observational incompleteness can be challenging as the incompleteness depends on at least galaxy color and size. Also, the survey incompleteness does not correspond to a simple surface brightness limit (see \autoref{fig:sb-mag}). However, if a forward modeling approach is taken and the observational incompleteness is applied to simulated data, then the SAGA data being compared should contain only confirmed satellites.
    \item Satellite properties and uncertainties: Derived satellite properties, such as stellar masses and star formation rates, tend to have higher uncertainties than directly observed properties, such as luminosities and colors. For example, the estimated stellar masses in this data release have a typical uncertainty of about 0.2 dex. Conversely, the observed properties modeled in simulations may have higher systematic uncertainties. If modeling the observed properties is not the main point of a comparison, we recommend using the derived properties (e.g., stellar masses) because with this approach, the main uncertainty consistently comes from the SAGA data; this is likely  to enable more robust comparisons among different simulations.
\end{enumerate}

\subsection{Follow-up Studies and Observations}
\label{sec:follow-up}

In \autoref{sec:qf}, we discussed the satellite quenched fraction as a function of stellar mass.
We refer readers to \paperfour{} for a more detailed study of the satellite quenched fraction and star formation rates. Specifically, in \paperfour{}, we inspect the quenched fraction as a function of projected distance and also compare the satellite quenched fraction between the SAGA and ELVES surveys. We also study satellite gas-phase metallicities and gas content in \paperfour{}, repeating our host correlation analysis for the quenched fraction of satellites.

On the theory side, in \autoref{sec:nsat} we only briefly compare the satellite abundance with the UM-SAGA model. In \paperfive{}, we compare the SAGA DR3 results with the model predictions in detail; generally, we find good agreements between SAGA results and simulations. The use of an empirical model such as the UM-SAGA model allows us to study the model ingredients required  to reproduce not only the satellite SMFs but also the quenched fraction. We refer readers to \paperfive{} for those discussions.

In \autoref{sec:correlation}, we explored the correlations between satellite abundance and various satellite or host properties.  Because correlation does not imply causation, we will need to compare the observed correlations with simulations to properly interpret them. We also need to account for the inter-correlations among the different properties that we explored. We leave these directions to a follow-up study.

In addition, we have several planned follow-up observations. Currently, we do not have \textsc{Hi} measurements for all the SAGA satellites (Section~2.5 of \paperfour{}). Efforts have been made \citep[e.g.,][]{2311.02152} to carry out \textsc{Hi} observation for some SAGA satellites, and we plan to continue this effort for the remaining satellites. \textsc{Hi} measurements of the sample will provide a complete picture of how host--satellite interaction affects the gas content and star formation of the satellites.

We are in the process of obtaining spatially resolved high-resolution spectroscopy for the bright SAGA satellites to investigate gas kinematics and rotation curves. These data will also be used to estimate dynamical masses.
We are working toward constructing complete satellite mass functions down to $M_\star \ge 10^{7.5}\,\msun$ to enable further small-scale tests of \LCDM{} such as quantifying the variance in mass functions for host halos of a given mass. For instance, directly measured satellite mass functions will enable tests of whether the mass function of the MW and the gap in the maximum rotation velocity distribution for MW satellites is consistent with being drawn from the distribution of SAGA mass functions.

Beyond the satellite population, the background galaxy redshifts in the SAGA footprint will enable a range of studies, including several that have already been published \citep{xSAGA,DESI-LOWZ,SOM-lensing,2401.16469}. Planned studies include the star-forming main sequence of the background galaxies (E.~Kado-Fong et al. 2024, in preparation) and AGN searches.
We also plan to curate NUV star formation rates for low-redshift, low-mass background galaxy redshifts in our future work.

\section{Summary}
\label{sec:summary}

We present the third data release of the SAGA Survey.
This data release includes a census of satellite galaxies in \nhosts{} Milky Way-mass systems and also a redshift catalog of galaxies that are in the sky footprint of these systems.
As in the last two data releases, the host galaxies (also called primary or central galaxies) in the SAGA systems have distances of 25--40.75 Mpc and are selected by their $K$-band luminosities and certain environment conditions.
The survey design, including the host selection, is detailed in \autoref{sec:survey}.

\subsection{Comparison with SAGA DR2}

For readers familiar with our second data release (DR2; as described in \papertwo{}), we compare these two data releases to highlight the similarities and differences.

\begin{enumerate}
    \item A total of \nhosts{} systems has been surveyed, including the 36 systems presented in DR2. Sixty-five systems are new in DR3. The host selection criteria remain the same (see \autoref{sec:hosts}).
    \item The photometric catalog in SAGA DR3 is based on DESI Legacy Imaging Surveys DR9 (see \autoref{sec:photometry}). We updated our catalog to use DESI Imaging DR9 for all SAGA systems, including those already published in \papertwo{}. The optical magnitudes reported in this work are based on DECam filters. We also include GALEX photometry in our DR3 catalog (see \autoref{sec:galex}).
    \item We follow the same target selection strategy as in \papertwo{}. We devote about half of our spectroscopic resources to the primary targeting region (a photometric region that almost all satellites are in) and the remaining to exploration mode (see \autoref{sec:targets}).
    \item As in \papertwo{}, we primarily use AAT and MMT for obtaining redshifts (see \autoref{sec:spec-multifiber}). A small fraction of redshifts were obtained with single-slit observations on Palomar, SALT, and Keck; the latter two are new in DR3 (see \autoref{sec:spec-single}).
    \item We also include literature redshifts available in the SAGA footprint (see \autoref{sec:spec-literature}). The majority of the literature redshifts are for distant background galaxies in the SAGA footprint.  We have included more surveys in DR3, the most notable being the DESI EDR. We also incorporated HI measurements from HI surveys.
    \item In addition to the satellite catalog, in DR3 we also publish a galaxy redshift catalog for all the galaxy redshifts we obtained and compiled. The DR3 data products are described in \autoref{sec:dr3} and are included as machine-readable tables.
    \item While the SAGA Survey has a designed depth of $r_o < 20.7$, we define three science samples based on stellar masses in DR3, corresponding to different levels of survey completeness: Gold ($\mstar \geq 10^{7.5} \msun$; highly complete), Silver ($10^{6.75} \leq \mstar/\msun < 10^{7.5}$; highly complete for star-forming satellites), and Participation ($\mstar < 10^{6.75} \msun$; not highly complete). These three science samples allow us to present our results more clearly; see \autoref{sec:samples} for details.
\end{enumerate}

\subsection{SAGA DR3 Main Findings}

Our main findings from this data release are summarized as follows. These findings were discussed in Sections~\ref{sec:smf}--\ref{sec:sat-pairs}. Overall, they are consistent with those reported in \papertwo{}, but with higher significance or robustness.
\begin{enumerate}
    \item The number of confirmed satellites in the \nhosts{} SAGA systems ranges from zero to 13 (\autoref{fig:saga-all-sats}). There exists significant system-to-system scatter in the individual satellite stellar mass function (SMF) of the \nhosts{} SAGA systems, and the numbers of satellites at different stellar masses are highly correlated (\autoref{fig:sat_smf_rainbow}). The average SAGA satellite stellar mass function is consistent with that of the ELVES Survey, and its amplitude and slope are consistent with MW's SMF (\autoref{fig:sat_lf}).
    \item The satellite quenched fraction for $\mstar < 10^{8.5} \msun$ among the SAGA systems is about $1\sigma$ lower than that in the Local Group (\autoref{fig:quenched_frac}). In particular, on the low mass end ($\mstar < 10^{7.5} \msun$), we identified 135 satellites with confirmed redshifts, of which 98 are star forming.
    \item Whether a satellite is star forming or quenched strongly correlates with the color of the satellite; however, no simple color cut can cleanly separate star-forming and quenched satellites (\autoref{fig:size-color-mass}, left). The fraction of satellites that are quenched also correlate (negatively) with the projected distance to the host galaxy (\autoref{fig:size-color-mass}, right).
    \item The galaxy size--mass relation of the SAGA satellites follows that of the MW and M\,31 satellites, with no visible difference between star-forming or quenched satellites (\autoref{fig:size-color-mass}, middle).
    \item The average satellite radial distribution of the \nhosts{} SAGA systems is less concentrated than that of the MW, but the 25\% most concentrated SAGA systems show a similar radial distribution to the MW (\autoref{fig:sat_radial}).
    \item While the numbers of satellites in the SAGA systems have a wide range, the distribution of the satellite abundance is consistent with \LCDM{} predictions (\autoref{fig:nsat-dist}). We find that the satellite abundance correlates most strongly with the stellar mass of the most massive satellite in each system, among a large set of quantities we examined  (\autoref{fig:nsat_correlation}).
    \item We find no evidence for corotating planes of satellites in the \nhosts{} SAGA satellite systems; however, there is a hint that physically close satellite pairs tend to have similar velocities (\autoref{fig:sat_coroating_pairs}).
\end{enumerate}

We discussed MW's satellite system in the context of the SAGA result in \autoref{sec:mw-context}.
In short, the existence of a recently accreted LMC in the MW system may explain many differences between the MW and the median SAGA system.
We also discussed several upcoming and planned future works in \autoref{sec:follow-up}. In particular, in \paperfour{}, we investigate the SAGA satellites' star formation properties, and in \paperfive{}, we compare the SAGA results with an updated UniverseMachine model (an empirical galaxy formation model applied to gravity-only cosmological  simulations).

In addition to the satellite systems, the background galaxy redshift catalog included in this data release will enable a wide range of science exploration. As shown in Figures~\ref{fig:z_distribution} and \ref{fig:redshift-mass}, the \nzsaga{} galaxy redshifts obtained by the SAGA Survey fill in a low-mass ($ 10^7 < \mstar/\msun < 10^9 $; $18 < r_o < 21$), low-redshift ($z < 0.1$) regime where galaxy redshifts are still sparse.
The photometric selection method we have developed will also help future surveys to identify low-mass, low-redshift galaxy candidates.

This data release marks a major milestone of the SAGA Survey, as the \nhosts{} satellite systems will enable many statistical comparisons with other observations and simulations. In  \autoref{sec:sim-comparison}, we listed specific recommendations for comparing the SAGA results with simulations. We look forward to independent studies of the SAGA satellite systems in the future.

As more and more observational data for other MW-mass satellite systems become available, not only do we gain a better understanding of galaxy formation physics, but also the unique history of our own MW satellite system.
While one can say any individual system is unique, the SAGA Survey has provided a perspective to examine in what ways the MW satellite system stands out. With this knowledge in hand, future observations of the MW satellite system will become even more powerful to enhance our understanding of galaxy formation and the nature of dark matter.

%%%%%%%%%%%%%%%%%%%%%%%%%%%%%%%%%%%%%%%%%
%acknowledgments
\medskip
\input{acknowledgments}
%%%%%%%%%%%%%%%%%%%%%%%%%%%%%%%%%%%%%%%%%

\facilities{AAT (2dF), MMT (Hectospec), Hale (DBSP), Keck:II (DEIMOS), SALT}

\software{%
Numpy \citep{numpy, 2020NumPy-Array},
SciPy \citep{2020SciPy-NMeth},
numexpr \citep{numexpr},
Matplotlib \citep{matplotlib},
IPython \citep{ipython},
Jupyter \citep{jupyter},
Astropy \citep{astropy},
statsmodels \citep{statsmodels},
astroplan \citep{astroplan2018},
healpy \citep{2005ApJ...622..759G, Zonca2019},
easyquery (\https{github.com/yymao/easyquery}),
adstex (\https{github.com/yymao/adstex}),
xfitfibs \citep{10.1117/12.316837},
HSRED \citep{10.1086/589642},
qplot (\https{github.com/bjweiner/qplot}),
2dfdr \citep{2015ascl.soft05015A},
Marz \citep{10.1016/j.ascom.2016.03.001}%
}

%%%%%%%%%%%%%%%%%%%%%%%%%%%%%%%%%%%%%%%%%
% REFERENCES
\let\Oldnewpage\newpage
\def\newpage{\bigskip}

\let\Oldthebibliography\thebibliography
\def\thebibliography{\Oldthebibliography{} \setlength{\baselineskip}{10pt plus 1pt}}

\bibliographystyle{aasjournal}
\bibliography{references}

\let\newpage\Oldnewpage
%%%%%%%%%%%%%%%%%%%%%%%%%%%%%%%%%%%%%%%%%

\appendix
\counterwithin{figure}{section}
\counterwithin{table}{section}

\section{Supplemental Analyses}
\label{app:additional-analyses}

%%%%%%%%%%%%%%%%%%%%%%%%%%%%%%%%%%%%%%%%%
\begin{figure*}[!tbp]
    \centering
    \includegraphics[width=\textwidth]{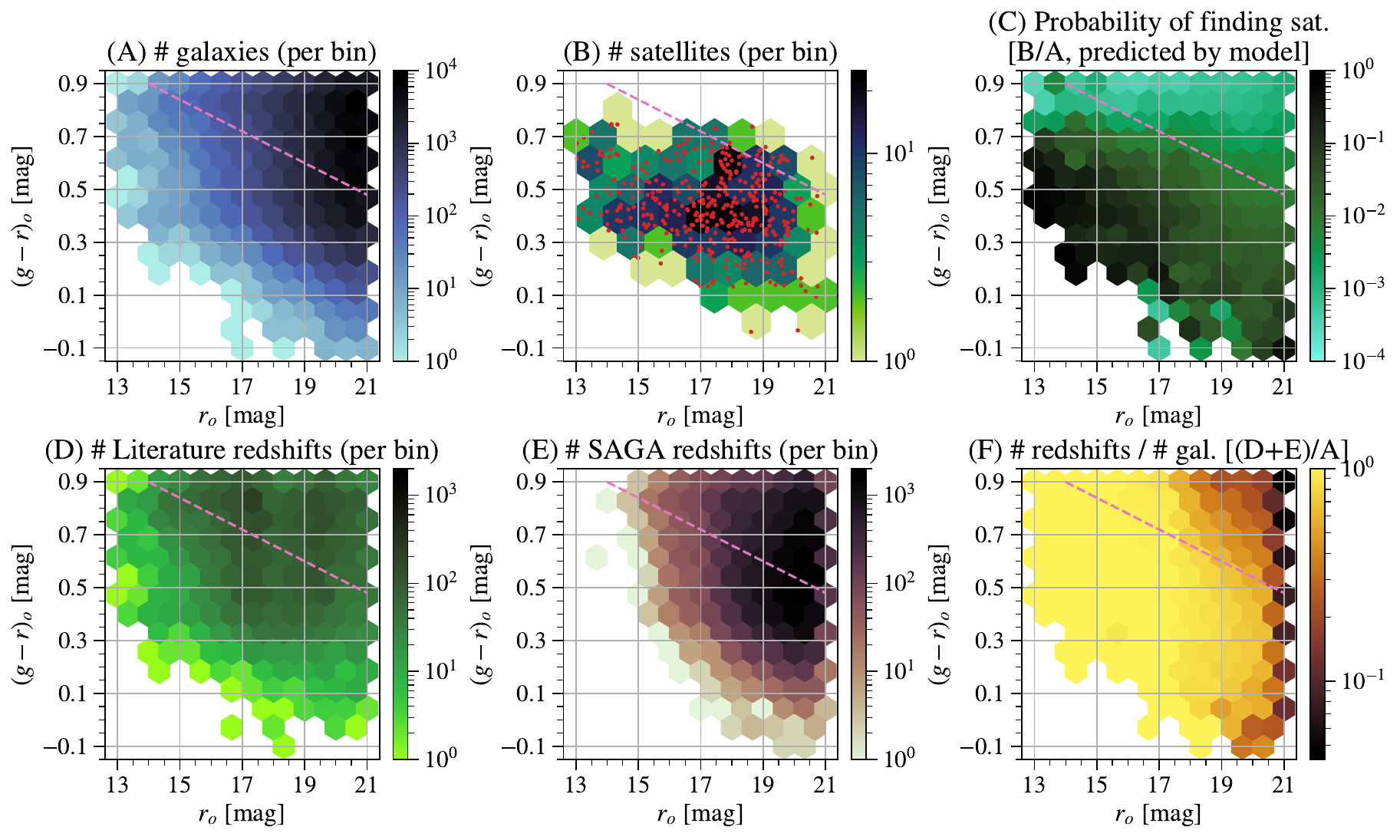}
    \caption{The color--magnitude distribution of (A) all galaxies (top left), (B) confirmed satellites (top center), (C) probability of finding a SAGA satellite (top right), (D) galaxies with literature redshifts (lower left), (E) galaxies with SAGA redshifts (lower center), and (F) the fraction of confirmed redshifts (lower right). Each panel is plotted in the $(g-r)_o$ color vs. $r_0$ magnitude photometric space; color indicates the number of galaxies per hexagon bin for Panels (A), (B), (D), and (E), and the fraction per bin for Panels (C) and (F).
    Panel (A) includes all galaxies in the cleaned photometric catalogs within 300 kpc of each SAGA host galaxy.
    Panel (B) shows the number of confirmed satellites; the red points indicate the individual satellites.
    Panel (C) shows the ratio of Panel (B) to Panel (A), with the incompleteness correction predicted by the model described in \autoref{app:model}.
    Panel (F) shows the ratio of the sum of Panels (D) and (E) to Panel (A).}
    \label{fig:targets}
\end{figure*}
%%%%%%%%%%%%%%%%%%%%%%%%%%%%%%%%%%%%%%%%
%%%%%%%%%%%%%%%%%%%%%%%%%%%%%%%%%%%%%%%%%
\begin{figure}[!htbp]
    \centering
    \includegraphics[width=\linewidth]{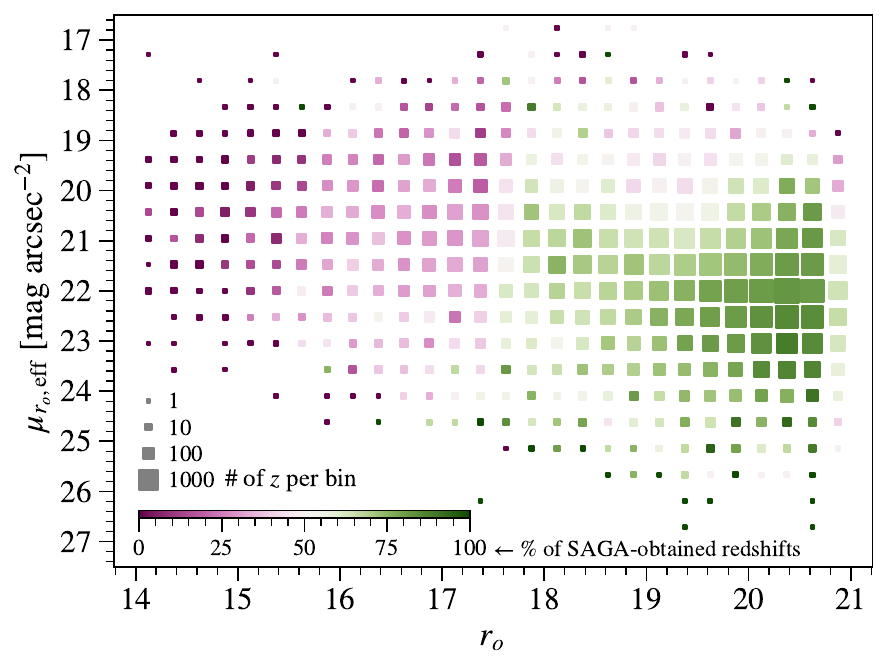}
    \caption{Same as the lower panel of \autoref{fig:z_distribution} but in bins of effective surface brightness ($x$-axis) and r-band magnitude ($x$-axis).}
    \label{fig:z_distribution_sb_mag}
\end{figure}
%%%%%%%%%%%%%%%%%%%%%%%%%%%%%%%%%%%%%%%%
%%%%%%%%%%%%%%%%%%%%%%%%%%%%%%%%%%%%%%%%%
\begin{figure*}[!tbp]
    \centering
    \includegraphics[width=\textwidth]{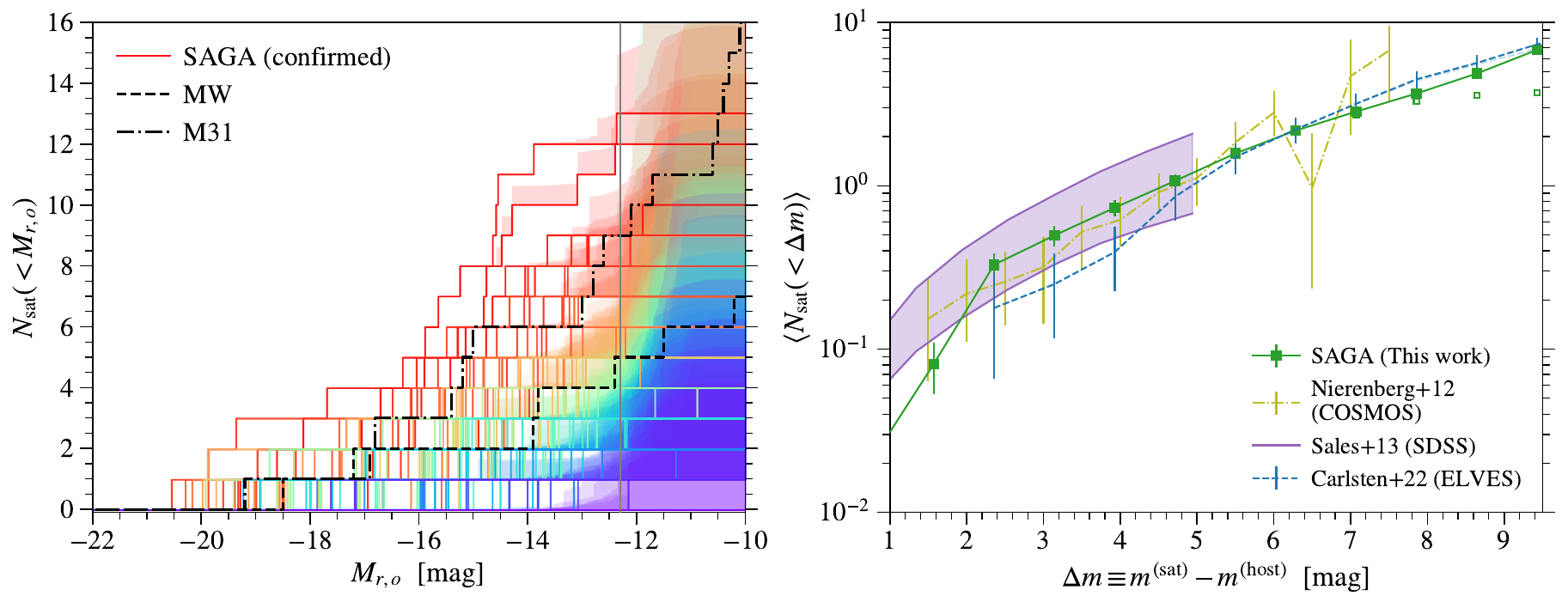}
    \caption{Reproduction of Figure~12 (satellite luminosity functions) of \papertwo{} from DR2 with the DR3 SAGA data. \textbf{Left}: Cumulative satellite $r$-band luminosity function for each of the \nhosts{} SAGA systems (colored solid lines), MW (black dashed line), and M\,31 (black dot-dashed line). The colored solid lines show only the confirmed SAGA satellites, and the shaded region shows incompleteness correction as discussed in \autoref{app:model}. The SAGA Survey's designed depth is about $M_{r,o} = -12.$, shown as the vertical gray line.
    \textbf{Right}: Average cumulative satellite $r$-band luminosity function, with satellite luminosity normalized by the luminosities of their respective host galaxies (calculated as differences in magnitudes). The average luminosity function of SAGA systems is shown as the green filled squares. The incompleteness correction is included in the filled squares, and uncorrected results are shown as green open squares. The yellow dot-dashed line shows the average luminosity function of systems in the COSMOS field (\citealt{Nierenberg2012}), the purple band SDSS (\citealt{Sales2013}), and the blue dashed line the incompleteness-corrected ELVES Survey \citep[the uncorrected version is shown as the faint blue dashed line that is almost identical to the corrected line]{2022ApJ...933...47C}. See the text in \autoref{app:additional-analyses} for the specific system selections. The error bars show Poisson noise on the luminosity functions.}
    \label{fig:sat_lf}
\end{figure*}
%%%%%%%%%%%%%%%%%%%%%%%%%%%%%%%%%%%%%%%%%
%%%%%%%%%%%%%%%%%%%%%%%%%%%%%%%%%%%%%%%%%
\begin{figure*}[!tbp]
    \centering
    \includegraphics[width=\textwidth]{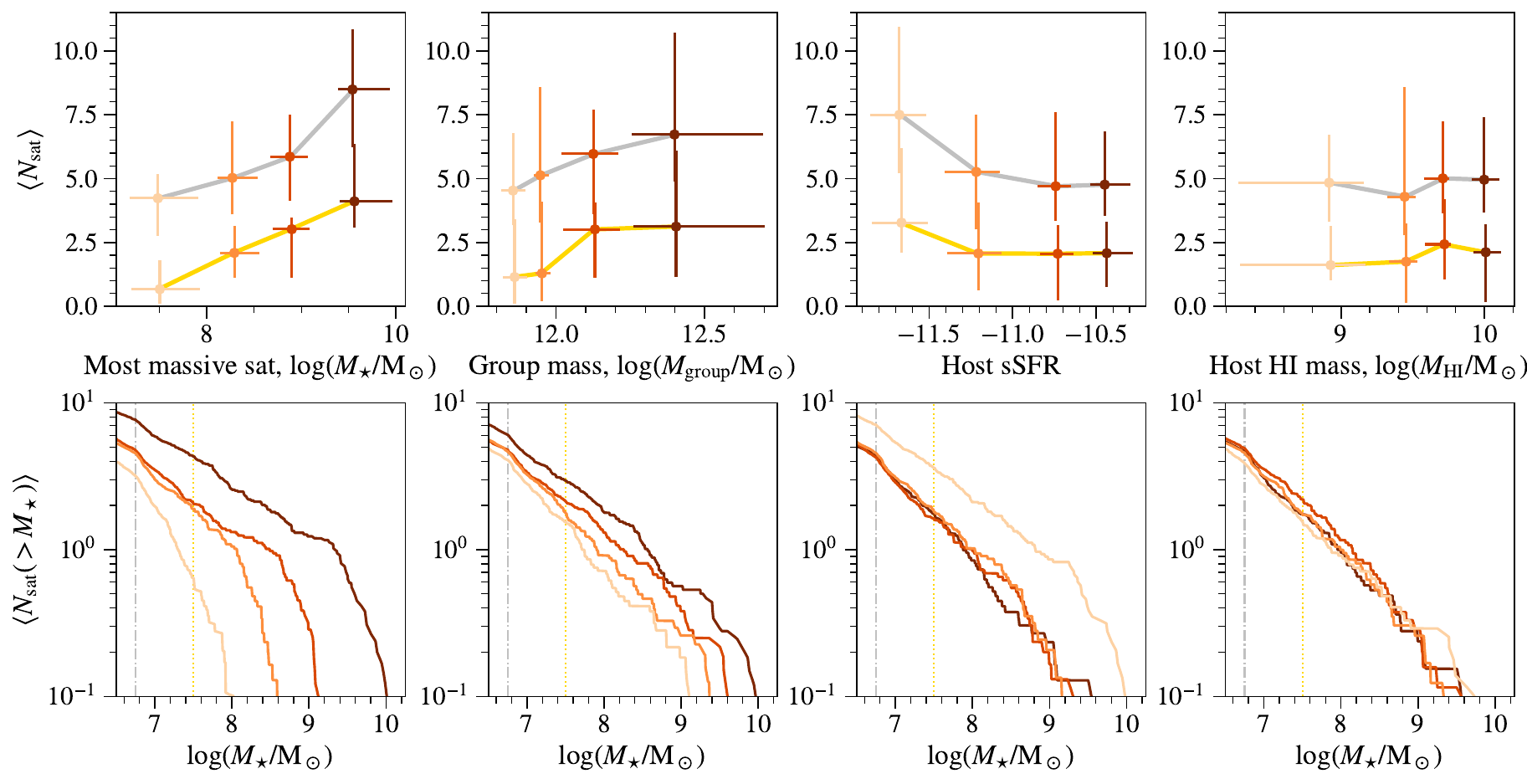}
    \caption{How SAGA satellite systems vary across host/system properties.  The upper panels are the average number of satellites per system, and the lower panels are satellite stellar mass functions. The SAGA systems are split into quartiles using four different host/system properties: (from left to right) the stellar mass of the most massive satellite, the group (halo) mass, the specific star formation rate of the host galaxy, and the HI mass of the host galaxy. Each panel in the upper row shows the average number of satellites per system as a function of the corresponding host/system property. The points connected by the gold line (lower line) show the average number of satellites in the Gold sample, and the higher points connected by the silver line (upper line) show that of the Gold$+$Silver sample. Each panel in the lower row shows how the average cumulative stellar mass function changes with the corresponding host/system property. The colors of the points in the upper panel correspond to the colors of the lines in the lower panel, with darker colors indicating higher quartiles.}
    \label{fig:smf_split}
\end{figure*}
%%%%%%%%%%%%%%%%%%%%%%%%%%%%%%%%%%%%%%%%%
%%%%%%%%%%%%%%%%%%%%%%%%%%%%%%%%%%%%%%%%%
\begin{figure}[!htbp]
    \centering
    \includegraphics[width=\linewidth]{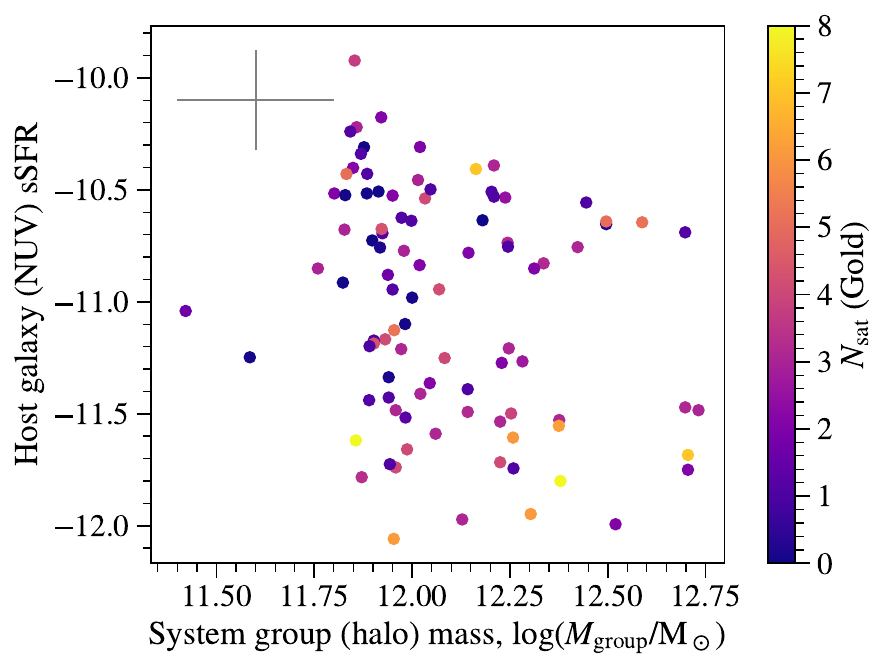}
    \caption{Same as \autoref{fig:nsat_two_properties} but the $y$-axis here shows the host galaxy's sSFR (inferred from NUV measurements; values shown are $\log[\dot{M}_\star / {M}_\star \cdot \text{yr}]$).}
    \label{fig:nsat_two_properties_ssfr}
\end{figure}
%%%%%%%%%%%%%%%%%%%%%%%%%%%%%%%%%%%%%%%%%
%%%%%%%%%%%%%%%%%%%%%%%%%%%%%%%%%%%%%%%%%
\begin{figure}[!htbp]
    \centering
    \includegraphics[width=\linewidth]{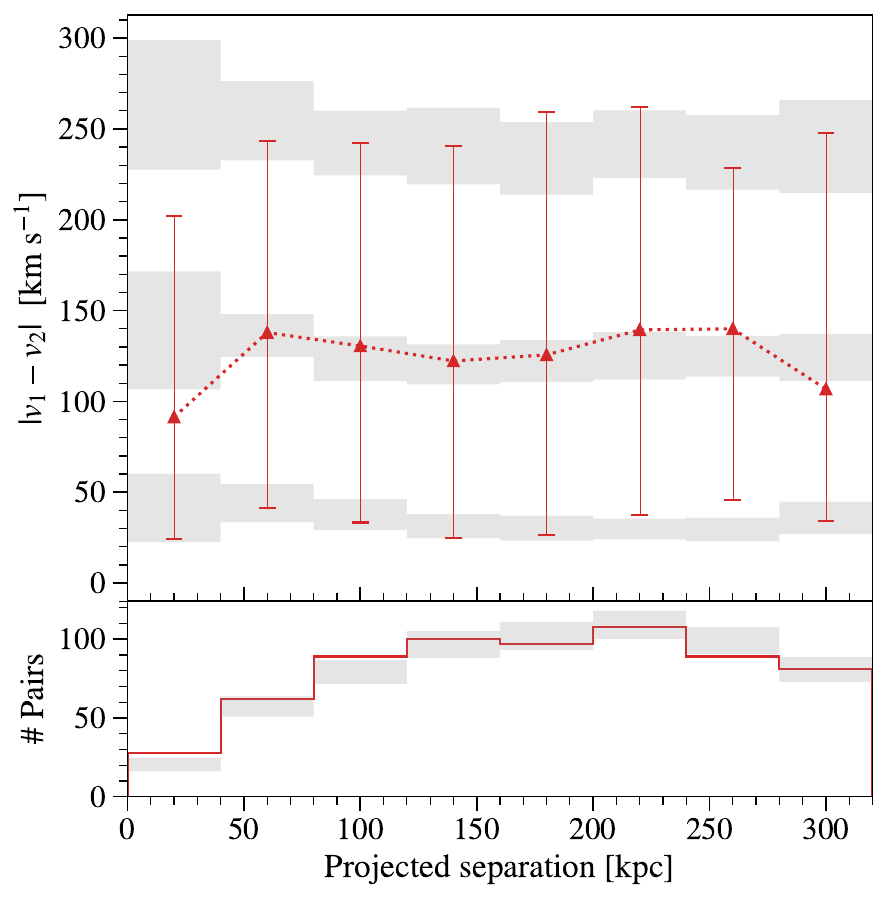}
    \caption{\textbf{Upper}: The velocity difference between all pairs of confirmed satellites in the SAGA Survey, binned by the projected separation between the two satellites.
    In each bin, the distribution of the velocity difference is shown as the median value (red triangles) and the 16th and 84th percentiles (lower and higher caps).
    The light gray bands show the $1\sigma$ distribution of the velocity difference when all satellites are assigned a position angle (relative to their hosts) uniformly at random.
    \textbf{Lower}: The number of pairs in bins of projected separation. The observed numbers are shown in red lines. The gray bands show the $1\sigma$ distribution of the number of pairs when all satellites are assigned a position angle (relative to their hosts) uniformly at random.}
    \label{fig:sat_close_pairs}
\end{figure}
%%%%%%%%%%%%%%%%%%%%%%%%%%%%%%%%%%%%%%%%%

In this section, we present several supplemental analyses that some readers may find interesting or useful.

In \autoref{fig:targets} we show the density of several galaxy samples in the color--magnitude plane (i.e., $(g-r)_o$ vs. $r_o$). The galaxy samples shown are
\begin{enumerate}
    \item Panel (A): All galaxies in the photometric catalog.
    \item Panel (B): Satellite galaxies (see \autoref{sec:dr3-satellites}) with confirmed redshifts. In addition to the density map, individual satellite galaxies are also shown as red points.
    \item Panel (C): All galaxies but weighted by $\mathcal{R}_\mathrm{sat}$ (see \autoref{app:model}); this is equivalent to probability of finding a SAGA satellite in the color--magnitude plane.
    \item Panel (D): Galaxies with literature redshifts.  One can observe two dark ridges around $r_o=17.7$ and 19.5, which correspond to the survey depths of SDSS and GAMA.
    \item Panel (E): Galaxies with redshifts that are first obtained by SAGA. Note that many of the SAGA redshifts are outside the primary targeting region, which is the region below the pink dashed line.
    \item Panel (F): This panel shows the fraction of galaxies with confirmed redshifts. In other words, it is the ratio of the combined density shown in Panels (D) and (E) to the density shown in Panel (A).
\end{enumerate}

\autoref{fig:z_distribution_sb_mag} summarizes the fraction of SAGA-obtained galaxy redshifts as a function of $r$-band magnitude and surface brightness, complementing the lower panel of \autoref{fig:z_distribution}. Similar to what we observed in \autoref{fig:z_distribution}, \autoref{fig:z_distribution_sb_mag} also shows that most redshifts for galaxies fainter than $r_o = 17.77$ (SDSS limit) were obtained by SAGA. It is worth noting that around $r_o = 17.5$, the SAGA Survey obtained many redshifts for galaxies with lower surface brightness.

In \autoref{fig:sat_lf}, we reproduce Figure~12 (satellite luminosity functions) of \papertwo{} with the DR3 SAGA data for readers who wish to make a direct comparison with our DR2 results. The left panel is similar to \autoref{fig:sat_smf_rainbow}; however, in this figure, to be consistent with \papertwo{}, the cumulative luminosity function for confirmed satellites is shown, with the shaded regions indicating incompleteness correction.

The right panel of \autoref{fig:sat_lf} shows the incompleteness-corrected cumulative luminosity function with respect to the host galaxies' luminosities. Overall, the DR2 and DR3 results are consistent, but the error bars on the DR3 data are significantly smaller, especially on the high-mass end. The SAGA satellite luminosity function has a dip at the bright (small $\Delta m$) end due to the isolation criteria we imposed on the SAGA hosts (see \autoref{sec:hosts}).

The COSMOS and SDSS lines are taken from \citet[][Figure~7, top left panel]{Nierenberg2012} and \citet[Figure~3, right panel]{Sales2013}, respectively, and are identical to the version shown in \papertwo{}.
The COSMOS hosts are selected from $0.1 <z<0.4$ and $10.5 < \log[\mstar/\msun] < 11$.
The SDSS hosts selected from $z<0.055$, and the lower and upper boundaries of the purple band represent the samples of $10 < \log(\mstar/\msun)<10.5$ and $10.5 < \log(\mstar/\msun)<11$, respectively.
The Local Volume line, on the other hand, is updated to the ELVES Survey result \citep{2022ApJ...933...47C}, using only the 14 hosts that are consistent with SAGA host selection (see \autoref{sec:smf}). The ELVES survey areas and survey incompleteness are taken into account when calculating the average.

To supplement the correlation results presented in \autoref{sec:correlation} and \autoref{fig:nsat_correlation}, we select four host/system properties and show how satellite abundance varies with these properties in \autoref{fig:smf_split}. The four host/system properties shown are the stellar mass of the most massive satellite, the group (halo) mass, the specific star formation rate of the host galaxy, and the HI mass of the host galaxy. The first three of these properties have strong correlation with satellite abundance as discussed in \autoref{sec:correlation}. The HI mass of the host galaxy is shown here as an example of a system property that lacks correlation with satellite abundance.

For each system property, we split the SAGA systems into four groups based on the property quartiles. Each group has a quarter of SAGA systems ($\sim$\,25).  Then, we plot the number of satellites as a function of the property (upper row) and the cumulative SMFs for the four groups (lower row). This visualization allows us to understand what might be driving the correction (or lack of). For example, the third column shows that the SAGA hosts that are quenched (lowest sSFR) tend to have more satellites, likely due to those systems' higher intrinsic masses. On the other hand, the satellite abundance around star-forming hosts does not have a strong dependence on the host sSFR, which is consistent with the lack of dependence on the host HI gas mass (fourth column).

To further inspect the correlation between host sSFR and satellite abundance, we show how satellite abundance depends on host sSFR and group mass in \autoref{fig:nsat_two_properties_ssfr}. We can observe a weak negative correlation between the host sSFR and system group mass, which contributes to the correlation between host sSFR and satellite abundance. However, at the fixed group mass, the most quenched host galaxies still tend to have more satellites.

To supplement the result presented in \autoref{sec:sat-pairs}, we further examine if close satellite pairs are more likely to have similar velocities. In \autoref{fig:sat_close_pairs}, we directly investigate how the velocity difference in each pair varies with the projected separation. In bins of projected separation, we show the median and 16th and 84th percentiles of the velocity difference of the two satellites in pairs. We also repeat the same random position angle analysis 5,000 times and show the 1$\sigma$ interval (for each statistic) in gray.
We can observe that in the leftmost bin (pairs with a projected separation less than 40\,kpc), the velocity difference distribution skews toward the lower end. In particular, the median and 16th and 84th percentiles are all $\gtrsim 1\sigma$ lower compared to the random distribution.
The lower panel of \autoref{fig:sat_close_pairs} also shows that there are more close pairs when compared to the random sample (again, at about $1\sigma$ level).
While these results are not statistically significant enough to draw a strong conclusion, they do indicate that the close pairs in SAGA systems contribute to the signal in the last bin of \autoref{fig:sat_coroating_pairs}.

\section{Survey Completeness}
\label{sec:completeness}

Throughout the paper, we use the word ``incompleteness'' to refer to any satellite in SAGA systems that we have not identified. There are three categories that may contribute to the survey incompleteness:
\begin{enumerate}
    \item The satellite is not detected as a source in the photometric catalog.
    \item The satellite is detected in the photometric catalog but has an underestimated luminosity and fell out from our sample.
    \item The satellite is present in the photometric catalog but does not have a confirmed redshift.
\end{enumerate}

We believe the first category is extremely rare and negligible for the purpose of this study. The DESI Legacy Imaging Surveys have a typical depth of $r=23.9$ ($5\sigma$ point-spread function depth; \citealt{Dey2019}).  Given the SAGA Survey's magnitude limit of $r_o=20.7$, we expect SAGA's galaxy sources to be well characterized in the DESI Imaging photometric catalog.  As discussed in \autoref{sec:photometry}, we compared our photometric catalog with the SMUDGes catalog \citep{2023ApJS..267...27Z} and found no galaxies missing in our catalog. In addition, we conducted a careful visual inspection of the full SAGA footprint to search for any low surface brightness galaxies that are not identified as sources in the photometric catalog, and we did not find any.

Some very low surface brightness galaxies may be in the second category. In our comparison with the SMUDGes catalog, we found 10 objects in this category (among all $r_0 < 20.7$ galaxies in the SAGA footprint); that is, they have an $r$-band magnitude brighter than 20.7 in the SMUDGes catalog but not in the original DESI Imaging catalog, which the SAGA Survey uses. However, given that biases in the measured luminosities of faint objects exist in all photometric surveys, we choose not to correct this issue. The underestimated luminosities affect only a small portion of our satellite population, and have little impact on our Gold science sample ($\mstar > 10^{7.5} \msun$).

Unlike the first two categories, the third category (sources that we did not obtain confirmed redshifts) can have a significant impact on our science results. Hence, we carefully quantify how the third category contributes to our survey's incompleteness. For objects in the third category, we can use the available photometric information to estimate how many satellites we might have missed in the photometric catalog. This estimation is then used as our incompleteness correction as the first two categories have negligible or unimportant contribution to our survey incompleteness. We describe this estimation in detail below.

\subsection{Incompleteness Correction Model}
\label{app:model}

To estimate how many satellites we might have missed (that is, did not have a confirmed redshift) in the photometric catalog, we first want to construct a model that can predict how likely each galaxy in the photometric catalog is to be a satellite as a function of selected photometric properties.
The model we use here is similar to the one introduced in Appendix~B of \papertwo{}, but two significant modifications are made.

First, the photometric properties we choose in this work are $(g_o, r_o, r_{\text{fiber},o})$, where $r_{\text{fiber},o}$ is the magnitude with a fiber of diameter $1.''5$ in 1$''$ Gaussian seeing. In addition, these three photometric properties only enter the model in the specific combination of $(r_{\text{fiber},o} - 0.65 r_o, \, g_o-r_o)$.
The choice of this particular combination will become apparent as we introduce the second modification below. To summarize, the model can be expressed as follows:
\begin{align*}
& \mathcal{R}_\mathrm{sat}(g_o, r_o, r_{\text{fiber},o}) \\
& \quad = \frac{\mathcal{R}_\mathrm{max}}{1+\exp\{-[\beta_0 + \beta_1 (r_{\text{fiber},o} - 0.65 r_o) + \beta_2 (g_o-r_o)]\}}.
\label{eq:p_sat_formula}
\end{align*}
Here, $\mathcal{R}_\mathrm{sat}$ is the rate (or probability) of finding a SAGA satellite at a given location of the photometric space $(g_o, r_o, r_{\text{fiber},o})$, and $\mathcal{R}_\mathrm{max}$ is the maximum value  $\mathcal{R}_\mathrm{sat}$ can reach.

The second modification is the data used to fit the model. To account for the possibility that the galaxy redshifts we obtained in the faint end (roughly $r > 19.5$) may not be representative (for example, we may preferentially obtain more redshifts for bluer objects), we only use the most complete part of the survey to fit the model. We select 44 systems in which no galaxy brighter than $r_o = 19.5$ has a spectrum that we failed to measure redshift from.

In addition, we also only use galaxies brighter than $r_o = 19.5$ to fit the model. Since the satellite rate is a strong function of the magnitude ($r_o$), we first find a linear combination of the model parameters $(g_o, r_o, r_{\text{fiber},o})$ such that the satellite rate has the least evolution with respect to magnitude. This procedure results in the specific combination we used here. In other words, we assume that when the satellite rate is modeled to depend on $(r_{\text{fiber},o} - 0.65 r_o, \, g_o-r_o)$, there will be no additional dependence on $r_o$. We verified this assumption by fitting the model to several magnitude bins.

Once the model is specified, the fitting procedure to obtain $(\beta_0, \beta_1, \beta_2, \mathcal{R}_\mathrm{max})$ is the same as \papertwo{}: we use linear regression to maximize the likelihood function. It is worth noting that we only use the redshift data we collected and compiled to fit the model. There are \textit{no} extra external constraints such as the form of the SMF.
The best-fit model parameters are $(\beta_0=-25.6, \beta_1=2.62, \beta_2=-3.84, \mathcal{R}_\text{max}=0.75)$
Panel (C) of \autoref{fig:targets} shows the prediction of the fitted model in the color--magnitude panel.

After obtaining the best-fit model, we define the incompleteness correction as follows.
For each galaxy source in our photometric catalog that does not have a confirmed redshift, we assign an  incompleteness correction weight of $\mathcal{R}_\mathrm{sat}(g_o, r_o, r_{\text{fiber},o})$ with the galaxy's photometric properties and our best-fit model. One can think of the incompleteness correction weight as the probability of that particular galaxy being a satellite. One exception is that when a galaxy has a color $(g-r)_o > 1.25$, it will be assigned no weight as it is physically too red to be within $z < 0.013$ (the furthest SAGA system). Galaxy sources that have confirmed redshifts are not assigned incompleteness correction weight, as they are either a satellite (see \autoref{sec:dr3-satellites}) or not.

To calculate the incompleteness correction for a given stellar mass bin, we can simply sum up the incompleteness correction weights of all galaxies in that stellar mass bin. Note that since we do not have the redshift measurements for these galaxies, the stellar masses used here are calculated assuming these galaxies are at the respective hosts' distances.

\section{Data Tables}
\label{app:tables}

In this section, we provide the tables for the SAGA DR3 data products. For large tables, only the table schema is shown here; their contents are available in a machine-readable format at \https{sagasurvey.org/data} and \https{doi.org/10.3847/1538-4357/ad64c4}.
\begin{enumerate}
    \item \autoref{tab:hosts} lists the schema of the SAGA DR3 Host Catalog (see \autoref{sec:hosts}).
    \item \autoref{tab:redshifts} lists the schema of the full SAGA DR3 Redshift Catalog, which includes both satellite and background galaxies (see \autoref{sec:dr3-redshifts} and \autoref{sec:dr3-background}).
    \item \autoref{tab:sats} lists the schema of the SAGA DR3 Satellite Catalog (see \autoref{sec:dr3-satellites}).
    \item \autoref{tab:candidates} lists the schema of the SAGA DR3 Likely Satellite Candidate Catalog (see \autoref{sec:dr3-candidates}).
    \item \autoref{tab:smf} provides a tabulated SMFs, including incompleteness corrected and uncorrected versions (see \autoref{sec:dr3-candidates} and \autoref{sec:smf}).  They are the same values as those shown in \autoref{fig:sat_smf} left panel.
    \item \autoref{tab:fquench} provides a tabulated satellite quenched function of stellar mass (see \autoref{sec:qf}). They are the same values as those shown in \autoref{fig:quenched_frac}.
\end{enumerate}

\setlength{\belowdeluxetableskip}{-14pt}

%%%%%%%%%%%%%%%%%%%%%%%%%%%%%%%%%%%%%%%%
% Host table schema
%%%%%%%%%%%%%%%%%%%%%%%%%%%%%%%%%%%%%%%%
\begin{deluxetable*}{llll}
\tabletypesize{\scriptsize}
\tablecaption{Schema of the SAGA DR3 Host Catalog\label{tab:hosts}}
\tablehead{\colhead{Num} & \colhead{Units} & \colhead{Label} & \colhead{Explanations}}
\startdata
\input{tables/schema_hosts}
\enddata
\tablecomments{The full contents of this table (\nhosts{} rows) are available in a machine-readable format at \https{sagasurvey.org/data} and \https{doi.org/10.3847/1538-4357/ad64c4}.}
\end{deluxetable*}
%%%%%%%%%%%%%%%%%%%%%%%%%%%%%%%%%%%%%%%%

%%%%%%%%%%%%%%%%%%%%%%%%%%%%%%%%%%%%%%%%
% Redshift table schema
%%%%%%%%%%%%%%%%%%%%%%%%%%%%%%%%%%%%%%%%
\begin{deluxetable*}{llll}
\tabletypesize{\scriptsize}
\tablecaption{Schema of SAGA DR3 Redshift Catalog\label{tab:redshifts}}
\tablehead{\colhead{Num} & \colhead{Units} & \colhead{Label} & \colhead{Explanations}}
\startdata
\input{tables/schema_redshifts}
\enddata
\tablecomments{The full contents of this table (\nztotal{} rows) are available in a machine-readable format at \https{sagasurvey.org/data} and \https{doi.org/10.3847/1538-4357/ad64c4}.}
\end{deluxetable*}
%%%%%%%%%%%%%%%%%%%%%%%%%%%%%%%%%%%%%%%%

%%%%%%%%%%%%%%%%%%%%%%%%%%%%%%%%%%%%%%%%
% Satellite table schema
%%%%%%%%%%%%%%%%%%%%%%%%%%%%%%%%%%%%%%%%
\begin{deluxetable*}{llll}
\tabletypesize{\scriptsize}
\tablecaption{Schema of SAGA DR3 Confirmed Satellite Catalog\label{tab:sats}}
\tablehead{\colhead{Num} & \colhead{Units} & \colhead{Label} & \colhead{Explanations}}
\startdata
\input{tables/schema_sats}
\enddata
\tablecomments{The full contents of this table (\nsats{} rows) are available in a machine-readable format at \https{sagasurvey.org/data} and \https{doi.org/10.3847/1538-4357/ad64c4}.}
\end{deluxetable*}
%%%%%%%%%%%%%%%%%%%%%%%%%%%%%%%%%%%%%%%%

%%%%%%%%%%%%%%%%%%%%%%%%%%%%%%%%%%%%%%%%
% Candidate table schema
%%%%%%%%%%%%%%%%%%%%%%%%%%%%%%%%%%%%%%%%
\begin{deluxetable*}{llll}
\tabletypesize{\scriptsize}
\tablecaption{Schema of SAGA DR3 Satellite Candidate Catalog\label{tab:candidates}}
\tablehead{\colhead{Num} & \colhead{Units} & \colhead{Label} & \colhead{Explanations}}
\startdata
\input{tables/schema_candidates}
\enddata
\tablecomments{The full contents of this table (\ncandidates{} rows) are available in a machine-readable format at \https{sagasurvey.org/data} and \https{doi.org/10.3847/1538-4357/ad64c4}.}
\end{deluxetable*}
%%%%%%%%%%%%%%%%%%%%%%%%%%%%%%%%%%%%%%%%

%%%%%%%%%%%%%%%%%%%%%%%%%%%%%%%%%%%%%%%%
% SMF data table
%%%%%%%%%%%%%%%%%%%%%%%%%%%%%%%%%%%%%%%%
\begin{deluxetable*}{c|cc|cc|cc}
\tabletypesize{\scriptsize}
\tablecaption{Average SAGA Satellite Stellar Mass Function\label{tab:smf}}
\tablehead{
   \colhead{} &
   \multicolumn{2}{c}{Total} &
   \multicolumn{2}{c}{Star Forming} &
   \multicolumn{2}{c}{Quenched} \\
   \cline{2-7}
  \colhead{$\log(\mstar/\msun)$} &
  \colhead{Corrected} & \colhead{Uncorrected} &
  \colhead{Corrected} & \colhead{Uncorrected} &
  \colhead{Corrected} & \colhead{Uncorrected}
}
\tablecolumns{7}
\startdata
\input{tables/data_smf}
\enddata
\tablecomments{Columns (2)--(7) show the SMF in $d \, \langle N_\text{sat} \rangle\, / \, d \, \log \mstar$.}
\end{deluxetable*}
%%%%%%%%%%%%%%%%%%%%%%%%%%%%%%%%%%%%%%%%

%%%%%%%%%%%%%%%%%%%%%%%%%%%%%%%%%%%%%%%%
% f_quenched data table
%%%%%%%%%%%%%%%%%%%%%%%%%%%%%%%%%%%%%%%%
\begin{deluxetable*}{c|ccc}
\tabletypesize{\scriptsize}
\tablecaption{Average SAGA Satellite Quenched Fractions\label{tab:fquench}}
\tablehead{
  \colhead{$\log(\mstar/\msun)$} &
  \colhead{$f_\text{quenched}$} &
  \colhead{$f_\text{quenched}^\text{(uncorrected)}$} &
  \colhead{$\sigma(f_\text{quenched})$}
}
\startdata
\input{tables/data_quenched_frac}
\enddata
\tablecomments{The quenched fraction ($f_\text{quenched}$) is calculated as $N_q/N$, where $N_q$ and $N$ are the numbers of quenched and all satellites in that mass bin. The errors are binomial (Poisson) errors $\sigma = \sqrt{p'(1-p')/N}$, where $p'$ is calculated using Maximum a Posteriori $(N_q+1)/(N+2)$.}
\end{deluxetable*}
%%%%%%%%%%%%%%%%%%%%%%%%%%%%%%%%%%%%%%%%

\end{document}

%% file: acknowledgments.tex
This work was supported by National Science Foundation (NSF) collaborative grants AST-1517148 and AST-1517422 awarded to M.G. and R.H.W. and by the Heising--Simons Foundation grant 2019-1402.
Additional support was provided by the Kavli Institute for Particle Astrophysics and Cosmology  at Stanford and SLAC.
% Personal
Support for Y.-Y.M.\ during 2019--2022 was in part provided by NASA through the NASA Hubble Fellowship grant no.\ HST-HF2-51441.001 awarded by the Space Telescope Science Institute, which is operated by the Association of Universities for Research in Astronomy, Incorporated, under NASA contract NAS5-26555. %YYM

The authors thank Tom Abel, Jenny Greene, Ananthan Karunakaran, Jose Sebastian Monzon, Ekta Patel, Frank van den Bosch, and Andrew Wetzel for helpful discussions and feedback that have improved this manuscript;
Rebecca Bernstein, Yu Lu, Phil Marshall, and Emily Sandford for contributions to the early stages of the survey;
Dustin Lang for developing and maintaining the Legacy Surveys Viewer;
Chris Lidman for guidance and support of AAT observing;
and the Center for Computational Astrophysics at the Flatiron Institute for hosting several SAGA team meetings.
Our gratitude also goes to all essential workers that support our lives and work, especially during the period of the COVID-19 pandemic when a large portion of this work was conducted.

% Observations
This work is in part based on data acquired at the Anglo-Australian Telescope (AAT), under programs A/3000 and NOAO\,0144/0267. We acknowledge the traditional custodians of the land on which the AAT stands, the Gamilaraay people, and pay our respects to elders past and present.
Observations reported here were in part obtained at the MMT Observatory, a joint facility of the Smithsonian Institution and the University of Arizona.
Some of the data presented here were obtained with the Hale Telescope at Palomar Observatory.
Some of the data presented here were obtained at W.~M.~Keck Observatory, which is a private 501(c)3 non-profit organization operated as a scientific partnership among the California Institute of Technology, the University of California, and the National Aeronautics and Space Administration. The Observatory was made possible by the generous financial support of the W.~M.~Keck Foundation. The authors wish to recognize and acknowledge the very significant cultural role and reverence that the summit of Maunakea has always had within the Native Hawaiian community. We are most fortunate to have the opportunity to conduct observations from this mountain.
This paper uses observations made with the Southern African Large Telescope (SALT) at the South African Astronomical Observatory (SAAO).

% Computing
This research made use of computational resources at SLAC National Accelerator Laboratory, a U.S.\ Department of Energy Office, and at the Sherlock
cluster at the Stanford Research Computing Center (SRCC); Y.-Y.M., R.H.W., Y.W., and E.O.N.\ are thankful for the support of the SLAC and SRCC computational teams. %SLAC and Shelock

% Legacy Imagaing
This work used public data from the Legacy Surveys.
The Legacy Surveys consist of three individual and complementary projects: the Dark Energy Camera Legacy Survey (DECaLS; Proposal ID \#2014B-0404; PIs: David Schlegel and Arjun Dey), the Beijing-Arizona Sky Survey (BASS; NOAO Prop. ID \#2015A-0801; PIs: Zhou Xu and Xiaohui Fan), and the Mayall z-band Legacy Survey (MzLS; Prop. ID \#2016A-0453; PI: Arjun Dey). DECaLS, BASS and MzLS together include data obtained, respectively, at the Blanco telescope, Cerro Tololo Inter-American Observatory, NSF's NOIRLab; the Bok telescope, Steward Observatory, University of Arizona; and the Mayall telescope, Kitt Peak National Observatory, NOIRLab. Pipeline processing and analyses of the data were supported by NOIRLab and the Lawrence Berkeley National Laboratory (LBNL). The Legacy Surveys project is honored to be permitted to conduct astronomical research on Iolkam Du'ag (Kitt Peak), a mountain with particular significance to the Tohono O'odham Nation.

NOIRLab is operated by the Association of Universities for Research in Astronomy (AURA) under a cooperative agreement with the National Science Foundation. LBNL is managed by the Regents of the University of California under contract to the U.S. Department of Energy.

This project used data obtained with the Dark Energy Camera (DECam), which was constructed by the Dark Energy Survey (DES) collaboration. Funding for the DES Projects has been provided by the U.S. Department of Energy, the U.S. National Science Foundation, the Ministry of Science and Education of Spain, the Science and Technology Facilities Council of the United Kingdom, the Higher Education Funding Council for England, the National Center for Supercomputing Applications at the University of Illinois at Urbana-Champaign, the Kavli Institute of Cosmological Physics at the University of Chicago, Center for Cosmology and Astro-Particle Physics at the Ohio State University, the Mitchell Institute for Fundamental Physics and Astronomy at Texas A\&M University, Financiadora de Estudos e Projetos, Fundacao Carlos Chagas Filho de Amparo, Financiadora de Estudos e Projetos, Fundacao Carlos Chagas Filho de Amparo a Pesquisa do Estado do Rio de Janeiro, Conselho Nacional de Desenvolvimento Cientifico e Tecnologico and the Ministerio da Ciencia, Tecnologia e Inovacao, the Deutsche Forschungsgemeinschaft and the Collaborating Institutions in the Dark Energy Survey. The Collaborating Institutions are Argonne National Laboratory, the University of California at Santa Cruz, the University of Cambridge, Centro de Investigaciones Energeticas, Medioambientales y Tecnologicas-Madrid, the University of Chicago, University College London, the DES-Brazil Consortium, the University of Edinburgh, the Eidgenossische Technische Hochschule (ETH) Zurich, Fermi National Accelerator Laboratory, the University of Illinois at Urbana-Champaign, the Institut de Ciencies de l'Espai (IEEC/CSIC), the Institut de Fisica d'Altes Energies, Lawrence Berkeley National Laboratory, the Ludwig Maximilians Universitat Munchen and the associated Excellence Cluster Universe, the University of Michigan, NSF's NOIRLab, the University of Nottingham, the Ohio State University, the University of Pennsylvania, the University of Portsmouth, SLAC National Accelerator Laboratory, Stanford University, the University of Sussex, and Texas A\&M University.

BASS is a key project of the Telescope Access Program (TAP), which has been funded by the National Astronomical Observatories of China, the Chinese Academy of Sciences (the Strategic Priority Research Program “The Emergence of Cosmological Structures” Grant \#XDB09000000), and the Special Fund for Astronomy from the Ministry of Finance. The BASS is also supported by the External Cooperation Program of Chinese Academy of Sciences (Grant \#114A11KYSB20160057), and Chinese National Natural Science Foundation (Grant \#12120101003, \#11433005).

The Legacy Survey team makes use of data products from the Near-Earth Object Wide-field Infrared Survey Explorer (NEOWISE), which is a project of the Jet Propulsion Laboratory/California Institute of Technology. NEOWISE is funded by the National Aeronautics and Space Administration.

The Legacy Surveys imaging of the DESI footprint is supported by the Director, Office of Science, Office of High Energy Physics of the U.S. Department of Energy under Contract No.\ DE-AC02-05CH1123, by the National Energy Research Scientific Computing Center, a DOE Office of Science User Facility under the same contract; and by the U.S. National Science Foundation, Division of Astronomical Sciences under Contract No.\ AST-0950945 to NOAO.

% SGA
The Siena Galaxy Atlas was made possible by funding support from the U.S. Department of Energy, Office of Science, Office of High Energy Physics under Award Number DE-SC0020086 and from the National Science Foundation under grant AST-1616414.

% DESI
This research used data obtained with the Dark Energy Spectroscopic Instrument (DESI). DESI construction and operations is managed by the Lawrence Berkeley National Laboratory. This material is based upon work supported by the U.S. Department of Energy, Office of Science, Office of High-Energy Physics, under Contract No.\ DE–AC02–05CH11231, and by the National Energy Research Scientific Computing Center, a DOE Office of Science User Facility under the same contract. Additional support for DESI was provided by the U.S. National Science Foundation (NSF), Division of Astronomical Sciences under Contract No.\ AST-0950945 to the NSF's National Optical-Infrared Astronomy Research Laboratory; the Science and Technology Facilities Council of the United Kingdom; the Gordon and Betty Moore Foundation; the Heising-Simons Foundation; the French Alternative Energies and Atomic Energy Commission (CEA); the National Council of Science and Technology of Mexico (CONACYT); the Ministry of Science and Innovation of Spain (MICINN), and by the DESI Member Institutions: www.desi.lbl.gov/collaborating-institutions. The DESI collaboration is honored to be permitted to conduct scientific research on Iolkam Du'ag (Kitt Peak), a mountain with particular significance to the Tohono O'odham Nation. Any opinions, findings, and conclusions or recommendations expressed in this material are those of the author(s) and do not necessarily reflect the views of the U.S. National Science Foundation, the U.S. Department of Energy, or any of the listed funding agencies.

%SDSS
This work used public data from the Sloan Digital Sky Survey IV.
Funding for the Sloan Digital Sky Survey IV has been provided by the  Alfred P. Sloan Foundation, the U.S.  Department of Energy Office of  Science, and the Participating  Institutions.

SDSS-IV acknowledges support and  resources from the Center for High  Performance Computing  at the  University of Utah. The SDSS  website is www.sdss4.org. SDSS-IV is managed by the

Astrophysical Research Consortium for the Participating Institutions  of the SDSS Collaboration including  the Brazilian Participation Group,  the Carnegie Institution for Science,  Carnegie Mellon University, Center for  Astrophysics | Harvard \&  Smithsonian, the Chilean Participation  Group, the French Participation Group,  Instituto de Astrof\'isica de  Canarias, The Johns Hopkins  University, Kavli Institute for the  Physics and Mathematics of the  Universe (IPMU) / University of  Tokyo, the Korean Participation Group,  Lawrence Berkeley National Laboratory,  Leibniz Institut f\"ur Astrophysik  Potsdam (AIP),  Max-Planck-Institut  f\"ur Astronomie (MPIA Heidelberg),  Max-Planck-Institut f\"ur  Astrophysik (MPA Garching),  Max-Planck-Institut f\"ur  Extraterrestrische Physik (MPE),  National Astronomical Observatories of  China, New Mexico State University,  New York University, University of  Notre Dame, Observat\'ario  Nacional / MCTI, The Ohio State  University, Pennsylvania State  University, Shanghai  Astronomical Observatory, United  Kingdom Participation Group,  Universidad Nacional Aut\'onoma  de M\'exico, University of Arizona,  University of Colorado Boulder,  University of Oxford, University of  Portsmouth, University of Utah,  University of Virginia, University  of Washington, University of  Wisconsin, Vanderbilt University,  and Yale University.

% GALEX
This work used public data from the GALEX Survey.
GALEX is operated for NASA by the California Institute of Technology under NASA contract NAS5-98034. 

% GAMA
This work used public data from the GAMA Survey.
GAMA is a joint European-Australasian project based around a spectroscopic campaign using the Anglo-Australian Telescope. The GAMA input catalogue is based on data taken from the Sloan Digital Sky Survey and the UKIRT Infrared Deep Sky Survey. Complementary imaging of the GAMA regions is being obtained by a number of independent survey programmes including GALEX MIS, VST KiDS, VISTA VIKING, WISE, Herschel-ATLAS, GMRT and ASKAP providing UV to radio coverage. GAMA is funded by the STFC (UK), the ARC (Australia), the AAO, and the participating institutions. The GAMA website is http://www.gama-survey.org/ .

% PRIMUS
This work used public data from PRIMUS.
Funding for PRIMUS is provided by NSF (AST-0607701, AST-0908246, AST-0908442, AST-0908354) and NASA (Spitzer-1356708, 08-ADP08-0019, NNX09AC95G).

% VIPERS
This paper uses data from the VIMOS Public Extragalactic Redshift Survey (VIPERS). VIPERS has been performed using the ESO Very Large Telescope, under the "Large Programme" 182.A-0886. The participating institutions and funding agencies are listed at http://vipers.inaf.it

% HETDEX
This work used public data from HETDEX.
HETDEX is led by the University of Texas at Austin McDonald Observatory and Department of Astronomy with participation from the Ludwig-Maximilians-Universit\"at M\"unchen, Max-Planck-Institut f\"ur Extraterrestrische Physik (MPE), Leibniz-Institut f\"ur Astrophysik Potsdam (AIP), Texas A\&M University, Pennsylvania State University, Institut f\"ur Astrophysik G\"ottingen, The University of Oxford, Max-Planck-Institut f\"ur Astrophysik (MPA), The University of Tokyo and Missouri University of Science and Technology.

Observations for HETDEX were obtained with the Hobby-Eberly Telescope (HET), which is a joint project of the University of Texas at Austin, the Pennsylvania State University, Ludwig-Maximilians-Universit\"at M\"unchen, and Georg-August-Universit\"at G\"ottingen. The HET is named in honor of its principal benefactors, William P. Hobby and Robert E. Eberly. The Visible Integral-field Replicable Unit Spectrograph (VIRUS) was used for HETDEX observations. VIRUS is a joint project of the University of Texas at Austin, Leibniz-Institut f\"ur Astrophysik Potsdam (AIP), Texas A\&M University, Max-Planck-Institut f\"urExtraterrestrische Physik (MPE), Ludwig-Maximilians-Universit\"at M\"unchen, Pennsylvania State University, Institut f\"ur Astrophysik G\"ottingen, University of Oxford, and the Max-Planck-Institut fur Astrophysik (MPA).

The authors acknowledge the Texas Advanced Computing Center (TACC) at The University of Texas at Austin for providing high performance computing, visualization, and storage resources that have contributed to the research results reported within this paper. URL: http://www.tacc.utexas.edu

Funding for HETDEX has been provided by the partner institutions, the National Science Foundation, the State of Texas, the US Air Force, and by generous support from private individuals and foundations.

% Others
This work used public data from the NASA-Sloan Altas, the WiggleZ Dark Energy Survey, the 2dF Galaxy Redshift Survey, the HectoMAP Redshift Survey, the 6dF Galaxy Survey, the Hectospec Cluster Survey, the Australian Dark Energy Survey, the 2-degree Field Lensing
Survey, and the Las Campanas Redshift Survey.
The authors acknowledge the people who contributed to conducting these surveys and making the resulting data public, and the funding agencies who supported the work.

% Databases
We acknowledge the usage of the HyperLeda database (\http{leda.univ-lyon1.fr}) and the Extragalactic Distance Database (EDD; \http{edd.ifa.hawaii.edu}). Support for the development of content for the EDD is provided by the National Science Foundation under Grant No.\ AST09-08846.
This research has made use of NASA's Astrophysics Data System. % ADS

%% file: tables/schema_hosts.tex
1 & $\cdots$ & Name & Host common name \\
2 & $\cdots$ & HOSTID & Unique host identifier that can be used to join the satellite table \\
3 & $\cdots$ & PGC & PGC number from HyperLEDA \\
4 & deg & RAdeg & Right ascension in decimal degrees (J2000) \\
5 & deg & DEdeg & Declination in decimal degrees (J2000) \\
6 & km/s & HRV & Heliocentric recession velocity of the host \\
7 & Mpc & Dist & Distance to the host \\
8 & mag & DistMod & Distance modulus \\
9 & mag & KsMag & Absolute $K_s$-band luminosity, extinction corrected and K-corrected \\
10 & dex(\msun{}) & log(Mhalo) & Log of halo mass from \citet{2017MNRAS.470.2982L} \\
11 & dex(\msun{}) & log(MHI) & Log of HI mass \\
12 & mag & rmag & Apparent $r$-band magnitude, extinction corrected, derived from DESI Legacy Imaging \\
13 & mag & gr & $g-r$ color, extinction corrected, derived from DESI Legacy Imaging \\
14 & mag/arcsec$^2$ & sb & Effective surface brightness, derived from DESI Legacy Imaging \\
15 & $\cdots$ & ba & Galaxy axis ratio, taken from DESI Legacy Imaging \\
16 & deg & PA & Galaxy position angle, taken from DESI Legacy Imaging \\
17 & $\cdots$ & Sersic & Galaxy Sersic index, taken from DESI Legacy Imaging \\
18 & dex(\msun{}) & log(M*) & Log of estimated galaxy stellar mass \\
19 & dex(\msun{}/yr) & log(sfr) & Log of star formation rate using NUV measurements \\
20 & $\cdots$ & HOSTID-MW & HOSTID of nearest MW-mass host \\
21 & Mpc & sep-MW & Physical distance to nearest MW-mass host \\
22 & $\cdots$ & count-MW & Number of MW-mass hosts within 1 Mpc, not including self \\
23 & Mpc & sep-massive & Physical distance to the nearest massive host \\
24 & $\cdots$ & nz-saga & Number of galaxy redshifts obtained by SAGA within 300 kpc in projection \\
25 & $\cdots$ & nz-total & Number of total galaxy redshifts within 300 kpc in projection \\
26 & $\cdots$ & nsat-G & Number of satellites in the Gold sample, with incompleteness correction \\
27 & $\cdots$ & nsat-GS & Number of satellites in the Gold$+$Silver sample, with incompleteness correction \\
28 & $\cdots$ & nsat-Gc & Number of confirmed satellites in the Gold sample \\
29 & $\cdots$ & nsat-GSc & Number of confirmed satellites in the Gold$+$Silver sample \\
30 & $\cdots$ & nsat-GSPc & Number of confirmed satellites in all three samples: Gold, Silver, Participation \\
31 & dex(\msun{}) & log(M*)-1s & Log of stellar mass for most massive confirmed satellite 

%% file: tables/schema_redshifts.tex
1 & $\cdots$ & OBJID & Unique galaxy identifier \\
2 & $\cdots$ & HOSTID & Unique host identifier, can be used to join the host table \\
3 & $\cdots$ & PGC & PGC number of galaxy, if available \\
4 & deg & RAdeg & Right ascension in decimal degrees (J2000) \\
5 & deg & DEdeg & Declination in decimal degrees (J2000) \\
6 & mag & rmag & Apparent $r$-band magnitude, extinction corrected, derived from DESI Legacy Imaging \\
7 & mag & e\_rmag & Uncertainty in rmag, derived from DESI Legacy Imaging \\
8 & mag & gr & g-r color, extinction corrected, derived from DESI Legacy Imaging \\
9 & mag & rmag-fiber & Apparent $r$-band fiber magnitude, 1.5-arcsec diameter, extinction corrected, derived from DESI Legacy Imaging \\
10 & mag/arcsec$^2$ & sb & Effective surface brightness, derived from DESI Legacy Imaging \\
11 & $\cdots$ & ba & Galaxy axis ratio, taken from DESI Legacy Imaging \\
12 & deg & PA & Galaxy position angle, taken from DESI Legacy Imaging \\
13 & $\cdots$ & Sersic & Galaxy Sersic index, taken from DESI Legacy Imaging \\
14 & $\cdots$ & TELNAME & Primary source for spectroscopic redshift \\
15 & $\cdots$ & z & Spectroscopic redshift, $-1$ if not measured \\
16 & dex(\msun{}) & log(M*) & Log of estimated galaxy stellar mass \\
17 & $\cdots$ & sample & If greater than 0, this entry is a satellite and is repeated in the satellite table.

%% file: tables/schema_sats.tex
1 & $\cdots$ & OBJID & Unique satellite identifier \\
2 & $\cdots$ & HOSTID & Unique host identifier, can be used to join the host table \\
3 & $\cdots$ & PGC & PGC number of satellite, if available \\
4 & deg & RAdeg & Right ascension in decimal degrees (J2000) \\
5 & deg & DEdeg & Declination in decimal degrees (J2000) \\
6 & kpc & Rhost & Projected radial distance of the object to the host \\
7 & mag & rmag & Apparent $r$-band magnitude, extinction corrected, derived from DESI Legacy Imaging \\
8 & mag & e\_rmag & Uncertainty in rmag, derived from DESI Legacy Imaging \\
9 & mag & gr & g-r color, extinction corrected, derived from DESI Legacy Imaging \\
10 & mag & rmag-fiber & Apparent $r$-band fiber magnitude, 1.5-arcsec diameter, extinction corrected, derived from DESI Legacy Imaging \\
11 & mag/arcsec$^2$ & sb & Effective surface brightness, derived from DESI Legacy Imaging \\
12 & $\cdots$ & ba & Galaxy axis ratio, taken from DESI Legacy Imaging \\
13 & deg & PA & Galaxy position angle, taken from DESI Legacy Imaging \\
14 & $\cdots$ & Sersic & Galaxy Sersic index, taken from DESI Legacy Imaging \\
15 & $\cdots$ & TELNAME & Primary source for spectroscopic redshift \\
16 & $\cdots$ & z & Spectroscopic redshift, $-1$ if not measured \\
17 & km/s & DVhost & Difference in heliocentric velocity with respect to the host galaxy \\
18 & dex(\msun{}) & log(M*) & Log of estimated stellar mass \\
19 & 10$^{-10}$ m & EW-Halpha & H$_\alpha$ equivalent width, measured from SAGA spectra \\
20 & 10$^{-10}$ m & e\_EW-Halpha & Uncertainty in EW-Halpha \\
21 & dex(\msun{}/yr) & log(sfr)Halpha & Log of star formation rate using H$_\alpha$ measurements \\
22 & dex(\msun{}/yr) & e\_log(sfr)Halpha & Uncertainty in log(sfr)Halpha \\
23 & mag & NUVmag & Apparent GALEX NUV magnitude, extinction corrected \\
24 & mag & e\_NUVmag & Uncertainty in NUVmag \\
25 & $\cdots$ & f\_NUVmag & Flag on NUVmag: 0 = undetected (upper limit only); 1 = detected; $-1$ = no GALEX coverage. \\
26 & dex(\msun{}/yr) & log(sfr)NUV & Log of star formation rate using NUV measurements \\
27 & dex(\msun{}/yr) & e\_log(sfr)NUV & Uncertainty in log(sfr)NUV \\
28 & $\cdots$ & quenched & Quenched or star forming: 0 = star forming; 1 = quenched. \\
29 & dex(\msun{}) & log(MHI) & Log of HI mass \\
30 & $\cdots$ & HIsource & Source of HI flux measurement \\
31 & 10$^{-19}$ W/(nm m$^2$) & flux-Halpha & Flux of H$_\alpha$ line \\
32 & 10$^{-19}$ W/(nm m$^2$) & e\_flux-Halpha & Uncertainty in flux-Halpha (standard deviation) \\
33 & 10$^{-19}$ W/(nm m$^2$) & flux-Hbeta & Flux of H$_\beta$ line \\
34 & 10$^{-19}$ W/(nm m$^2$) & e\_flux-Hbeta & Uncertainty in flux-Hbeta (standard deviation) \\
35 & 10$^{-19}$ W/(nm m$^2$) & flux-NII & Flux of [N II] 6853\,\AA{} line \\
36 & 10$^{-19}$ W/(nm m$^2$) & e\_flux-NII & Uncertainty in flux-NII (standard deviation) \\
37 & 10$^{-19}$ W/(nm m$^2$) & flux-OIII & Flux of [O III] 5007\,\AA{} line \\
38 & 10$^{-19}$ W/(nm m$^2$) & e\_flux-OIII & Uncertainty in flux-OIII (standard deviation) \\
39 & 10$^{-19}$ W/(nm m$^2$) & flux-SII6717 & Flux of [S II] 6717\,\AA{} line \\
40 & 10$^{-19}$ W/(nm m$^2$) & e\_flux-SII6717 & Uncertainty in flux-SII6717 (standard deviation) \\
41 & 10$^{-19}$ W/(nm m$^2$) & flux-SII6731 & Flux of [S II] 6731\,\AA{} line \\
42 & 10$^{-19}$ W/(nm m$^2$) & e\_flux-SII6731 & Uncertainty in flux-SII6731 (standard deviation) \\
43 & $\cdots$ & flux-note & Note on line flux: 0 = unverified flux calibration; 1 = flux calibration verified in the \citet{2401.16469} pipeline. \\
44 & $\cdots$ & sample & DR3 sample: 1 = Gold; 2 = Silver; 3 = Participation. 

%% file: tables/schema_candidates.tex
1 & $\cdots$ & OBJID & Unique galaxy identifier \\
2 & $\cdots$ & HOSTID & Unique host identifier, can be used to join the host table \\
3 & deg & RAdeg & Right ascension in decimal degrees (J2000) \\
4 & deg & DEdeg & Declination in decimal degrees (J2000) \\
5 & kpc & Rhost & Projected radial distance of the object to the host \\
6 & mag & rmag & Apparent $r$-band magnitude, extinction corrected, derived from DESI Legacy Imaging \\
7 & mag & e\_rmag & Uncertainty in rmag, derived from DESI Legacy Imaging \\
8 & mag & gr & g-r color, extinction corrected, derived from DESI Legacy Imaging \\
9 & mag & rmag-fiber & Apparent $r$-band fiber magnitude, 1.5 arcsec diameter, extinction corrected, derived from DESI Legacy Imaging \\
10 & mag/arcsec$^2$ & sb & Effective surface brightness, derived from DESI Legacy Imaging \\
11 & $\cdots$ & ba & Galaxy axis ratio, taken from DESI Legacy Imaging \\
12 & deg & PA & Galaxy position angle, taken from DESI Legacy Imaging \\
13 & $\cdots$ & Sersic & Galaxy Sersic index, taken from DESI Legacy Imaging \\
14 & dex(\msun{}) & log(M*) & Log of estimated galaxy stellar mass, if candidate is at the same distance as the host \\
15 & $\cdots$ & quenched & Star forming or quenched, estimated purely based on its color: 0 = star forming; 1 = quenched. \\
16 & $\cdots$ & Psat & Probability of being a satellite, based on the incompleteness correction model 

%% file: tables/data_smf.tex
10.31 & 0.026 & 0.026 & 0.000 & 0.000 & 0.026 & 0.026 \\
\phn9.94 & 0.185 & 0.185 & 0.106 & 0.106 & 0.079 & 0.079 \\
\phn9.56 & 0.475 & 0.475 & 0.449 & 0.449 & 0.026 & 0.026 \\
\phn9.19 & 0.475 & 0.475 & 0.449 & 0.449 & 0.027 & 0.026 \\
\phn8.81 & 0.820 & 0.818 & 0.713 & 0.713 & 0.107 & 0.106 \\
\phn8.44 & 1.041 & 1.030 & 0.818 & 0.818 & 0.222 & 0.211 \\
\phn8.06 & 1.524 & 1.452 & 1.135 & 1.135 & 0.389 & 0.317 \\
\phn7.69 & 2.304 & 1.954 & 1.417 & 1.399 & 0.887 & 0.554 \\
\phn7.31 & 3.111 & 1.927 & 1.312 & 1.241 & 1.798 & 0.686 \\
\phn6.94 & 5.950 & 1.162 & 1.218 & 0.950 & 4.732 & 0.211 \\
\phn6.56 & 4.376 & 0.317 & 1.306 & 0.238 & 3.071 & 0.079 

%% file: tables/data_quenched_frac.tex
9.38 & 0.135 & 0.135 & 0.049 \\
8.62 & 0.160 & 0.157 & 0.053 \\
8.15 & 0.259 & 0.235 & 0.061 \\
7.83 & 0.335 & 0.269 & 0.063 \\
7.57 & 0.408 & 0.275 & 0.061 \\
7.31 & 0.562 & 0.333 & 0.056 \\
6.95 & 0.792 & 0.250 & 0.026 